\documentclass[aps,prc,preprint,groupedaddress]{revtex4-2}

\usepackage{graphicx}
\usepackage{bm}
\usepackage{amsmath}
\usepackage{amssymb}
\usepackage{wasysym}

\usepackage{soulpos}
\usepackage[svgnames]{xcolor}
\ulposdef{\hlst}{%
    \rlap{\textcolor{yellow}{\rule[-.75ex]{\ulwidth}{2.5ex}}}%
    \rule[.45ex]{\ulwidth}{.1ex}%
}

\usepackage{graphicx}
\usepackage{dcolumn}
\usepackage{bm}

\begin{document}

\preprint{00000}

\title{Determination of proton and neutron contributions to the $0_{gs}^+ \rightarrow 2_1^+$ excitations in $^{42}$Si and $^{44}$S using inelastic proton scattering in inverse kinematics and intermediate energy Coulomb excitation} 

\author{L.A. Riley}
\affiliation{Department of Physics and Astronomy, Ursinus College, Collegeville, Pennsylvania 19426}%

\author{P. D. Cottle} \affiliation{Department of Physics, Florida State University, Tallahassee, Florida 32306, USA}

\author{M. Spieker} \affiliation{Department of Physics, Florida State University, Tallahassee, Florida 32306, USA}

\author{A. Volya} \affiliation{Department of Physics, Florida State University, Tallahassee, Florida 32306, USA}

\author{I. Conroy}
\affiliation{Department of Physics and Astronomy, Ursinus College, Collegeville, Pennsylvania 19426}

\author{A. M. Himmelreich}
\affiliation{Department of Physics and Astronomy, Ursinus College, Collegeville, Pennsylvania 19426}

\author{Sk M. Ali}
\affiliation{Facility for Rare Isotope Beams, Michigan State University, East Lansing, Michigan 48824, USA}

\author{T. Beck}
\affiliation{Facility for Rare Isotope Beams, Michigan State University, East Lansing, Michigan 48824, USA}

\author{J. Chung-Jung}
\affiliation{Facility for Rare Isotope Beams and Department of Physics and Astronomy, Michigan State University, East Lansing, Michigan 48824, USA}

\author{A. L. Conley} \affiliation{Department of Physics, Florida State University, Tallahassee, Florida 32306, USA}

\author{P. Farris}
\affiliation{Facility for Rare Isotope Beams and Department of Physics and Astronomy, Michigan State University, East Lansing, Michigan 48824, USA}

\author{A. Gade}
\affiliation{Facility for Rare Isotope Beams and Department of Physics and Astronomy, Michigan State University, East Lansing, Michigan 48824, USA}

\author{S. A. Gillespie}
\affiliation{Facility for Rare Isotope Beams, Michigan State University, East Lansing, Michigan 48824, USA}

\author{G. Grauvogel}
\affiliation{Facility for Rare Isotope Beams and Department of Physics and Astronomy, Michigan State University, East Lansing, Michigan 48824, USA}

\author{M. Hausmann}
\affiliation{Facility for Rare Isotope Beams, Michigan State University, East Lansing, Michigan 48824, USA}

\author{M. Heinze}
\affiliation{Department of Physics and Astronomy, Ursinus College, Collegeville, Pennsylvania 19426}

\author{A. M. Hill}
\affiliation{Facility for Rare Isotope Beams and Department of Physics and Astronomy, Michigan State University, East Lansing, Michigan 48824, USA}

\author{D. Houlihan} \affiliation{Department of Physics, Florida State University, Tallahassee, Florida 32306, USA}

\author{B. Kelly} \affiliation{Department of Physics, Florida State University, Tallahassee, Florida 32306, USA}

\author{K. W. Kemper} \affiliation{Department of Physics, Florida State University, Tallahassee, Florida 32306, USA}

\author{J. Kosa}
\affiliation{Department of Physics and Astronomy, Ursinus College, Collegeville, Pennsylvania 19426}

\author{B. Longfellow}
\affiliation{Lawrence Livermore National Laboratory, Livermore, California 94550, USA}

\author{B. McNulty}
\affiliation{Department of Physics and Astronomy, Ursinus College, Collegeville, Pennsylvania 19426}

\author{S. Noji}
\affiliation{Facility for Rare Isotope Beams, Michigan State University, East Lansing, Michigan 48824, USA}

\author{N. D. Pathirana}
\affiliation{Facility for Rare Isotope Beams, Department of Physics and Astronomy, and Joint Institute for Nuclear Astrophysics - Center for the   Evolution of the Elements, Michigan State University, East Lansing, MI 48824, USA}

\author{J. Pereira}
\affiliation{Facility for Rare Isotope Beams, Michigan State University, East Lansing, Michigan 48824, USA}

\author{Z. Rahman}
\affiliation{Facility for Rare Isotope Beams and Department of Physics and Astronomy, Michigan State University, East Lansing, Michigan 48824, USA}

\author{D. Weisshaar}
\affiliation{Facility for Rare Isotope Beams, Michigan State University, East Lansing, Michigan 48824, USA}

\author{R. G. T. Zegers}
\affiliation{Facility for Rare Isotope Beams and Department of Physics and Astronomy, Michigan State University, East Lansing, Michigan 48824, USA}

\date{July 24, 2025} 

\begin{abstract}
We have measured the $0_{\mathrm{g.s.}}^+ \rightarrow 2_1^+$ transition in the neutron rich $N=28$ isotope $^{42}$Si using the probes of intermediate energy Coulomb excitation and inelastic proton scattering in inverse kinematics at the Facility for Rare Isotope Beams with beam particle rates of $\approx 5$ particles/s.  The results of these two measurements allowed us to determine $M_n/M_p$, the ratio of the neutron and proton transition matrix elements for the $0_{\mathrm{g.s.}}^+ \rightarrow 2_1^+$ transition.  In addition, we have measured the $0_{\mathrm{g.s.}}^+ \rightarrow 2_1^+$ transition in the isotone $^{44}$S using inverse kinematics inelastic proton scattering.  By comparing the $^{44}$S proton scattering result with a recent intermediate energy Coulomb excitation result on the same transition, we were able to determine $M_n/M_p$ for the $0_{\mathrm{g.s.}}^+ \rightarrow 2_1^+$ transition in this nucleus as well.  This work strengthens the evidence that $^{42}$Si has a stable quadrupole deformation in its ground state and that $^{44}$S does not. Both conclusions are further supported by shell model calculations carried out with the FSU interaction.    
\end{abstract}

\maketitle


\section{Introduction}

While $N=28$ is a major shell closure in stable nuclei, Werner \textit{et al.} \cite{We94} predicted in 1996 that the $N=28$ shell closure would become less effective in neutron rich nuclei and that $N=28$ isotopes including $^{44}$S would be collective.  This prediction fueled an experimental drive to measure the structure of $N=28$ isotones with $Z<20$.  

Shortly after the prediction of Werner \textit{et al.} was published, Glasmacher \textit{et al.} \cite{Gl97} measured both the energy and the $B(E2;0_{\mathrm{g.s.}}^+ \rightarrow 2_1^+)$ value of the $2_1^+$ state in $^{44}$S using the technique of intermediate energy Coulomb excitation, although the $B(E2;0_{\mathrm{g.s.}}^+ \rightarrow 2_1^+)$ value was corrected significantly downward more recently \cite{Lo21}.  The energy of the $2_1^+$ state and $B(E2;0_{\mathrm{g.s.}}^+ \rightarrow 2_1^+)$ value measured by Glasmacher \textit{et al.} were consistent with collectivity in this nucleus and seemed to confirm the prediction of Werner \textit{et al.}. A subsequent measurement of the energy of the $2_1^+$ state in the $N=28$ isotone $^{42}$Si \cite{Ba07} implied that this nucleus is even more collective than $^{44}$S.  In addition to these early studies, a tremendous amount of experimental effort has been invested in probing $^{42}$Si, $^{44}$S and their neighbors (for example, see Ref. \cite{Gad19} and references cited therein).  

The $B(E2;0_{\mathrm{g.s.}}^+ \rightarrow 2_1^+)$ value provides information on the contribution of protons to the $0_{\mathrm{g.s.}}^+ \rightarrow 2_1^+$ transition (that is, the proton transition matrix element $M_p$).  However, it does not give the neutron contribution, the neutron transition matrix element $M_n$.  Knowing both $M_p$ and $M_n$ is important because the ratio $M_n/M_p$ provides important nuclear structure information that $M_p$ cannot by itself provide.  For example, this ratio gives a way of distinguishing between collective open-shell nuclei and single closed shell nuclei.  As Bernstein, Brown and Madsen \cite{Be81} pointed out, a nucleus in which the $2_1^+$ state is a collective vibration with the proton and neutron fluids oscillating with the same amplitude has $M_n/M_p=N/Z$.  A nucleus with a closed proton shell will have $M_n/M_p > N/Z$ because valence neutrons contribute disproportionately to the $0_{\mathrm{g.s.}}^+ \rightarrow 2_1^+$ transition.  Likewise, a nucleus with a closed neutron shell will have $M_n/M_p < N/Z$.  

According to Ref. \cite{Be81}, $M_n$ for the $0_{\mathrm{g.s.}}^+ \rightarrow 2_1^+$ transition can be determined by using a hadronic probe, such as inelastic proton scattering, to excite the $2_1^+$ state.  The hadronic probe drives both neutron and proton contributions in the excitation, so a comparison of the deformation length for a hadronic probe to the $M_p$ value taken from an electromagnetic measurement allows the determination of $M_n/M_p$.

In the present article, we describe in detail the techniques deployed on the two measurements of $^{42}$Si, both of which were challenged by low beam rates of less than 10 particles per second.  In addition, we report on an inverse kinematics proton scattering experiment on the $0_{\mathrm{g.s.}}^+ \rightarrow 2_1^+$ excitation in $^{44}$S, which together with the intermediate energy Coulomb excitation measurement reported in Ref. \cite{Lo21} allows a determination of $M_n/M_p$.

\section{Experiments}

The experiments were performed at the Facility for Rare Isotope Beams at Michigan State University (FRIB)~\cite{FRIB}. All secondary beams were produced by fragmentation of a $^{48}$Ca$^{20+}$ primary beam.  The production target consisted of an 8-mm thick graphite wheel rotating at 500 rpm. Each secondary beam was separated in the Advanced Rare Isotope Separator (ARIS)~\cite{ARIS1, ARIS2} with an aluminum wedge and a degrader.  Details about the production of the beams (including primary beam energy and power), wedge and degrader thicknesses as well as the momentum spread, purity, midtarget energy and speed of the secondary beams are given in Table \ref{tab:beams}. 

\begin{table*}
  \caption{\label{tab:beams} Energy and power of the primary beam and effective thicknesses of the aluminum wedges and degraders used in ARIS, momentum spreads $\Delta p/p$, purities, midtarget energies, velocities used in Doppler reconstruction of $\gamma$-ray spectra, and total yields of the secondary beams. The measurements described in the present work were made in two parts: (A) Coulomb excitation and (B) inverse kinematics proton scattering.}
  \begin{ruledtabular}
    \begin{tabular}{ccc|ccccccc}
            & Primary Energy &  Power & Secondary & Wedge & Degrader & $\Delta p/p$ & Purity & Midtarget Energy & Doppler \\
      Part & (MeV/nucleon)  & (kW)   &   Beam    & (mm)  &   (mm)   &     (\%)     &  (\%)  & (MeV/nucleon)     & $v/c$ \\\hline
      A     &     217        &  5     & $^{42}$Si & 3.00  &  11.64   &  2.0         &   65   & 76.0              & 0.357 \\ \hline
      B     &     225        & 10     & $^{42}$Si & 3.27  &  11.77   &  4.4         &   22   & 91.2              & 0.391 \\ 
            &                &        & $^{44}$S  & 1.49  &   9.69   &  0.74        &   93   & 82.3              & 0.397 \\
    \end{tabular}
  \end{ruledtabular}
\end{table*}

For the intermediate energy Coulomb excitation measurement, the $^{42}$Si secondary beam was delivered to a $980~\mathrm{mg/cm^2}$ $^{209}$Bi reaction target.  The beam rate was $\approx 3$~particles/s and a total of $1.25 \times 10^6$ $^{42}$Si nuclei were delivered to the reaction target.  

The NSCL/Ursinus College liquid hydrogen target was used for the inverse kinematics proton scattering measurements.  The target consisted of a cylindrical aluminum liquid hydrogen cell, with 125-$\mu$m Kapton entrance and exit windows and a nominal thickness of 30 mm, which was mounted on a cryocooler.  The target cell and the cryocooler were surrounded by a 1-mm-thick aluminum radiation shield with entrance and exit windows covered by 5-$\mu\mathrm{m}$ aluminized Mylar foil.  For the $^{42}$Si($p,p'$) measurement, the $^{42}$Si beam rate was $\approx 7$~particles/s.  A total of $1.04 \times 10^6$ $^{42}$Si nuclei were delivered to the target.  The $^{44}$S beam rate was $\approx 5000$~particles/s, and a total of $1.89 \times 10^7$ $^{44}$S nuclei impinged on the target.

\begin{figure}
  \includegraphics[scale=0.55]{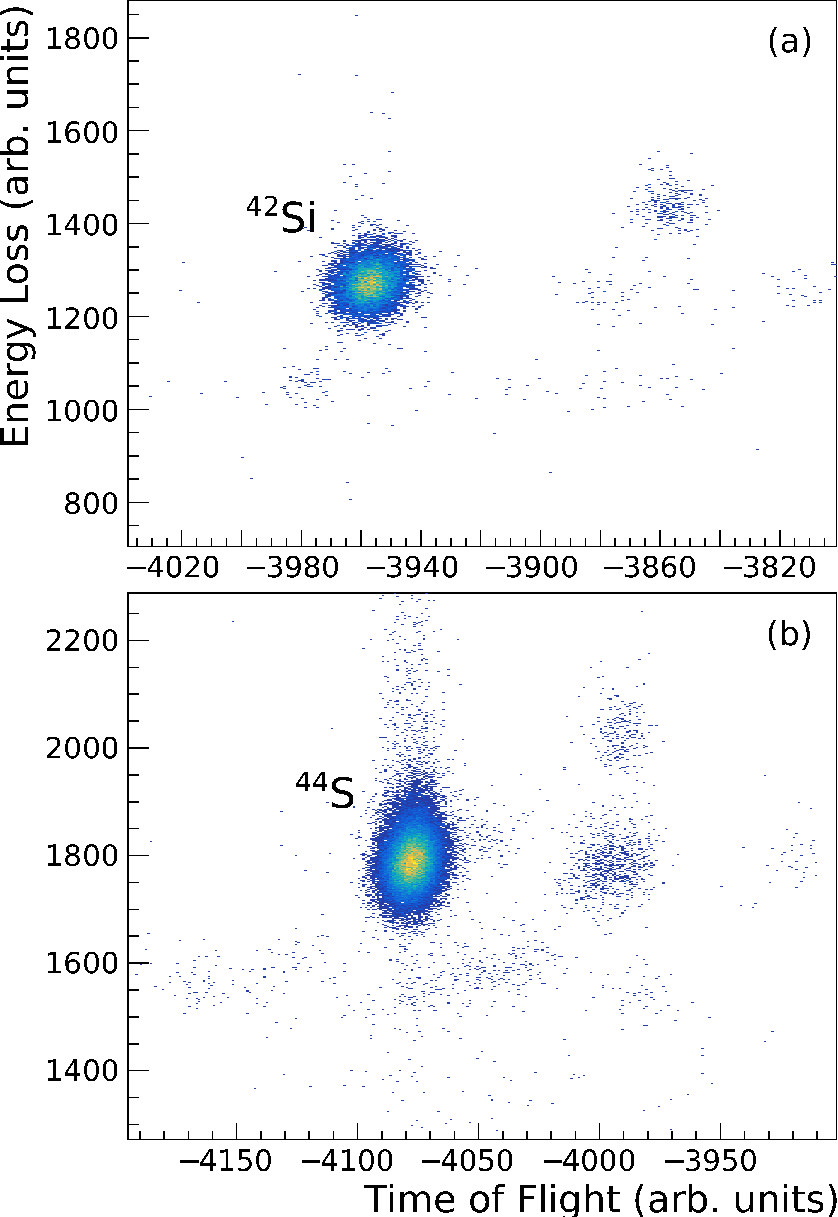}
  \caption{\label{fig:pid} (Color online) Particle identification spectrum of the outgoing beam-like reaction products from the inverse kinematics proton scattering measurements of $^{42}$Si (a) and $^{44}$S in coincidence with incoming particle identification cuts.}
\end{figure}

Particle identification was performed upstream of the reaction target to ensure that reaction products from the intermediate energy Coulomb excitation and inelastic proton scattering reactions could be distinguished from products of proton knockout reactions from other isotopes in the beam cocktail.  To do this, secondary beam particles were identified using times of flight from plastic scintillator timing detectors located at the final focal plane of ARIS and at the object of the analysis line of the S800 spectrograph (S800)~\cite{S800}. Beam-like reaction products were identified by time of flight from the S800 object scintillator and energy loss in the S800 ionization chamber. To limit the count rate of triggers, all timing measurements were started by a plastic scintillator detector in the focal plane of the S800 and stopped by delayed signals from upstream detectors. Particle identification spectra of the $^{42}$Si and $^{44}$S beams collected in coincidence with incoming particle identification cuts appear in Figure~\ref{fig:pid}.

In $^{44}$S, there is a $0^+$ 2.6-$\mu s$ isomer at 1365 keV that decays both by an $E0$ transition to the ground state and a 36-keV $\gamma$ ray to the $2_1^+$ state \cite{Fo10}.  The branching ratio for the $\gamma$-ray decay to the $2_1^+$ state is 14\%.  To properly understand the yield of $2_1^+ \rightarrow 0_{\mathrm{g.s.}}^+$ $\gamma$-rays in the $^{44}$S($p,p'$) measurement, we must understand the population of this isomer in the $^{44}$S beam.  A reexamination of the data from the $^{44}$S intermediate-energy Coulomb excitation measurement reported in Ref. \cite{Lo21} provides a means to gain some insight about the isomeric content of the beam in that experiment \cite{LoPC}.  During the experiment of Ref. \cite{Lo21}, a hodoscope in the focal plane of the S800 magnetic spectrograph was used to collect delayed $\gamma$-rays.  If 1\% of the $^{44}$S nuclei in the beam were in the isomeric state, then the experimenters would have expected approximately 2000 $2_1^+ \rightarrow 0_{\mathrm{g.s.}}^+$ 1329 keV $\gamma$-ray counts in the hodoscope spectrum.  Such a peak would have been easily discerned in that spectrum.  However, no such peak was evident.  Therefore, we can conclude that the isomeric content of the $^{44}$S beam in the experiment of Ref. \cite{Lo21} was less than 1\%.  The $^{44}$S beam energy in the experiment of Ref. \cite{Lo21} was 73 MeV/nucleon, not far from the 82.3 MeV/nucleon $^{44}$S beam energy in the present experiment.  Therefore, we assume that the isomeric content of the $^{44}$S in the present experiment was less than 1\%, as it was in the experiment of \cite{Lo21}.

\begin{figure}
  \includegraphics[scale=0.62]{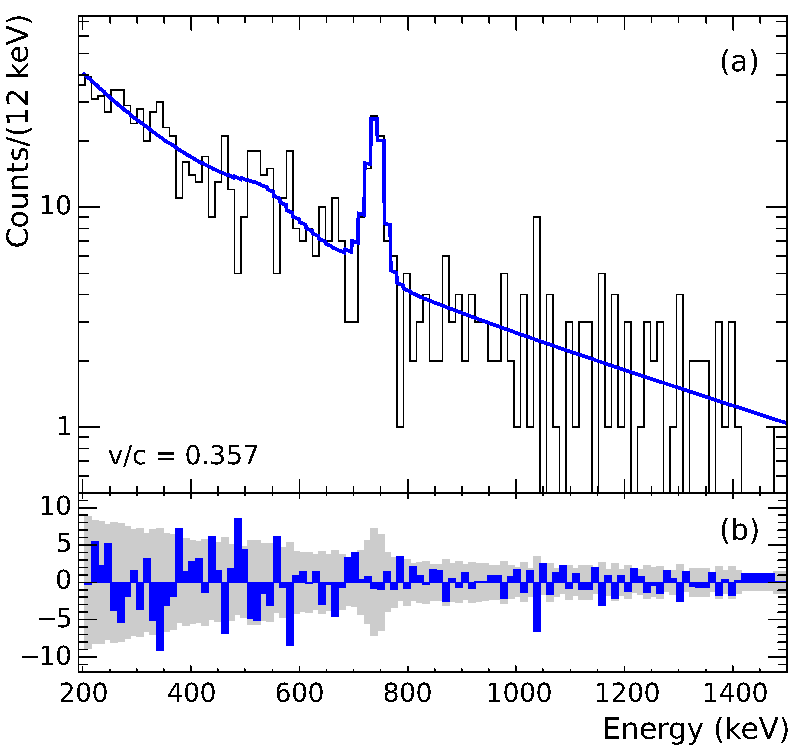}
  \caption{\label{fig:coulex_spectrum} (Color online) (a) Doppler corrected spectrum of $\gamma$ rays collected in coincidence with incoming and outgoing $^{42}$Si particles passing through the $^{209}$Bi reaction target for the intermediate energy Coulomb excitation experiment. The solid curve is the best fit obtained with UCGretina \cite{Ril21} described in the text. (b) The spectrum of residuals between the fit and the measured spectrum. The shaded region covers $\pm$ the square root of the sum of the fit and the measured counts in each bin.}
\end{figure}

During both the intermediate energy Coulomb excitation and inelastic proton scattering measurements, the GRETINA $\gamma$-ray tracking array~\cite{GRETINA,GRETINA2} was used. All 12 GRETINA modules were used during the intermediate energy Coulomb excitation measurement of $^{42}$Si.  During that measurement, four modules were centered at $58^\circ$ and eight modules were centered at $90^\circ$ with respect to the beam axis. Figure~\ref{fig:coulex_spectrum} displays the Doppler reconstructed $\gamma$-ray spectrum collected in coincidence with both incoming and outgoing $^{42}$Si particle identification cuts and a prompt timing cut between GRETINA and a timing scintillator in the S800 focal plane.

\begin{figure}
  \includegraphics[scale=0.62]{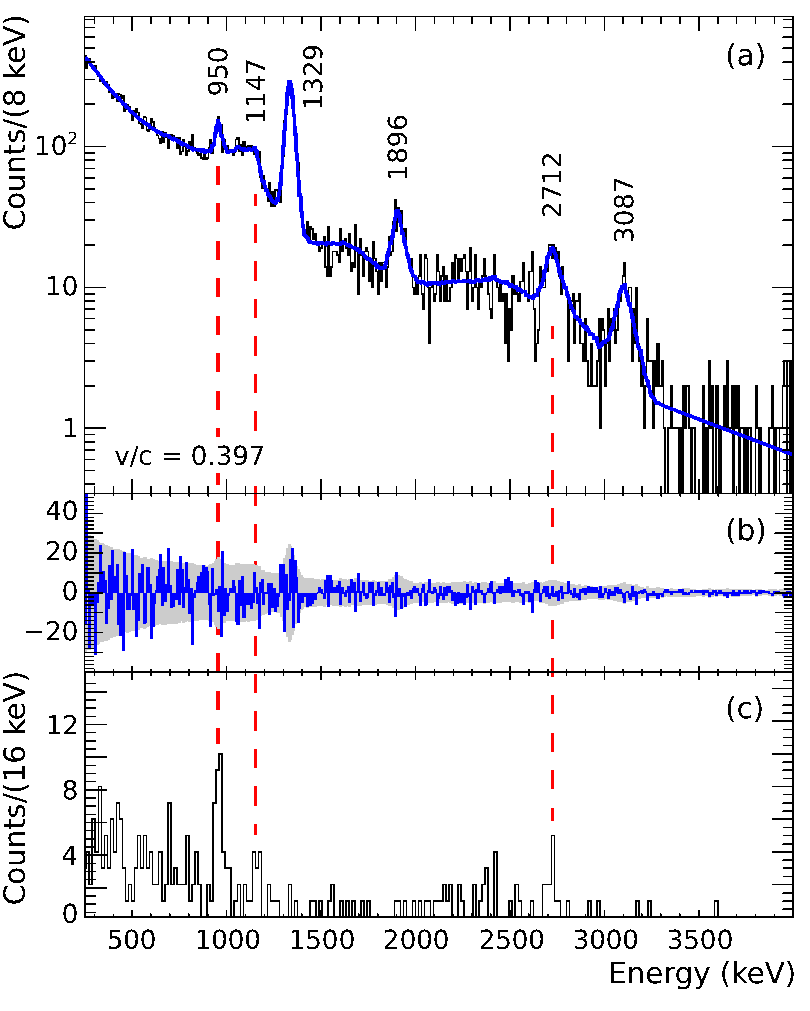}
  \caption{\label{fig:s44_ppprime_spectrum} (Color online) (a), (b) Same as Fig. \ref{fig:coulex_spectrum} but for the $^{44}$S inelastic proton scattering experiment. (c) Spectrum of $\gamma$ rays collected in coincidence with the $2^+_1 \rightarrow 0^+_{g.s.}$ transition.}
\end{figure}

\begin{figure}
  \includegraphics[scale=0.62]{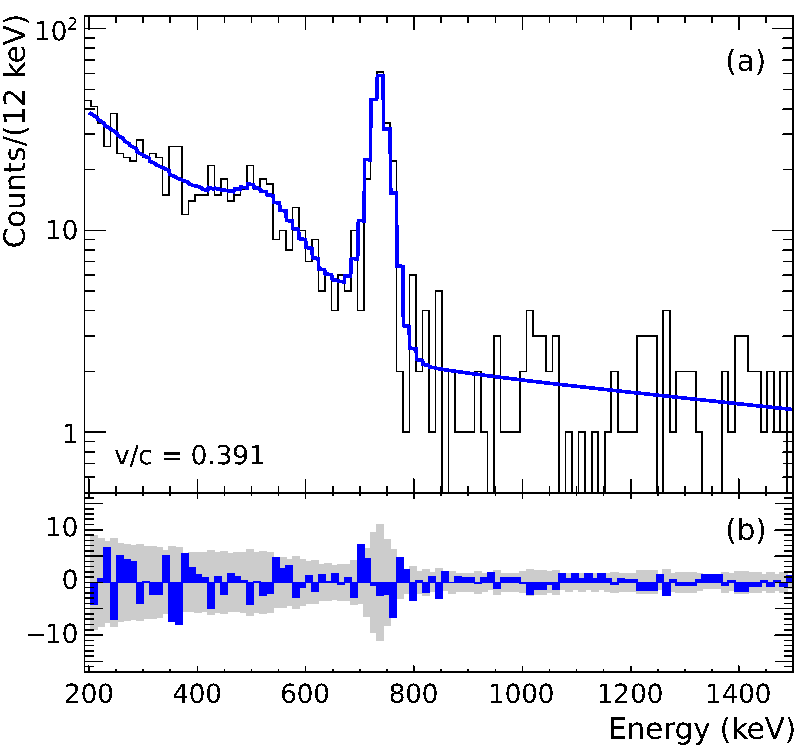}
  \caption{\label{fig:si42_ppprime_spectrum} (Color online) Same as Fig. \ref{fig:coulex_spectrum} but for the $^{42}$Si inelastic proton scattering experiment.}
\end{figure}

During the inverse kinematics proton scattering measurements, a different configuration of GRETINA was used to accommodate the NSCL/Ursinus College liquid hydrogen target.  Modules were installed only in the northern hemisphere of the GRETINA mounting shell, with two modules centered at $58^\circ$, four at $90^\circ$, and two at $122^\circ$ with respect to the beam axis. Figures~\ref{fig:s44_ppprime_spectrum} and \ref{fig:si42_ppprime_spectrum} show Doppler reconstructed $\gamma$-ray spectra collected in coincidence with both incoming and outgoing $^{44}$S and $^{42}$Si particle identification cuts, respectively, and a prompt timing cut between GRETINA and a timing scintillator in the S800 focal plane.

\section{Data Analysis}

\subsection{$^{42}$Si}

We analyzed the intermediate energy Coulomb excitation data using the distorted wave Born approximation (DWBA) instead of the conventional Alder-Winther relativistic Coulomb excitation analysis \cite{Win79}.  The Alder-Winther analysis requires the identification of a ``safe" angle, corresponding to a minimum impact parameter; that is, the scattering angle below which we can be confident that the interaction between the $^{42}$Si beam nucleus and the $^{209}$Bi target is entirely electromagnetic. In Sec.~\ref{sec:alder-winther}, we present an Alder-Winther analysis and show that there are insufficient statistics in our measurement to determine a ``safe-angle'' empirically and further that the Alder-Winther predicted ``safe-angle'' cut, further reduced by the experimental angular resolution, excludes more than half of the observed statistics. 

The proton deformation length $\delta_p$ is determined mainly by the inelastic cross section measured via intermediate energy Coulomb excitation, and the proton scattering deformation length $\delta_{(p,p')}$ is determined mainly by the inelastic proton scattering cross section. However, there is some ``crosstalk" between these analyses. The DWBA analysis of the Coulomb excitation measurement required knowledge of $M_n/M_p$ since the calculation involves the deformation lengths of both the Coulomb and nuclear potentials. The DWBA analysis of the $^{42}$Si($p,p'$) measurement also required the $B(E2)$ value to set the deformation length of the Coulomb potential. Therefore, an iterative process was implemented to simultaneously analyze the intermediate energy Coulomb excitation and inelastic proton scattering data for $^{42}$Si. We can assess the magnitude of the ``crosstalk" between results by comparing DWBA calculations of the Coulomb excitation with and without the deformed nuclear potentials and similarly by comparing calculations of the proton scattering with and without the deformed Coulomb potential. The nuclear contribution to the calculated inelastic Coulomb excitation cross section is at the 4\% level. For the proton scattering analysis, the nuclear potentials dominate, and the deformed Coulomb potential affects the calculated inelastic cross section at a 6\% level.


The $^{42}$Si $\gamma$-ray spectra for both intermediate energy Coulomb excitation (Fig.~\ref{fig:coulex_spectrum}) and inelastic proton scattering (Fig.~\ref{fig:si42_ppprime_spectrum}) experiments each have only one apparent $\gamma$ ray, the $2_1^+ \rightarrow 0_{\mathrm{g.s.}}^+$ $\gamma$ ray near 740 keV. To assess the potential impact of unobserved feeding, we included a 1430~keV $\gamma$ ray deexciting the $\approx$2170~keV state observed in both two-proton and one-proton removal reactions~\cite{Tak12,Gad19} in the fit to the inelastic proton scattering spectrum. The resulting yield was not statistically significant, and the corresponding statistical uncertainty places an upper limit on the feeding correction of 4\%. 

\begin{figure}
  \includegraphics[scale=0.55]{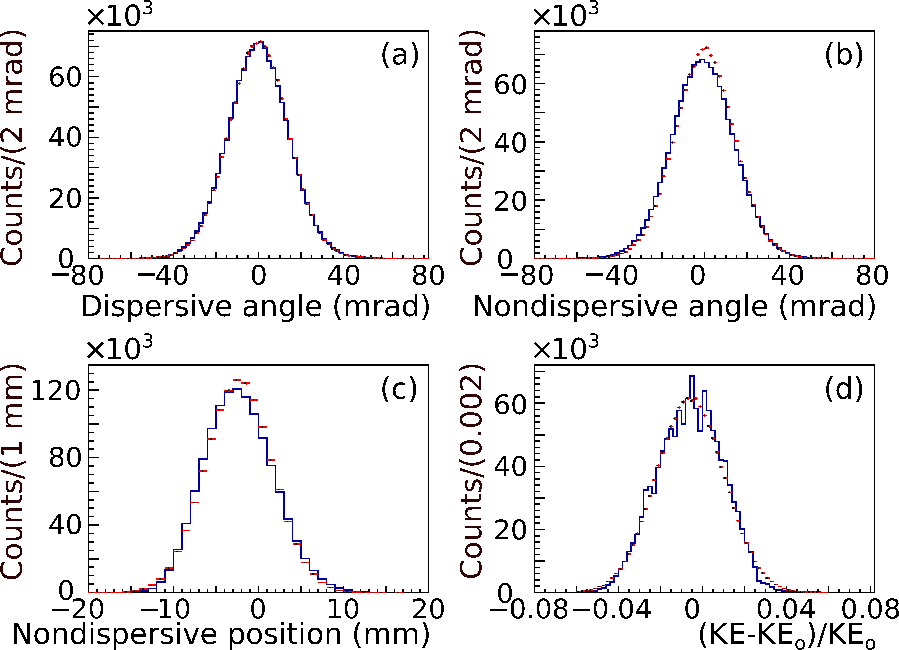}
  \caption{\label{fig:coulex_s800} (Color online) Spectra of the (a) dispersive angle, (b) nondispersive angle, (c) nondispersive position, and (d) kinetic energy relative to the S800 magnetic rigidity measured in the S800 focal plane in coincidence with both incoming and outgoing $^{42}$Si particle identification cuts. Simulated spectra are shown as red markers.}
\end{figure}

We used the \textsc{geant4}~\cite{Geant4} simulation program UCGretina~\cite{Ril21} to simulate the full response of GRETINA to $\gamma$ rays emitted by beam-like reaction products excited in the target and tracked as they traveled downstream. The momentum and position distributions of beam-like reaction products which emit $\gamma$ rays in flight have significant impacts on  $\gamma$-ray line shapes in Doppler reconstructed spectra. We varied simulation parameters that determine the momentum and position distributions of the incoming beam to fit the angular, nondispersive position, and kinetic energy distributions of outgoing reaction products measured in the S800 focal plane. These measured and simulated distributions are compared in Fig.~\ref{fig:coulex_s800}.  Scattering-angle distributions predicted by the reaction theories used for the intermediate energy Coulomb excitation and inverse kinematics proton scattering reactions were included in the simulations used to fit to the measured $\gamma$-ray spectrum. The solid curves shown in Figures~\ref{fig:coulex_spectrum} and \ref{fig:si42_ppprime_spectrum} are linear combinations of double exponentials describing the prompt background and the best-fit simulations determined by the fitting procedures described below.

The low energy of the $2_1^+$ state in $^{42}$Si and the $B(E2;0_{\mathrm{g.s.}}^+ \rightarrow 2_1^+)$ values of isotopes in the neighborhood of this nucleus~\cite{Pri16,Lo21} suggest that the lifetime of the $2_1^+$ state may be tens of picoseconds.  If this is the case, then $2_1^+ \rightarrow 0_{\mathrm{g.s.}}^+$ $\gamma$ rays would be emitted in both experiments not only from inside the targets but also significantly downstream.  In turn, this would affect line shapes and the observed energy centroids in the Doppler reconstructed $\gamma$-ray spectra.  This distribution of $\gamma$-ray emission vertices would also affect the scattering and absorption of these $\gamma$ rays by the reaction targets.   

\begin{figure}
  \includegraphics[scale=0.6]{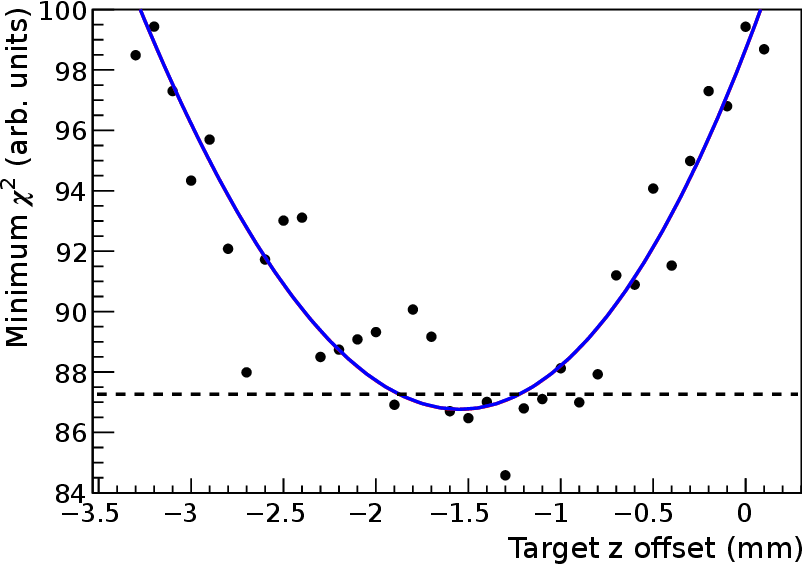}
  \caption{\label{fig:target_z} (Color online) Minimum $\chi^2$ values from fits in the region of the full-energy peak of the 1329~keV $\gamma$ ray in $^{44}$S plotted for a range of positions of the target along the beam axis during the proton scattering experiment.}
\end{figure}

Because offsets of the targets in both measurements along the beam axes relative to the center of GRETINA would also affect the energy centroids of the full-energy peaks, a laser alignment system was used to determine the position of the reaction target along the beam axis relative to the center of GRETINA in the intermediate energy Coulomb excitation experiment. The position relative to the center of GRETINA was found to be $z = 1.4(10)$~mm. We determined the position of the liquid-hydrogen target along the beam axis relative to the center of GRETINA by fitting simulations to the measured $\gamma$-ray spectrum of $^{44}$S over a broad range of target z positions. A plot of the minimum $\chi^2$ from the fits to the $^{44}$S inelastic proton scattering spectrum in the region of the 1329~keV transition, which is known to a precision of $\pm 0.5$~keV~\cite{Che23}, appears in Fig.~\ref{fig:target_z}, constraining the target offset along the beam axis to $z = -1.6(3)$~mm. We ran simulations covering the uncertainty ranges of the target offsets in both measurements of $^{42}$Si and included the corresponding variation as a component of the systematic uncertainties in our $2^+_1$ state energy, half-life, and inelastic cross section results.

In addition, the analysis of the proton scattering data for both $^{42}$Si and $^{44}$S required a determination of the geometry of the liquid hydrogen target.  The Kapton entrance and exit windows of the liquid-hydrogen target bulge outwards due to the difference in pressure between the target cell and the beam line vacuum. The geometry of the bulging was determined using the $^{44}$S beam because its rate was much higher than that for $^{42}$Si. The bulge thickness was determined to be 1.5~mm based on the energy loss of the $^{44}$S beam in the target by comparison of the kinetic energy distributions, measured in the S800 focal plane, of the beam passing through the full and empty target cell. The curved windows produce a target profile presenting a thickness dependent on the trajectory of the beam. A simulation of the $^{44}$S beam passing through the target with a 1.5-mm window-bulge thickness and a realistic beam profile yielded an effective target thickness of 32.8~mm. The pressure and temperature of the target cell were monitored throughout the experiment and remained in the ranges $16 \leq T \leq 19$~K and $700 \leq P \leq 836$~Torr. Based on the measured temperature and pressure, the time-weighted average density of the target was $73.41~\mathrm{mg/cm^3}$~\cite{Lem24}, giving an areal target density of $241~\mathrm{mg/cm^2}$.

\begin{figure}
  \includegraphics[scale=0.52]{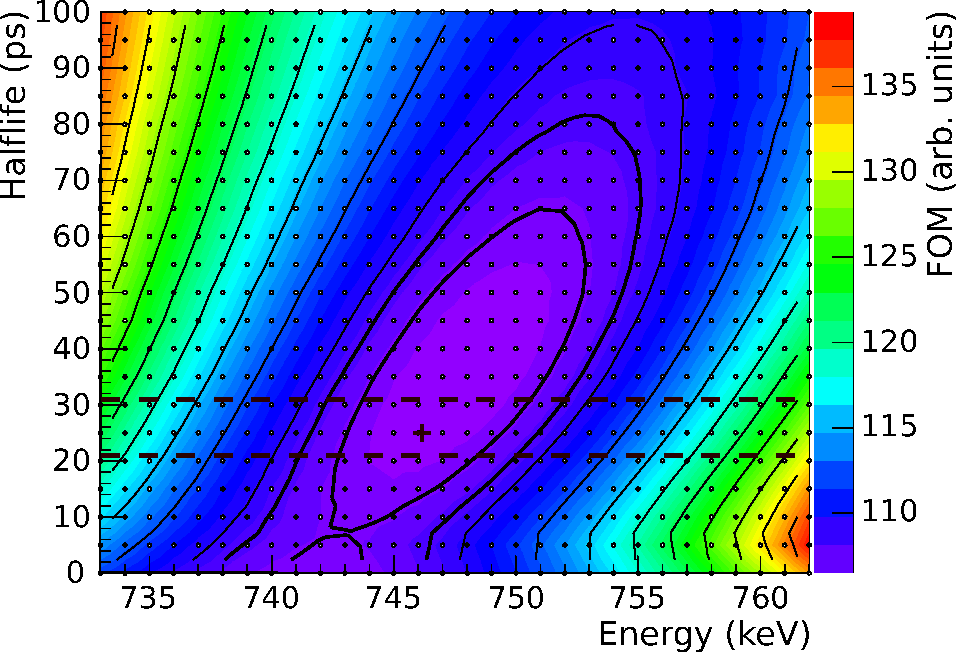}
  \caption{\label{fig:coulex_fcns} (Color online) The minimum figure of merit (FOM) from log-likelihood fits of simulations to the measured $\gamma$-ray spectrum collected following Coulomb excitation of $^{42}$Si over a range of $2^+_1$-state half-lives and deexcitation $\gamma$-ray energies. The final result of the fitting process described in the text is marked with a $\boldsymbol{+}$, and the half-life values corresponding to the uncertainty limits of the $B(E2; 0^+_{g.s.} \rightarrow 2^+_1)$ result are marked with dashed horizontal lines.}
\end{figure}

We assessed the correlation between the energy and half-life of the $2^+_1$ state in the intermediate energy Coulomb excitation measurement by performing simulations over a broad range of energies and half-lives and fitting them to the measured Doppler reconstructed $\gamma$-ray spectrum. A surface plot of the minimum figure of merit (FOM) from log-likelihood fits in the energy-half-life space is displayed in Fig.~\ref{fig:coulex_fcns}. The heavy contours correspond to 70\% and 90\% confidence regions bounded by figures of merit at 1.2 (70\%) and 2.3 (90\%) above the surface minimum~\cite{Jam81}. Starting with the energy and half-life at the surface minimum, we used an iterative process, determining the excitation cross section, deducing a $B(E2; 0^+_{g.s.} \rightarrow 2^+_1)$ value and corresponding half-life, and refitting using the minimum-FOM energy with this half-life. This iterative process converged within two cycles.

\begin{figure}
  \includegraphics[scale=0.52]{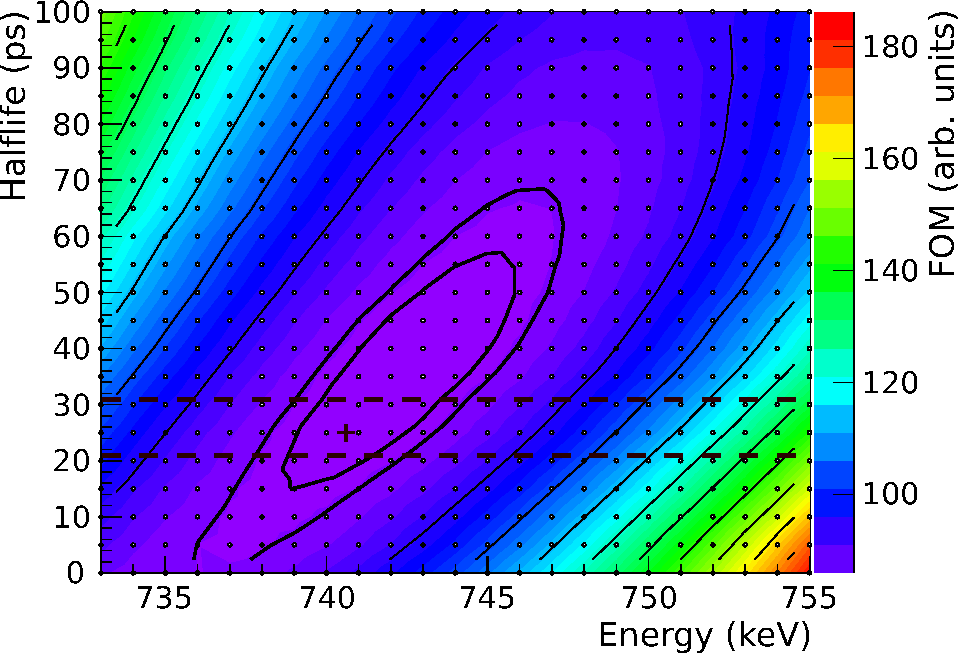}
  \caption{\label{fig:ppprime_fcns} (Color online) The minimum figure of merit (FOM) from log-likelihood fits of simulations to the measured $\gamma$-ray spectrum collected following inverse kinematics proton scattering from $^{42}$Si over a range of $2^+_1$-state half-lives and deexcitation $\gamma$-ray energies. The energy and half-life pair used for the final fit is marked with a $\boldsymbol{+}$, and the half-life values corresponding to the uncertainty limits of the $B(E2; 0^+_{g.s.} \rightarrow 2^+_1)$ result are marked with dashed horizontal lines.}
\end{figure}

In the analysis of the inelastic proton scattering data, as in the Coulomb excitation analysis, we ran simulations of the response of GRETINA to the $\gamma$ ray deexciting the $2^+_1$ state over a range of $2^+_1$-state energies and half-lives, and used them in fits to the measured spectrum. This process yielded the figure of merit surface in Fig.~\ref{fig:ppprime_fcns}, which shows a similar energy-half-life correlation to that observed in the Coulomb excitation analysis. We used the half-life corresponding to the $B(E2; 0^+_{g.s.} \rightarrow 2^+_1)$ value determined in that analysis and the corresponding $2^+_1$-state energy along the ``valley'' in the figure of merit surface to produce the final fit shown as the solid curve in Fig.~\ref{fig:si42_ppprime_spectrum}(a). 

The $B(E2; 2_1^+ \rightarrow 0_{\mathrm{g.s.}}^+)$ value that resulted from this process was then used as a starting point for the iterative DWBA analyses of the data from intermediate energy Coulomb excitation and inelastic proton scattering experiments described in the following paragraphs.  The analysis used the DWBA code \textsc{fresco} \cite{FRESCO} and macroscopic form factors for both reactions.  Global optical model potentials were used for both the reactions.  The analysis of the intermediate energy Coulomb excitation reaction used a global optical model potential for heavy-ion scattering \cite{Fur12}, while the global potential of Ref. \cite{Kon03} was used for the inelastic proton scattering reaction.  The excited $^{42}$Si nuclei were aligned in both reactions, and the \textsc{fresco} analysis provided information on the resulting angular distributions of the $\gamma$ rays.  Accounting for the angular distributions of $\gamma$ rays reduced our cross section results in the intermediate energy Coulomb excitation experiment by about 14\% compared to the results we obtain if we assume that $\gamma$-ray emission is isotropic.  For the proton scattering experiment, the $\gamma$-ray angular distributions from alignment parameters generated in the \textsc{fresco} analysis reduced the cross section result by 8\%.  

The effect of scattering and absorption of $\gamma$ rays in the reaction targets was accounted for by the inclusion of models of the reaction targets in the \textsc{geant4} simulations used to fit the measured $\gamma$-ray spectra. This was an 11\% effect in the Coulomb excitation measurement and 15\% for the proton-scattering measurements.

To calculate cross sections for a heavy-ion reaction like the one we used for our intermediate energy Coulomb excitation study of $^{42}$Si, the DWBA code requires two deformation lengths.  The first is the proton (Coulomb) deformation length of the Coulomb component of the deformed optical potential.  This deformation length, $\delta_p$, is related to $B(E2; 0^+_{g.s.} \rightarrow 2^+_1)$ by
\begin{equation}
\label{eq:delta_p}
\delta_p = \frac{4 \pi}{3 Z} \frac{1}{r_C A^{1/3}} \sqrt{\frac{B(E2; 0^+_{g.s.} \rightarrow 2^+_1)}{e^2}} 
\end{equation}
where $r_C = 1.2$~fm is the radius parameter of the Coulomb potential.

The second deformation length is that of the nuclear component of the deformed optical potential, $\delta_N$.  To specify the relationship between $\delta_p$ and $\delta_N$, we start with Eq.~(7) in Ref.~\cite{Be83} to obtain
\begin{equation}
\label{eq:mnmp}
\delta(F) = \left( \frac{1 + \frac{b_n}{b_p} \frac{M_n}{M_p}}{1 + \frac{b_n}{b_p} \frac{N}{Z}} \right) \delta_p,
\end{equation}
where $\delta(F)$ is the deformation length measured with probe F, $M_n/M_p$ is the ratio of neutron to proton transition matrix elements, and $\frac{b_n}{b_p}$ is the ratio of the sensitivities of the experimental probe $F$ to neutrons and protons.  For the present purpose, $\delta(F)=\delta_N$. For heavy-ion scattering, we use $\frac{b_n}{b_p} = 1$.


If we have a proton deformation length $\delta_p$ and a deformation length from the inelastic proton scattering reaction, $\delta_{(p,p')}$, a rearrangement of Eq.~\ref{eq:mnmp} provides this equation to calculate $M_n/M_p$:
\begin{equation}
\label{eq:mnmp2}
\frac{M_n}{M_p} = \frac{b_p}{b_n} \left[\frac{\delta(F)}{\delta_p} \left( 1 + \frac{N}{Z} \frac{b_n}{b_p} \right) - 1 \right],
\end{equation}
where $\delta(F) = \delta_{(p,p')}$. We used this relation in the iterative process that led to the $B(E2;0_{\mathrm{g.s.}}^+ \rightarrow 2_1^+)$, $\delta_{(p,p')}$ and $M_n/M_p$ results reported here.


The value of $\frac{b_n}{b_p}$ for inelastic proton scattering varies with the incident energy of the proton \cite{Be81}.  At incident energies of 50 MeV and below, $\frac{b_n}{b_p}=3$.  At 1 GeV, $\frac{b_n}{b_p} = 0.95$.  For the inverse kinematics proton scattering reactions reported here, the midtarget beam energies are 91.2 MeV/nucleon for $^{42}$Si and 82.3 MeV/nucleon for $^{44}$S. At these energies, we expect $\frac{b_n}{b_p}$ to be in between the values for 50 MeV and 1 GeV incident energies, but we cannot be more precise than that.  So we adopt $\frac{b_n}{b_p} = 2 \pm 1$.


To evaluate the systematic uncertainties inherent in the analysis described above, the iterative analysis described above was then repeated using the code \textsc{ECIS97} \cite{ECIS97} for both intermediate energy Coulomb excitation and proton scattering measurements. Two statistical uncertainties were considered, a 15\% contribution from the fit to the $\gamma$-ray spectrum and a 5\% contribution due to the total variation in cross section corresponding to the 90\% confidence contours in the energy-half-life FOM surface in Fig.~\ref{fig:coulex_fcns}. Systematic errors from the uncertainty in the empirically-determined $M_n/M_p$ value (5.5\%), discrepancies between \textsc{fresco} and \textsc{ECIS97} results (4\%), and uncertainties in UCGretina-simulated $\gamma$-ray collection efficiencies (5\%) were combined in quadrature with the statistical uncertainties to arrive at the uncertainty in the final $B(E2; 0^+_{g.s.} \rightarrow 2^+_1)$ result.

\begin{figure}
  \includegraphics[scale=0.55]{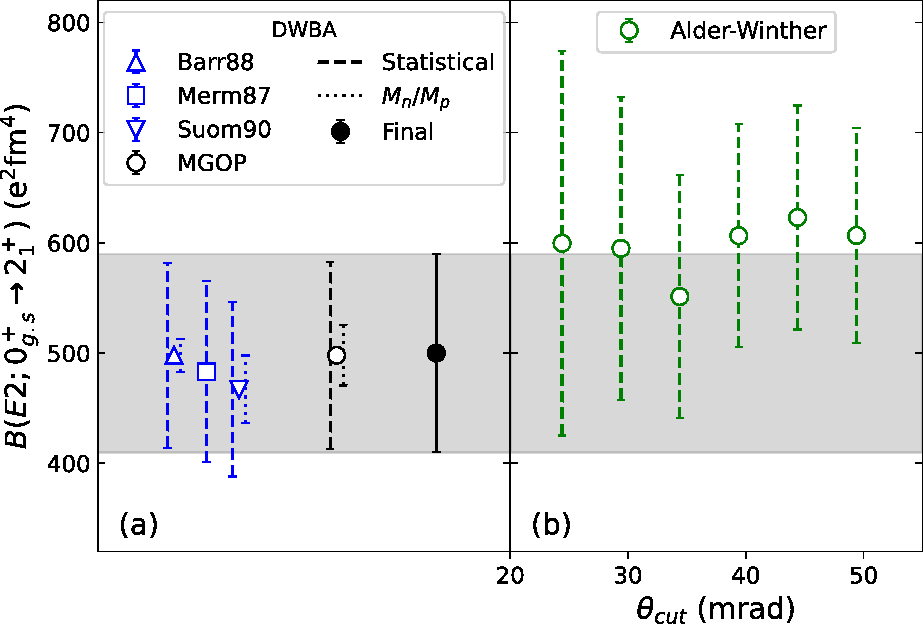}
  \caption{\label{fig:be2} (Color online) $B(E2; 0^+_{g.s.} \rightarrow 2^+_1)$ results for $^{42}$Si from the (a) DWBA analysis and (b) Alder-Winther analysis described in the text. The shaded region corresponds to the uncertainty range of the final result.}
\end{figure}

The DWBA results were also evaluated using three other heavy-ion optical model potentials:  one from $^{208}\mathrm{Pb}(^{17}\mathrm{O},^{17}\mathrm{O}')$ at 84 MeV/nucleon~\cite{Bar88} (Barr88), a second from $^{208}\mathrm{Pb}(^{16}\mathrm{O},^{16}\mathrm{O}')$ at 49.6 MeV/nucleon~\cite{Mer87} (Merm87), and a third from $^{208}\mathrm{Pb}(^{40}\mathrm{Ar},^{40}\mathrm{Ar}')$ at 40~MeV/nucleon~\cite{Suo90}.  The results with these potentials are compared to the results using the global optical model potential (MGOP) in Fig.~\ref{fig:be2}, which is the final result of $B(E2; 0^+_{g.s.} \rightarrow 2^+_1) = 500(90)~e^2\,\mathrm{fm^4}$.  

The cross section for exciting the $2_1^+$ state of $^{42}$Si in the intermediate energy Coulomb excitation experiment was determined to be 165(28)~mb, giving the $B(E2;0_{\mathrm{g.s.}}^+ \rightarrow 2_1^+)=500(90)~e^2\,\mathrm{fm^4}$ result from the previous paragraph.  The corresponding half-life of the $2^+_1$ state is $t_{1/2} = 25^{+6}_{-4}$~ps.  For the proton scattering experiment on $^{42}$Si, the cross section and deformation length for exciting the $2_1^+$ state were $\sigma = 21(2)$~mb and $\delta_{(p,p')} = 1.23(7)$~fm.  The result for $M_n/M_p$ was 1.34(32).  The value of $M_n/M_p$ expected for a quadrupole deformed rotor composed of a homogeneous neutron-proton fluid is $N/Z$, so it is worth noting that the present $M_n/M_p$ result is $0.67(16) (N/Z)$, which is significantly below the homogeneous rotor expectation.

Khan \cite{Kh22} developed a procedure for calculating $M_n/M_p$ from the results of one electromagnetic probe and one hadronic probe that is more general than Eq.~\ref{eq:mnmp2} in that it includes separate radius parameters for protons and neutrons and separate diffuseness parameters for protons and neutrons.  If we use the $B(E2;0_{\mathrm{g.s.}}^+ \rightarrow 2_1^+)$ and $\delta_{(p,p')}$ results obtained here as inputs into the Khan procedure, the $M_n/M_p$ result is $1.16(56)$, which is consistent with our primary conclusion that $M_n/M_p$ is significantly different from $N/Z$, which is 2.00 for $^{42}$Si.   

\subsection{Alder-Winther Analysis of intermediate energy Coulomb excitation of $^{42}$Si}
\label{sec:alder-winther}
\begin{figure}
  \includegraphics[scale=0.73]{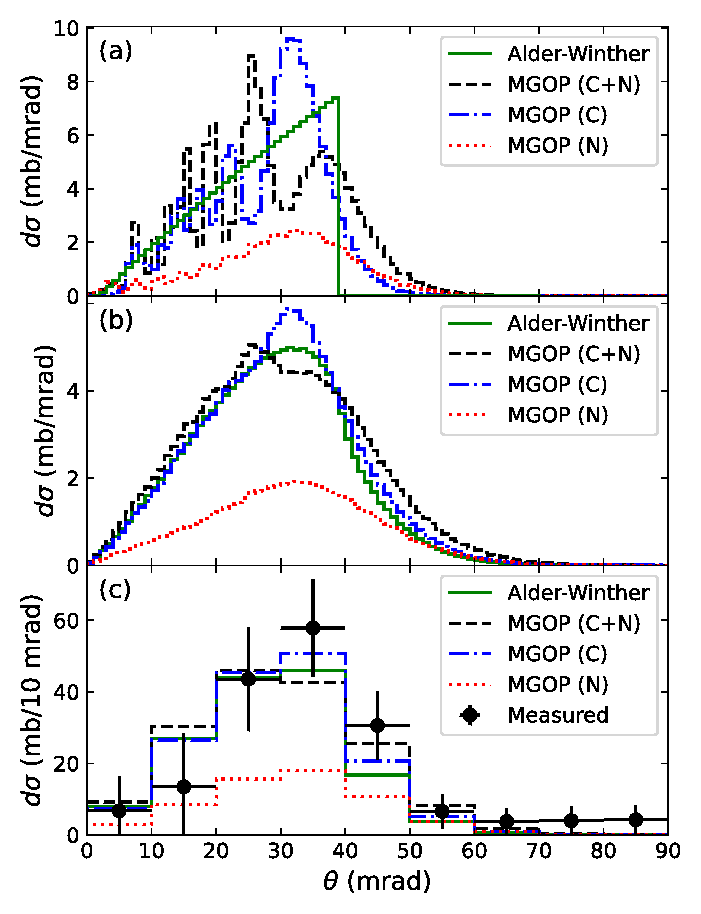}
  \caption{\label{fig:mgop_aw_dsigma} (Color online) (a) Laboratory-frame inelastic partial cross sections in 1 mrad bins calculated with the Alder-Winther formalism~\cite{Win79} and DWBA calculations made with \textsc{fresco} using the global microscopic-basis optical potential (MGOP)~\cite{Fur12} with Coulomb and nuclear (C+N), Coulomb only (C), and nuclear only (N) components of the potential. (b) The differential cross sections in panel (a) folded with the angular spread of the incoming beam, the angular straggling in the target, and the angular resolution of the S800. (c) The folded partial cross sections in panel (b) with a 10 mrad binning compared with measured partial cross sections.}
\end{figure}

We implemented a conventional Alder-Winther analysis of the $^{42}$Si intermediate energy Coulomb excitation data for comparison with the results of our DWBA analysis.  The Alder-Winther differential cross section, integrated into 1 mrad partial cross section bins through a black disk laboratory cutoff angle of 39~mrad, corresponding to the minimum impact parameter below which nuclear interactions take place in this semiclassical picture, is shown as the solid curve in Fig.~\ref{fig:mgop_aw_dsigma}(a). In order to accurately assess partial cross sections in scattering angle cuts applied to measured spectra, the theoretical partial cross section histograms in Fig.~\ref{fig:mgop_aw_dsigma}(a) have been folded with the 8.0~mrad angular spread of the incoming beam, the 8.7~mrad angular straggling in the target, and the 2.0~mrad angular resolution of the S800, added in quadrature, using the method described in detail in Ref.~\cite{Vaq18} to produce the partial cross sections in Fig.~\ref{fig:mgop_aw_dsigma}(b). In Fig.~\ref{fig:mgop_aw_dsigma}(c), folded partial cross sections from reaction theory are compared with measured partial cross sections calculated from the scattering angle spectrum gated on the $2^+_1 \rightarrow 0^+_{g.s.}$ full-energy peak.

It is evident in Fig.~\ref{fig:mgop_aw_dsigma}(b) that the folded Alder-Winther distribution departs from roughly linear behavior at 27~mrad, due to the smearing of the black disk cutoff by the empirical scattering angle resolution of the experiment. This scattering angle also marks a departure of the folded Alder-Winther distribution from the Coulomb-only DWBA distribution. A 27~mrad scattering-angle cut excludes more than half of the total inelastic cross section for populating the $2^+_1$ state. In Fig.~\ref{fig:be2}(b), $B(E2; 0^+_{g.s.} \rightarrow 2^+_1)$ values determined from partial inelastic cross sections in laboratory frame scattering angle cuts from $25 \leq \theta_{\mathrm{max}} \leq 50$~mrad are shown with error bars representing statistical uncertainties. The cross section for each scattering angle cut was determined from a fit to the $\gamma$-ray spectrum in coincidence with that cut with a simulation including angular momentum alignment parameters predicted for that cut by the Alder-Winther reaction theory. The $B(E2; 0^+_{g.s.} \rightarrow 2^+_1)$ values were then determined from the cross sections using the Alder-Winther theory. The results in Fig.~\ref{fig:be2}(b) are all compatible with each other within uncertainty. We do not have sufficient statistics in the measurement to support a choice of scattering angle cut empirically. However, the DWBA partial cross sections in Fig.~\ref{fig:mgop_aw_dsigma} show nuclear contributions even within the restrictive 27~mrad cut, increasing the inelastic cross section at lower scattering angles relative to the Alder-Winther and Coulomb-only DWBA partial cross sections, leading to the systematic discrepancy between the Alder-Winther and DWBA $B(E2; 0^+_{g.s.} \rightarrow 2^+_1)$ results.

\subsection{The energy of the $2^+_1$ state of $^{42}$Si}

A high-precision measurement of the $\gamma$ decay of the $2^+_1$ state of $^{42}$Si at rest in the laboratory has not yet been reported. The $\approx$25-$\mathrm{ps}$ half-life of the $2^+_1$ state of $^{42}$Si presents a challenge to measuring the energy of the deexcitation $\gamma$ ray in fast-beam experiments due to the correlation between mean lifetime and $\gamma$-ray energy evident in Figures~\ref{fig:coulex_fcns} and \ref{fig:ppprime_fcns}. The half-life corresponding to the final $B(E2; 0^+_{g.s.} \rightarrow 2^+_1)$ result from the present work, based on the DWBA analysis using the MGOP, is marked with a $\boldsymbol{+}$, and its uncertainty range is bounded by two horizontal dashed lines in Figures~\ref{fig:coulex_fcns} and \ref{fig:ppprime_fcns}. Using the part of the 90\% confidence regions falling within these boundaries and including a $\approx 1~\mathrm{keV/mm}$ variation due to uncertainties in the target positions along the beam axis to establish uncertainties, we find $E_{2^+_1} = 747(7)$~keV for the Coulomb excitation measurement and $E_{2^+_1} = 741(4)$~keV for the inverse kinematics proton scattering measurement. There are three prior measurements of the energy of the the $2^+_1$ state of $^{42}$Si via $\gamma$-ray spectroscopy with fast beams: 770(19)~keV from the $^{43}\mathrm{P}(^9\mathrm{Be},X)$ reaction made with $\mathrm{BaF_2}$ scintillator detectors~\cite{Ba07}, 742(8)~keV from the $^{44}\mathrm{S}(\mathrm{C},X)$ reaction made with the DALI2 array~\cite{Tak14} of NaI(Tl) scintillator detectors~\cite{Tak12}, and 737(8)~keV from the $^{43}~\mathrm{P}(^9\mathrm{Be},X)$ reaction made with GRETINA~\cite{Gad19}. Calculating the uncertainty-weighted average of these results and the two results from the present work yields a best value of 742(4)~keV. We have used this value in calculating $2^+_1$-state half-lives from measured $B(E2)$ values.

\subsection{$^{44}$S}

Because a recent intermediate energy Coulomb excitation result \cite{Lo21} is available, to determine $M_n/M_p$ for the 
$0_{\mathrm{g.s.}}^+ \rightarrow 2_1^+$ transition it was only necessary to measure $^{44}$S($p,p'$) in inverse kinematics.  As we did for $^{42}$Si($p,p'$), we extracted a deformation parameter $\delta_{(p,p')}$ using \textsc{fresco} \cite{FRESCO}, a macroscopic form factor, and the global optical model parameters of Ref. \cite{Kon03}.  

\begin{figure}
  \includegraphics[scale=0.52]{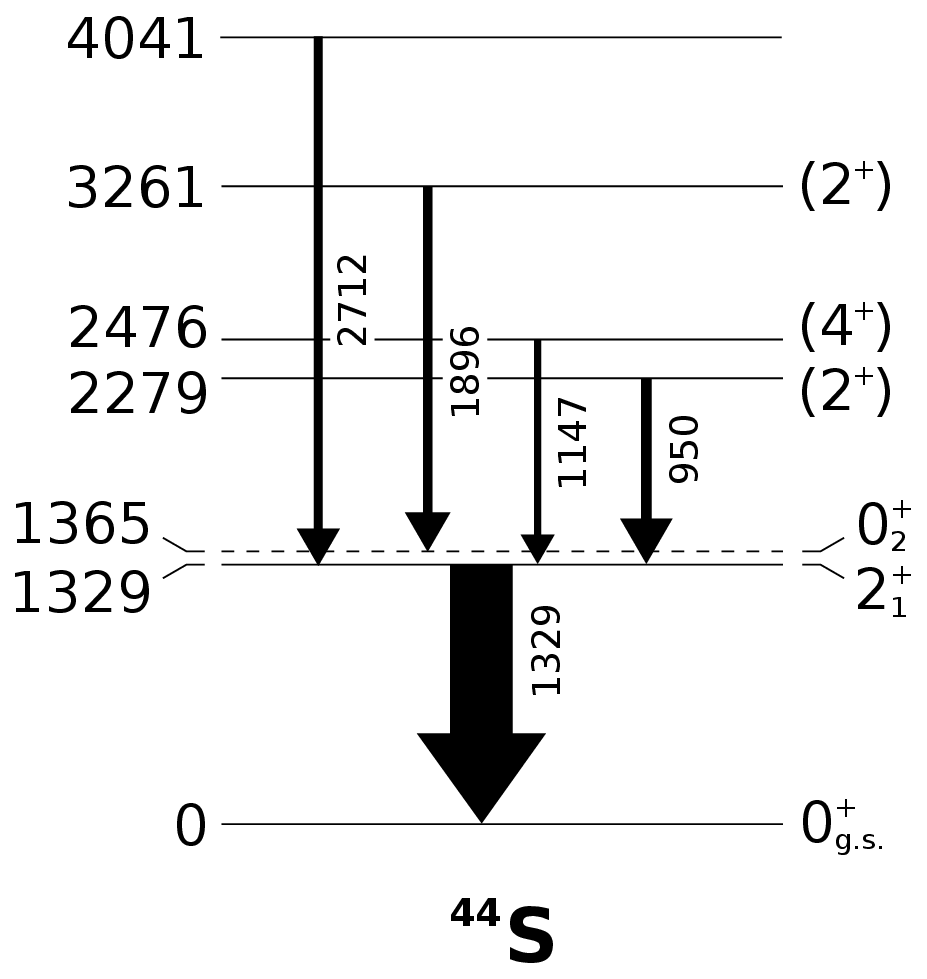}
  \caption{\label{fig:s44_level} (Color online) Partial level scheme of $^{44}$S showing states populated in the present work. Arrow widths are proportional to measured $\gamma$-ray intensities.}
\end{figure}

While the $\gamma$-ray spectra for $^{42}$Si showed there was no significant feeding of the $2_1^+$ state from higher-lying states, the situation was much different in the $^{44}$S($p,p'$) spectrum, which is shown in Fig. \ref{fig:s44_ppprime_spectrum}.  The $^{44}$S $\gamma$-rays observed in the present ($p,p'$) experiment are listed in Table~\ref{tab:s44_gammas}, and a partial level scheme of $^{44}$S showing the states populated in the present measurement is displayed in Fig. \ref{fig:s44_level}.  As a result, the observed yield, which was already adjusted for the angular distribution of the $\gamma$-rays (as was done for $^{42}$Si($p,p'$)), also had to be corrected for feeding.  In the end, the inelastic cross section for directly populating the $2^+_1$ state was determined to be $\sigma = 9.9(8)$~mb.  With \textsc{fresco}, we concluded that this cross section corresponds to a deformation length $\delta_{(p,p')} = 0.78(3)$~fm.  The $1\sigma$ experimental uncertainty range on the present $\delta_{(p,p')}$ result does not overlap with the corresponding range for the previously reported result of $\delta_{(p,p')} = 1.07(16)$~fm~\cite{Ril19}, although there is overlap at the $2\sigma$ level.

The $M_n/M_p$ result for $^{44}$S was determined by using Eq. \ref{eq:mnmp} with the present $\delta_{(p,p')}$ result, $\frac{b_n}{b_p}=2 \pm 1$ for proton scattering, and the Coulomb deformation length $\delta_p$ calculated from the $B(E2; 0^+_{g.s.} \rightarrow 2^+_1) = 221(28)~e^2\,\mathrm{fm^4}$ result of Ref.~\cite{Lo21} using Eq. \ref{eq:delta_p}.  The result is $M_n/M_p = 1.36(20) = 0.78(12) (N/Z)$.  

\begin{table}
\caption{\label{tab:s44_gammas} Level energies, spins and parities, and $\gamma$-ray energies from Ref.~\cite{Che23} and $\gamma$-ray energies, relative intensities, and cross sections from the inverse kinematics proton-scattering measurement of $^{44}$S.}
\begin{ruledtabular}
\begin{tabular}{rcrcrc}
$E_\mathrm{level}$ [keV] & $J^\pi$ [$\hbar$] & $E_\gamma$ [keV] &
  $E_\gamma$ [keV] & $I_\gamma$ [\%] & $\sigma$ [mb]
\\\hline\hline
& \multicolumn{2}{c}{Ref.~\cite{Che23}} & &  \\\cline{2-3}
1329      & $2^+$       & 1329.0(5) & 1329     & 100(8) &  9.9(8)\\
2279(2)   & $(2^+)$     &  952(4)   & 950(2)   &  16(2) &  2.5(3)\\
2476(3)   & $(4^+)$     & 1138(6)   & 1147(3)  &   8(2) &  1.3(3)\\    
3261(4)   & $(2^+)$     & 1897(6)   & 1896(4)  &  13(2) &  2.1(2)\\
4041(6)   &             & 2698(13)  & 2712(6)  &  13(2) &  2.1(2)\\\hline
          &             & ---       & 3087(7)  &  11(1) &\\
\end{tabular}
\end{ruledtabular}
\end{table}

\section{Discussion and Theory}

The prediction of Werner \textit{et al.} that the $N=28$ shell closure would narrow \cite{We94} and the subsequent measurement of collectivity in $^{44}$S \cite{Gl97} sparked a tremendous amount of experimental and theoretical work on neutron rich nuclei near $N=28$. Grevy \textit{et al.} \cite{Gr04} deduced that $^{42}$Si is deformed from their study of half-lives of several isotopes at and near $N=28$.  Fridmann \textit{et al.} \cite{Fr05,Fr06} disagreed with that conclusion on the basis of their study of $^{42}$Si via the two-proton knockout reaction, but shortly afterward Bastin \textit{et al.} \cite{Ba07} proved that $^{42}$Si is collective by identifying the $2_1^+$ state in $^{42}$Si at 770(19) keV. (Of course, this result has since been revised.)  Takeuchi \textit{et al.} \cite{Tak12} argued that a state in $^{42}$Si they identified at 2173(14) keV using the two-proton knockout reaction is the $4^+$ member of the ground state band.  If this were the case, it would provide further evidence for the stable deformation of $^{42}$Si because of the resulting $E(4_1^+)/E(2_1^+)$ ratio of 2.9.  However, Gade \textit{et al.} \cite{Gad19} used their one-proton knockout measurement of $^{42}$Si and the reaction model analysis of Tostevin, Brown and Simpson \cite{To13} to argue that the 2173 keV state (measured by Gade \textit{et al.} to occur at 2150(13) keV) may be the $0_2^+$ state instead of the $4_1^+$ state.

Following the observation of the $2_1^+$ state of $^{44}$S by Glasmacher \textit{et al.} \cite{Gl97}, Sohler \textit{et al.} \cite{So02} extended the $^{44}$S level scheme by observing $\gamma$ rays from the fragmentation of $^{48}$Ca.  Sohler \textit{et al.} concluded that this nucleus exhibits shape coexistence.  Force \textit{et al.} \cite{Fo10} identified the
$0_2^+$ state of $^{44}$S less than 50 keV above the $2_1^+$ state, placing the claim of shape coexistence in this nucleus on very firm ground.  Santiago-Gonzalez \textit{et al.} \cite{Sa11} found a third coexisting shape in $^{44}$S in what appeared then to be an isomeric $4_1^+$ state.  Several years later, Parker \textit{et al.} \cite{Pa17} measured the lifetime of this $4_1^+$ state, firmly establishing its isomeric character.  The $4^+$ state associated with the ground-state band has not yet been identified experimentally.

A series of theoretical predictions \cite{Su21,Ma22,Su22,Wa23,Yu24,Ma24} agree that $^{42}$Si is an oblate deformed rotor.  Furthermore, while they agree in general that $^{44}$S is a collective nucleus, they do not agree on the specific behavior of this isotope.  Instead, they predict a range of behaviors from soft vibrator to stable prolate deformation.

While three coexisting configurations have been observed at low energies in $^{44}$S, Utsuno \textit{et al.} \cite{Ut15} predicted that four configurations would coexist at low energies in this nucleus.  This prediction arises from the coupling of neutrons with projections of angular momentum of $\Omega=1/2$ and $7/2$ on the nuclear symmetry axis to the $K=1/2$ ground state and the $K=7/2$ isomer in $^{43}$S.  Furthermore, Utsuno \textit{et al.} concluded on the basis of calculations using the variation after angular-momentum projection (AM-VAP) beyond mean field method that the ground state of $^{44}$S is triaxial, with a triaxiality parameter $\gamma=33^\circ$.  In their picture, the band built on the ground state evolves toward a prolate shape with increasing spin, reaching $\gamma=13$ degrees by $J=6$.  But the $2_1^+$ state is still fairly triaxial, with $\gamma= 23^\circ$.

Longfellow \textit{et al.} \cite{Lo21} demonstrated that the SDPF-U and SDPF-MU shell model interactions cannot reproduce the experimental $B(E2;0_{\mathrm{g.s.}}^+ \rightarrow 2_1^+)$ values for the $N=28$ isotopes $^{44}$S and $^{46}$Ar.  Calculations with both of these interactions predict that the $B(E2;0_{\mathrm{g.s.}}^+ \rightarrow 2_1^+)$ values in the $N=28$ isotopes of S and Ar are either equal to or larger than (depending on the effective charges selected) the corresponding quantities in the $N=26$ isotopes $^{42}$S and $^{44}$Ar.  However, the experimental results show that for both elements $B(E2;0_{\mathrm{g.s.}}^+ \rightarrow 2_1^+)$ is significantly smaller in the $N=28$ isotope than in the $N=26$ isotope.

\begin{figure}
  \includegraphics[scale=0.75]{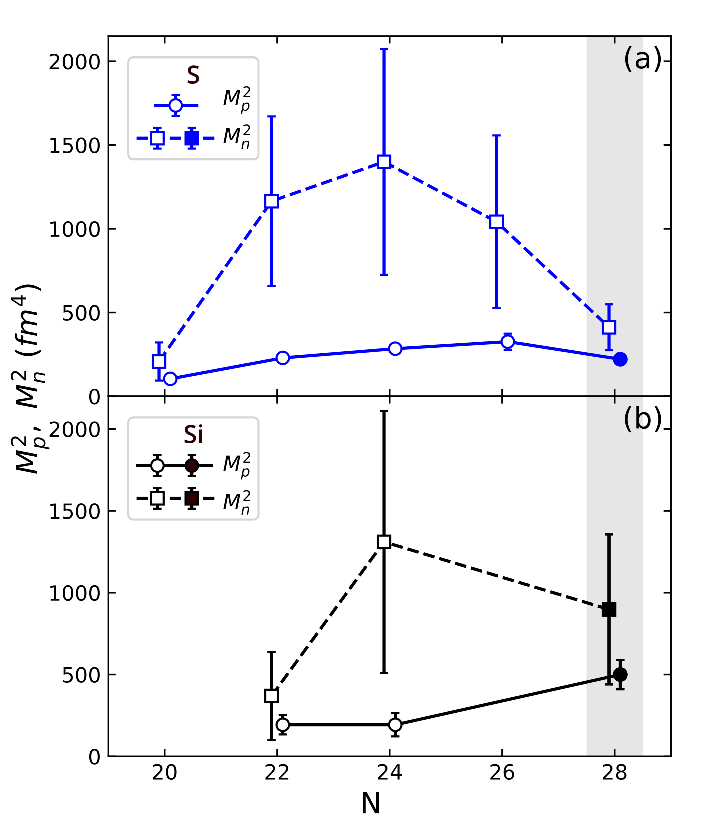}
  \caption{\label{fig:mn2_mp2} (Color online) The systematic behavior of $M_n^2$ and $M_p^2$ values for $0_{\mathrm{g.s.}}^+ \rightarrow 2_1^+$ excitations in the neutron rich even-even isotopes of S and Si.  The $M_p^2$ values are taken from Refs. \cite{Pri16,Lo21} ($\Circle$) and the present work ($\CIRCLE$).  $M_n^2$ values are deduced from Refs. \cite{Ma99,Ca07,Ril19} ($\square$) and the present work ($\blacksquare$).}
\end{figure}

Figure~\ref{fig:mn2_mp2} expands on the theme of the systematics presented by Longfellow \textit{et al.} \cite{Lo21} by plotting both $M_p^2$, which is proportional to $B(E2;0_{\mathrm{g.s.}}^+ \rightarrow 2_1^+)$, and $M_n^2$ for the $N=20-28$ even-even isotopes of S and Si.  Given the uncertainties in the measurements, the most striking feature of this figure is the increase in $M_p^2$ for $^{42}$Si compared to $^{38}$Si. (No measurement of $B(E2;0_{\mathrm{g.s.}}^+ \rightarrow 2_1^+)$ for $^{40}$Si has been reported.)  The plot of the S isotopes suggests that both $M_p^2$ and $M_n^2$ are lower at $N=28$ than in lighter isotopes.  In addition, the $M_n^2$ value for $^{42}$Si does not appear to be larger than the values for lighter isotopes.   

The present intermediate energy Coulomb excitation result for $B(E2;0_{\mathrm{g.s.}}^+ \rightarrow 2_1^+)$ in $^{42}$Si is more than twice as large as the result for the neighboring isotone $^{44}$S, which supports the predictions that $^{42}$Si has a stable quadrupole deformation.  However, the $M_n/M_p$ value for a rotational excitation in a stably deformed liquid drop composed of a homogeneous neutron-proton fluid is $N/Z$, the same value as for an isoscalar vibrational excitation.  The present result for $^{42}$Si, $M_n/M_p=1.34(32)=0.67(16) (N/Z)$, differs significantly from this simple expectation.    

The experimental value for $B(E2;0_{\mathrm{g.s.}}^+ \rightarrow 2_1^+)$ in $^{44}$S ($221(28)$ $e^2fm^4$), which contrasts sharply with the corresponding $^{42}$Si result, favors the soft vibrator picture for this nucleus over the prolate deformed interpretation.  However, the $M_n/M_p$ result of $1.36(20)=0.78(12)(N/Z)$ varies significantly from the expected $N/Z$ value for an isoscalar vibrational excitation.

To gain more insight about the collective behavior of $^{42}$Si and $^{44}$S, we performed shell-model calculations of these two isotopes using the FSU interaction \cite{Lu20,Lu19,cosmo} with the code \textsc{COSMO} \cite{cosmo}. The FSU Hamiltonian \cite{Lu20,Lu19,cosmo} represents a new generation of empirical interactions \cite{Brown2022} that effectively has no core and spans the $s\text{-}p\text{-}sd\text{-}pf$ valence space. It performs equally well across a broad range of nuclei, up to the mass region where contributions from $g$-shell configurations become significant. This approach employs a particle-hole $\hbar \omega$ hierarchy, allowing for control over spurious center-of-mass excitations. The effects of mixing between different $\hbar \omega$ configurations were negligible in this study. Unlike many limited Hamiltonians designed for specific, narrow regions, the FSU interaction provides binding energies across the nuclear chart with the same precision as it does excitation energies and spectra for individual nuclei. Changes in binding energies are among the most important indicators of shell inversion, deformation, and other collective effects \cite{Brown2006}. The FSU Hamiltonian predicts the binding energy of $^{42}$Si relative to $^{28}$Si to be 73.307 MeV, while the experimental value is 73.072 MeV. This exceptional agreement suggests that the underlying physics is well captured.  More generally, calculations with the FSU interaction correctly reproduce the shell evolution and spectroscopy of $sd\text{-}pf$ nuclei.

The present shell-model calculation with the FSU interaction reproduces the experimental values for both $M_n/M_p$ and $B(E2; 0_{\mathrm{g.s.}}^+ \rightarrow 2_1^+)$ in $^{42}$Si. The isoscalar effective charges used for the transitions are $e_p = 1.5e$ and $e_n = 0.5e$. The calculation gives $M_n/M_p = 1.41$, which corresponds to $0.70(N/Z)$, closely matching the experimental value of $M_n/M_p = 1.34(32)$, or $0.67(16)(N/Z)$. The calculated $B(E2; 0_{\mathrm{g.s.}}^+ \rightarrow 2_1^+)$ is $492$ $e^2\,\text{fm}^4$, which is very close to the experimental result of $500(90)$ $e^2\,\text{fm}^4$. Moreover, the shell-model calculations provide additional insights into the nature of both the ground and $2_1^+$ states in $^{42}$Si.

The calculation for $^{42}$Si gives a quadrupole moment $Q_2$ for the $2_1^+$ state of $20.40~\mathrm{efm^2}$. Assuming a rotor model and that this state is part of a rotational band \cite{Zelevinsky2017}, we can extract an intrinsic quadrupole moment of $Q_0 = (-7/2) Q_2 = -71.4~e\,\mathrm{fm^2}$. In the rotor model, an intrinsic quadrupole moment of this value yields $B(E2; 0_{\mathrm{g.s.}}^+ \rightarrow 2_1^+) = \frac{5}{16 \pi} Q_0^2 = 507~e^2\,\mathrm{fm^4}$, which is quite close to the shell-model value of $492~e^2\,\mathrm{fm^4}$. The E2 transition strength for $0_{\mathrm{g.s.}}^+ \rightarrow 2_1^+$ nearly saturates the sum rule, highlighting its collective nature.

Further confirmation of the rotational nature of the calculated states comes from the quadrupole moment of the $4_1^+$ state. This value, $Q_4 = 27.5~e\,\mathrm{fm^2}$, corresponds to an intrinsic quadrupole moment in the rotational model of $Q_0 = (-11/4) Q_4 = -75.60~e\,\mathrm{fm^2}$. This is quite close to the $Q_0$ value for the $2_1^+$ state, reinforcing confidence in the rotational interpretation of the $0^+_{\mathrm{g.s.}}$, $2_1^+$, and $4_1^+$ band in $^{42}$Si, as supported by both the shell-model calculation and experimental results.

The FSU interaction shell model results, including results for the energies of the $2_1^+$ and $4_1^+$ states, are listed and compared to the present experimental results in Table~\ref{tab:si42_shellmodel}.  

In addition, the table includes results from a shell model calculation we performed with the SDPF-MU interaction \cite{Ut12} and COSMO that includes $0 \hbar \omega$, $2 \hbar \omega$ and $4 \hbar \omega$ excitations, with effective charges $e_p=1.5$ and $e_n=0.5$.  We also ran a calculation with only $0 \hbar \omega$ excitations, and another with $0 \hbar \omega$ and $2 \hbar \omega$ excitations.  The $0 \hbar \omega + 2 \hbar \omega$ calculation gave significantly different results than the $0 \hbar \omega$ calculation did.  For example, the $B(E2;0_{\mathrm{g.s.}}^+ \rightarrow 2_1^+)$ value from the $0 \hbar \omega$ calculation was 875 $e^2\,\mathrm{fm^4}$, while the $0 \hbar \omega + 2 \hbar \omega$ result was 684 $e^2\,\mathrm{fm^4}$.  The $0 \hbar \omega + 2 \hbar \omega + 4 \hbar \omega$ result, 638 $e^2\,\mathrm{fm^4}$, showed that the results were converging and that computationally intensive $0 \hbar \omega + 2 \hbar \omega + 4 \hbar \omega + 6 \hbar \omega$ calculation was not necessary. The SDPF-MU results for the energy of the $2_1^+$ state vary in a similar way.  The $0 \hbar \omega + 2 \hbar \omega$ result, 1132 keV, is quite different from the $0 \hbar \omega$ result, 821 keV.  In contrast, the $0 \hbar \omega + 2 \hbar \omega + 4 \hbar \omega$ result, 1232 keV, shows that calculation is converging.   

The $B(E2;0_{\mathrm{g.s.}}^+ \rightarrow 2_1^+)$ result from the $0 \hbar \omega + 2 \hbar \omega + 4 \hbar \omega$ SDPF-MU calculation, $638~\mathrm{e^2fm^4}$, is more than one standard deviation above the experimental value but nevertheless supports a rotational interpretation like the FSU interaction calculation does. As shown in Table~\ref{tab:si42_shellmodel} and as in the case of the calculations with the FSU interaction, the intrinsic quadrupole moments extracted from the calculated $B(E2)$ and $Q_2$ and $Q_4$ values in the SDPF-MU calculation are nearly identical, once again supporting a rotational interpretation. In short, the overarching conclusions of the shell model calculations with the FSU and SDPF-MU interactions are identical, even though there are quantitative differences in the matrix elements calculated using the two interactions.  

Finally, the $(M_n/M_p)/(N/Z)$ result from our $0 \hbar \omega + 2 \hbar \omega + 4 \hbar \omega$ SDPF-MU calculation was 0.77, which is close to the FSU interaction result (0.70) as well as being consistent with the experimental result of 0.67(16).

\begin{table}
\caption{\label{tab:si42_shellmodel} Results from shell model calculations on $^{42}$Si with FSU and SDPF-MU interactions compared to present experimental results.}
\begin{ruledtabular}
\begin{tabular}{cccc}
 & Expt. & FSU &
  SDPF-MU
\\\hline\hline
$B(E2;0_{\mathrm{g.s.}}^+ \rightarrow 2_1^+)$ ($e^2\,\mathrm{fm^4}$) & 500(90) & 493 & 638 \\ 
$(M_n/M_p)/(N/Z)$ & 0.67(16) & 0.70 & 0.77 \\
$E(2_1^+)$ (keV) & 742(4) & 1559 & 1232 \\
$E(4_1^+)$ (keV) & n/a & 2598 & 2834 \\
$Q_0$ from $B(E2)$ ($e\,\mathrm{fm^2}$) & 70.9(64) & 70.4 & 80.1 \\ 
$Q_0$ from $Q_2$ ($e\,\mathrm{fm^2}$) & n/a & $-$71.4 & $-$82.4 \\    
$Q_0$ from $B(E2)$ ($e\,\mathrm{fm^2}$) & n/a & 69.7 & 83.0 \\
$Q_0$ from $Q_4$ ($e\,\mathrm{fm^2}$) & n/a & $-$75.6 & $-$83.1 \\
\end{tabular}
\end{ruledtabular}
\end{table}

The situation in $^{44}$S is quite different. The result for 
$B(E2; 0_{\mathrm{g.s.}}^+ \rightarrow 2_1^+)$ calculated with the FSU interaction, $483~e^2\,\mathrm{fm^4}$ is much higher than the experimental value of 
$221(28)~e^2\,\mathrm{fm^4}$ reported in Ref. \cite{Lo21}. Neither the experimental nor the theoretical $B(E2; 0_{\mathrm{g.s.}}^+ \rightarrow 2_1^+)$ result for $^{44}$S supports a rotational interpretation for this nucleus. Theory shows that the intrinsic moment of the $2_1^+$ state is too small compared to the transitional $B(E2; 0_{\mathrm{g.s.}}^+ \rightarrow 2_1^+)$, suggesting a mixed configuration.

The present results provide a definitive answer to the question of whether the $N=28$ major shell closure still exists in $^{42}$Si.  The $B(E2;0_{\mathrm{g.s.}}^+ \rightarrow 2_1^+)$ result in this nucleus demonstrates that it has a significant stable axially symmetric quadrupole deformation in the ground state.  This cannot occur in a nucleus that has a shell closure for either protons or neutrons.  Therefore, the $N=28$ major shell closure is quenched in $^{42}$Si by the narrowing of the gap between the $f_{7/2}$ and $p_{3/2}$ neutron orbits.  However, the $M_n/M_p$ result for $^{42}$Si shows that the simple picture of a deformed nucleus in which the protons and neutrons are homogeneously distributed throughout the nucleus does not apply to $^{42}$Si.  That is, there are microscopic effects in this nucleus that do not allow such a homogeneous distribution to occur.  The shell model calculations presented here reproduce the experimental result for $M_n/M_p$ in $^{42}$Si; that is, these calculations provide a quantitative understanding of these microscopic effects in this deformed nucleus.    

\section{Summary}

In summary, we have reported on measurements of the $0_{\mathrm{g.s.}}^+ \rightarrow 2_1^+$ transition in $^{42}$Si via intermediate-energy Coulomb excitation and inelastic proton scattering in inverse kinematics.  The $^{42}$Si intermediate-energy Coulomb excitation experiment was performed with a beam rate of $\approx3$ particles/s, while the proton scattering reaction was measured with a rate of $\approx7$ particles/s.  To obtain the highest precision results with the modest numbers of counts in the $2_{1}^+\rightarrow0_{\mathrm{g.s.}}^+$ $\gamma$-ray peaks, the data from the two experiments were analyzed with the DWBA simultaneously using an iterative process that rapidly converged.  The result for the $M_n/M_p$ value, $1.34(32)=0.67(16)(N/Z)$, is not consistent with the value of $N/Z$ expected for a stably deformed rotor consisting of a homogeneous neutron-proton fluid.  We also performed a shell-model calculation for $^{42}$Si using the FSU interaction.  This calculation reproduces this experimental result and supports the interpretation of $^{42}$Si as an oblate deformed rotor.

In addition, we measured the $0_{\mathrm{g.s.}}^+ \rightarrow 2_1^+$ excitation in the isotone $^{44}$S via inverse kinematics inelastic proton scattering.  By comparing the present result with the result of the intermediate energy Coulomb excitation measurement reported in Ref. \cite{Lo21}, we determined that $M_n/M_p=1.36(20)=0.78(12)(N/Z)$ for this excitation.  While this $M_n/M_p$ result is similar to that in $^{42}$Si, the $B(E2;0_{\mathrm{g.s.}}^+ \rightarrow 2_1^+)$ value reported in Ref. \cite{Lo21} for $^{44}$S, $221(28)~\mathrm{e^2fm^4}$, is less than half the corresponding value reported here for $^{42}$Si, $500(90)~\mathrm{e^2fm^4}$.  In addition, the present shell model calculation does not support a stable axially-symmetric deformation in $^{44}$S.  We conclude on the basis of both the $B(E2;0_{\mathrm{g.s.}}^+ \rightarrow 2_1^+)$ value and the shell-model calculation that $^{44}$S is not stably deformed.  

\section{Acknowledgment}
This material is based upon work supported by the US Department of Energy (DOE), Office of Science, Office of Nuclear Physics and used resources of FRIB Operations, which is a DOE Office of Science User Facility (Award No. DE-SC0023633). This work was further supported by National Science Foundation Grants No. PHY-2012522, No. PHY-2412808, No. PHY-2208804, No. PHY-2209429 and U.S. Department of Energy Grant No. DE-SC0009883. 
 

\begin{thebibliography}{57}%
\makeatletter
\providecommand \@ifxundefined [1]{%
 \@ifx{#1\undefined}
}%
\providecommand \@ifnum [1]{%
 \ifnum #1\expandafter \@firstoftwo
 \else \expandafter \@secondoftwo
 \fi
}%
\providecommand \@ifx [1]{%
 \ifx #1\expandafter \@firstoftwo
 \else \expandafter \@secondoftwo
 \fi
}%
\providecommand \natexlab [1]{#1}%
\providecommand \enquote  [1]{``#1''}%
\providecommand \bibnamefont  [1]{#1}%
\providecommand \bibfnamefont [1]{#1}%
\providecommand \citenamefont [1]{#1}%
\providecommand \href@noop [0]{\@secondoftwo}%
\providecommand \href [0]{\begingroup \@sanitize@url \@href}%
\providecommand \@href[1]{\@@startlink{#1}\@@href}%
\providecommand \@@href[1]{\endgroup#1\@@endlink}%
\providecommand \@sanitize@url [0]{\catcode `\\12\catcode `\$12\catcode
  `\&12\catcode `\#12\catcode `\^12\catcode `\_12\catcode `\%12\relax}%
\providecommand \@@startlink[1]{}%
\providecommand \@@endlink[0]{}%
\providecommand \url  [0]{\begingroup\@sanitize@url \@url }%
\providecommand \@url [1]{\endgroup\@href {#1}{\urlprefix }}%
\providecommand \urlprefix  [0]{URL }%
\providecommand \Eprint [0]{\href }%
\providecommand \doibase [0]{https://doi.org/}%
\providecommand \selectlanguage [0]{\@gobble}%
\providecommand \bibinfo  [0]{\@secondoftwo}%
\providecommand \bibfield  [0]{\@secondoftwo}%
\providecommand \translation [1]{[#1]}%
\providecommand \BibitemOpen [0]{}%
\providecommand \bibitemStop [0]{}%
\providecommand \bibitemNoStop [0]{.\EOS\space}%
\providecommand \EOS [0]{\spacefactor3000\relax}%
\providecommand \BibitemShut  [1]{\csname bibitem#1\endcsname}%
\let\auto@bib@innerbib\@empty
\bibitem [{\citenamefont {Werner}\ \emph {et~al.}(1996)\citenamefont {Werner},
  \citenamefont {Sheikh}, \citenamefont {Misu}, \citenamefont {Nazarewicz},
  \citenamefont {Rikovska}, \citenamefont {Heeger}, \citenamefont {Umar},\ and\
  \citenamefont {Strayer}}]{We94}%
  \BibitemOpen
  \bibfield  {author} {\bibinfo {author} {\bibfnamefont {T.}~\bibnamefont
  {Werner}}, \bibinfo {author} {\bibfnamefont {J.}~\bibnamefont {Sheikh}},
  \bibinfo {author} {\bibfnamefont {M.}~\bibnamefont {Misu}}, \bibinfo {author}
  {\bibfnamefont {W.}~\bibnamefont {Nazarewicz}}, \bibinfo {author}
  {\bibfnamefont {J.}~\bibnamefont {Rikovska}}, \bibinfo {author}
  {\bibfnamefont {K.}~\bibnamefont {Heeger}}, \bibinfo {author} {\bibfnamefont
  {A.}~\bibnamefont {Umar}},\ and\ \bibinfo {author} {\bibfnamefont
  {M.}~\bibnamefont {Strayer}},\ }\href
  {https://doi.org/https://doi.org/10.1016/0375-9474(95)00476-9} {\bibfield
  {journal} {\bibinfo  {journal} {Nuclear Physics A}\ }\textbf {\bibinfo
  {volume} {597}},\ \bibinfo {pages} {327} (\bibinfo {year}
  {1996})}\BibitemShut {NoStop}%
\bibitem [{\citenamefont {Glasmacher}\ \emph {et~al.}(1997)\citenamefont
  {Glasmacher}, \citenamefont {Brown}, \citenamefont {Chromik}, \citenamefont
  {Cottle}, \citenamefont {Fauerbach}, \citenamefont {Ibbotson}, \citenamefont
  {Kemper}, \citenamefont {Morrissey}, \citenamefont {Scheit}, \citenamefont
  {Sklenicka},\ and\ \citenamefont {Steiner}}]{Gl97}%
  \BibitemOpen
  \bibfield  {author} {\bibinfo {author} {\bibfnamefont {T.}~\bibnamefont
  {Glasmacher}}, \bibinfo {author} {\bibfnamefont {B.}~\bibnamefont {Brown}},
  \bibinfo {author} {\bibfnamefont {M.}~\bibnamefont {Chromik}}, \bibinfo
  {author} {\bibfnamefont {P.}~\bibnamefont {Cottle}}, \bibinfo {author}
  {\bibfnamefont {M.}~\bibnamefont {Fauerbach}}, \bibinfo {author}
  {\bibfnamefont {R.}~\bibnamefont {Ibbotson}}, \bibinfo {author}
  {\bibfnamefont {K.}~\bibnamefont {Kemper}}, \bibinfo {author} {\bibfnamefont
  {D.}~\bibnamefont {Morrissey}}, \bibinfo {author} {\bibfnamefont
  {H.}~\bibnamefont {Scheit}}, \bibinfo {author} {\bibfnamefont
  {D.}~\bibnamefont {Sklenicka}},\ and\ \bibinfo {author} {\bibfnamefont
  {M.}~\bibnamefont {Steiner}},\ }\href
  {https://doi.org/https://doi.org/10.1016/S0370-2693(97)00077-4} {\bibfield
  {journal} {\bibinfo  {journal} {Physics Letters B}\ }\textbf {\bibinfo
  {volume} {395}},\ \bibinfo {pages} {163} (\bibinfo {year}
  {1997})}\BibitemShut {NoStop}%
\bibitem [{\citenamefont {Longfellow}\ \emph {et~al.}(2021)\citenamefont
  {Longfellow}, \citenamefont {Weisshaar}, \citenamefont {Gade}, \citenamefont
  {Brown}, \citenamefont {Bazin}, \citenamefont {Brown}, \citenamefont {Elman},
  \citenamefont {Pereira}, \citenamefont {Rhodes},\ and\ \citenamefont
  {Spieker}}]{Lo21}%
  \BibitemOpen
  \bibfield  {author} {\bibinfo {author} {\bibfnamefont {B.}~\bibnamefont
  {Longfellow}}, \bibinfo {author} {\bibfnamefont {D.}~\bibnamefont
  {Weisshaar}}, \bibinfo {author} {\bibfnamefont {A.}~\bibnamefont {Gade}},
  \bibinfo {author} {\bibfnamefont {B.~A.}\ \bibnamefont {Brown}}, \bibinfo
  {author} {\bibfnamefont {D.}~\bibnamefont {Bazin}}, \bibinfo {author}
  {\bibfnamefont {K.~W.}\ \bibnamefont {Brown}}, \bibinfo {author}
  {\bibfnamefont {B.}~\bibnamefont {Elman}}, \bibinfo {author} {\bibfnamefont
  {J.}~\bibnamefont {Pereira}}, \bibinfo {author} {\bibfnamefont
  {D.}~\bibnamefont {Rhodes}},\ and\ \bibinfo {author} {\bibfnamefont
  {M.}~\bibnamefont {Spieker}},\ }\href
  {https://doi.org/10.1103/PhysRevC.103.054309} {\bibfield  {journal} {\bibinfo
   {journal} {Phys. Rev. C}\ }\textbf {\bibinfo {volume} {103}},\ \bibinfo
  {pages} {054309} (\bibinfo {year} {2021})}\BibitemShut {NoStop}%
\bibitem [{\citenamefont {Bastin}\ \emph {et~al.}(2007)\citenamefont {Bastin},
  \citenamefont {Gr\'evy}, \citenamefont {Sohler}, \citenamefont {Sorlin},
  \citenamefont {Dombr\'adi}, \citenamefont {Achouri}, \citenamefont
  {Ang\'elique}, \citenamefont {Azaiez}, \citenamefont {Baiborodin},
  \citenamefont {Borcea}, \citenamefont {Bourgeois}, \citenamefont {Buta},
  \citenamefont {B\"urger}, \citenamefont {Chapman}, \citenamefont {Dalouzy},
  \citenamefont {Dlouhy}, \citenamefont {Drouard}, \citenamefont {Elekes},
  \citenamefont {Franchoo}, \citenamefont {Iacob}, \citenamefont {Laurent},
  \citenamefont {Lazar}, \citenamefont {Liang}, \citenamefont {Li\'enard},
  \citenamefont {Mrazek}, \citenamefont {Nalpas}, \citenamefont {Negoita},
  \citenamefont {Orr}, \citenamefont {Penionzhkevich}, \citenamefont
  {Podoly\'ak}, \citenamefont {Pougheon}, \citenamefont {Roussel-Chomaz},
  \citenamefont {Saint-Laurent}, \citenamefont {Stanoiu}, \citenamefont
  {Stefan}, \citenamefont {Nowacki},\ and\ \citenamefont {Poves}}]{Ba07}%
  \BibitemOpen
  \bibfield  {author} {\bibinfo {author} {\bibfnamefont {B.}~\bibnamefont
  {Bastin}}, \bibinfo {author} {\bibfnamefont {S.}~\bibnamefont {Gr\'evy}},
  \bibinfo {author} {\bibfnamefont {D.}~\bibnamefont {Sohler}}, \bibinfo
  {author} {\bibfnamefont {O.}~\bibnamefont {Sorlin}}, \bibinfo {author}
  {\bibfnamefont {Z.}~\bibnamefont {Dombr\'adi}}, \bibinfo {author}
  {\bibfnamefont {N.~L.}\ \bibnamefont {Achouri}}, \bibinfo {author}
  {\bibfnamefont {J.~C.}\ \bibnamefont {Ang\'elique}}, \bibinfo {author}
  {\bibfnamefont {F.}~\bibnamefont {Azaiez}}, \bibinfo {author} {\bibfnamefont
  {D.}~\bibnamefont {Baiborodin}}, \bibinfo {author} {\bibfnamefont
  {R.}~\bibnamefont {Borcea}}, \bibinfo {author} {\bibfnamefont
  {C.}~\bibnamefont {Bourgeois}}, \bibinfo {author} {\bibfnamefont
  {A.}~\bibnamefont {Buta}}, \bibinfo {author} {\bibfnamefont {A.}~\bibnamefont
  {B\"urger}}, \bibinfo {author} {\bibfnamefont {R.}~\bibnamefont {Chapman}},
  \bibinfo {author} {\bibfnamefont {J.~C.}\ \bibnamefont {Dalouzy}}, \bibinfo
  {author} {\bibfnamefont {Z.}~\bibnamefont {Dlouhy}}, \bibinfo {author}
  {\bibfnamefont {A.}~\bibnamefont {Drouard}}, \bibinfo {author} {\bibfnamefont
  {Z.}~\bibnamefont {Elekes}}, \bibinfo {author} {\bibfnamefont
  {S.}~\bibnamefont {Franchoo}}, \bibinfo {author} {\bibfnamefont
  {S.}~\bibnamefont {Iacob}}, \bibinfo {author} {\bibfnamefont
  {B.}~\bibnamefont {Laurent}}, \bibinfo {author} {\bibfnamefont
  {M.}~\bibnamefont {Lazar}}, \bibinfo {author} {\bibfnamefont
  {X.}~\bibnamefont {Liang}}, \bibinfo {author} {\bibfnamefont
  {E.}~\bibnamefont {Li\'enard}}, \bibinfo {author} {\bibfnamefont
  {J.}~\bibnamefont {Mrazek}}, \bibinfo {author} {\bibfnamefont
  {L.}~\bibnamefont {Nalpas}}, \bibinfo {author} {\bibfnamefont
  {F.}~\bibnamefont {Negoita}}, \bibinfo {author} {\bibfnamefont {N.~A.}\
  \bibnamefont {Orr}}, \bibinfo {author} {\bibfnamefont {Y.}~\bibnamefont
  {Penionzhkevich}}, \bibinfo {author} {\bibfnamefont {Z.}~\bibnamefont
  {Podoly\'ak}}, \bibinfo {author} {\bibfnamefont {F.}~\bibnamefont
  {Pougheon}}, \bibinfo {author} {\bibfnamefont {P.}~\bibnamefont
  {Roussel-Chomaz}}, \bibinfo {author} {\bibfnamefont {M.~G.}\ \bibnamefont
  {Saint-Laurent}}, \bibinfo {author} {\bibfnamefont {M.}~\bibnamefont
  {Stanoiu}}, \bibinfo {author} {\bibfnamefont {I.}~\bibnamefont {Stefan}},
  \bibinfo {author} {\bibfnamefont {F.}~\bibnamefont {Nowacki}},\ and\ \bibinfo
  {author} {\bibfnamefont {A.}~\bibnamefont {Poves}},\ }\href
  {https://doi.org/10.1103/PhysRevLett.99.022503} {\bibfield  {journal}
  {\bibinfo  {journal} {Phys. Rev. Lett.}\ }\textbf {\bibinfo {volume} {99}},\
  \bibinfo {pages} {022503} (\bibinfo {year} {2007})}\BibitemShut {NoStop}%
\bibitem [{\citenamefont {Gade}\ \emph {et~al.}(2019)\citenamefont {Gade},
  \citenamefont {Brown}, \citenamefont {Tostevin}, \citenamefont {Bazin},
  \citenamefont {Bender}, \citenamefont {Campbell}, \citenamefont {Crawford},
  \citenamefont {Elman}, \citenamefont {Kemper}, \citenamefont {Longfellow},
  \citenamefont {Lunderberg}, \citenamefont {Rhodes},\ and\ \citenamefont
  {Weisshaar}}]{Gad19}%
  \BibitemOpen
  \bibfield  {author} {\bibinfo {author} {\bibfnamefont {A.}~\bibnamefont
  {Gade}}, \bibinfo {author} {\bibfnamefont {B.~A.}\ \bibnamefont {Brown}},
  \bibinfo {author} {\bibfnamefont {J.~A.}\ \bibnamefont {Tostevin}}, \bibinfo
  {author} {\bibfnamefont {D.}~\bibnamefont {Bazin}}, \bibinfo {author}
  {\bibfnamefont {P.~C.}\ \bibnamefont {Bender}}, \bibinfo {author}
  {\bibfnamefont {C.~M.}\ \bibnamefont {Campbell}}, \bibinfo {author}
  {\bibfnamefont {H.~L.}\ \bibnamefont {Crawford}}, \bibinfo {author}
  {\bibfnamefont {B.}~\bibnamefont {Elman}}, \bibinfo {author} {\bibfnamefont
  {K.~W.}\ \bibnamefont {Kemper}}, \bibinfo {author} {\bibfnamefont
  {B.}~\bibnamefont {Longfellow}}, \bibinfo {author} {\bibfnamefont
  {E.}~\bibnamefont {Lunderberg}}, \bibinfo {author} {\bibfnamefont
  {D.}~\bibnamefont {Rhodes}},\ and\ \bibinfo {author} {\bibfnamefont
  {D.}~\bibnamefont {Weisshaar}},\ }\href
  {https://doi.org/10.1103/PhysRevLett.122.222501} {\bibfield  {journal}
  {\bibinfo  {journal} {Phys. Rev. Lett.}\ }\textbf {\bibinfo {volume} {122}},\
  \bibinfo {pages} {222501} (\bibinfo {year} {2019})}\BibitemShut {NoStop}%
\bibitem [{\citenamefont {Bernstein}\ \emph {et~al.}(1981)\citenamefont
  {Bernstein}, \citenamefont {Brown},\ and\ \citenamefont {Madsen}}]{Be81}%
  \BibitemOpen
  \bibfield  {author} {\bibinfo {author} {\bibfnamefont {A.}~\bibnamefont
  {Bernstein}}, \bibinfo {author} {\bibfnamefont {V.}~\bibnamefont {Brown}},\
  and\ \bibinfo {author} {\bibfnamefont {V.}~\bibnamefont {Madsen}},\ }\href
  {https://doi.org/https://doi.org/10.1016/0370-2693(81)90219-7} {\bibfield
  {journal} {\bibinfo  {journal} {Physics Letters B}\ }\textbf {\bibinfo
  {volume} {103}},\ \bibinfo {pages} {255} (\bibinfo {year}
  {1981})}\BibitemShut {NoStop}%
\bibitem [{\citenamefont {Wei}\ \emph {et~al.}(2022)\citenamefont {Wei},
  \citenamefont {Ao}, \citenamefont {Arend}, \citenamefont {Beher},
  \citenamefont {Bollen}, \citenamefont {Bultman}, \citenamefont {Casagrande},
  \citenamefont {Chang}, \citenamefont {Choi}, \citenamefont {Cogan},
  \citenamefont {Compton}, \citenamefont {Cortesi}, \citenamefont {Curtin},
  \citenamefont {Davidson}, \citenamefont {Du}, \citenamefont {Elliott},
  \citenamefont {Ewert}, \citenamefont {Facco}, \citenamefont {Fila},
  \citenamefont {Fukushima}, \citenamefont {Ganni}, \citenamefont {Ganshyn},
  \citenamefont {Gao}, \citenamefont {Glasmacher}, \citenamefont {Guo},
  \citenamefont {Hao}, \citenamefont {Hartung}, \citenamefont {Hasan},
  \citenamefont {Hausmann}, \citenamefont {Holland}, \citenamefont {Hseuh},
  \citenamefont {Ikegami}, \citenamefont {Jager}, \citenamefont {Jones},
  \citenamefont {Joseph}, \citenamefont {Kanemura}, \citenamefont {Kim},
  \citenamefont {Knudsen}, \citenamefont {Kortum}, \citenamefont {Kwan},
  \citenamefont {Larter}, \citenamefont {Laxdal}, \citenamefont {Larmann},
  \citenamefont {Laturkar}, \citenamefont {LeTourneau}, \citenamefont {Li},
  \citenamefont {Lidia}, \citenamefont {Machicoane}, \citenamefont {Magsig},
  \citenamefont {Manwiller}, \citenamefont {Marti}, \citenamefont {Maruta},
  \citenamefont {McCartney}, \citenamefont {Metzgar}, \citenamefont {Miller},
  \citenamefont {Momozaki}, \citenamefont {Morris}, \citenamefont {Mugerian},
  \citenamefont {Nesterenko}, \citenamefont {Nguyen}, \citenamefont
  {O’Brien}, \citenamefont {Openlander}, \citenamefont {Ostroumov},
  \citenamefont {Patil}, \citenamefont {Plastun}, \citenamefont {Popielarski},
  \citenamefont {Popielarski}, \citenamefont {Portillo}, \citenamefont
  {Priller}, \citenamefont {Rao}, \citenamefont {Reaume}, \citenamefont {Ren},
  \citenamefont {Saito}, \citenamefont {Smith}, \citenamefont {Steiner},
  \citenamefont {Stolz}, \citenamefont {Tarasov}, \citenamefont {Tousignant},
  \citenamefont {Walker}, \citenamefont {Wang}, \citenamefont {Wenstrom},
  \citenamefont {West}, \citenamefont {Witgen}, \citenamefont {Wright},
  \citenamefont {Xu}, \citenamefont {Xu}, \citenamefont {Yamazaki},
  \citenamefont {Zhang}, \citenamefont {Zhao}, \citenamefont {Zhao},
  \citenamefont {Dixon}, \citenamefont {Wiseman}, \citenamefont {Kelly},
  \citenamefont {Hosoyama},\ and\ \citenamefont {Prestemon}}]{FRIB}%
  \BibitemOpen
  \bibfield  {author} {\bibinfo {author} {\bibfnamefont {J.}~\bibnamefont
  {Wei}}, \bibinfo {author} {\bibfnamefont {H.}~\bibnamefont {Ao}}, \bibinfo
  {author} {\bibfnamefont {B.}~\bibnamefont {Arend}}, \bibinfo {author}
  {\bibfnamefont {S.}~\bibnamefont {Beher}}, \bibinfo {author} {\bibfnamefont
  {G.}~\bibnamefont {Bollen}}, \bibinfo {author} {\bibfnamefont
  {N.}~\bibnamefont {Bultman}}, \bibinfo {author} {\bibfnamefont
  {F.}~\bibnamefont {Casagrande}}, \bibinfo {author} {\bibfnamefont
  {W.}~\bibnamefont {Chang}}, \bibinfo {author} {\bibfnamefont
  {Y.}~\bibnamefont {Choi}}, \bibinfo {author} {\bibfnamefont {S.}~\bibnamefont
  {Cogan}}, \bibinfo {author} {\bibfnamefont {C.}~\bibnamefont {Compton}},
  \bibinfo {author} {\bibfnamefont {M.}~\bibnamefont {Cortesi}}, \bibinfo
  {author} {\bibfnamefont {J.}~\bibnamefont {Curtin}}, \bibinfo {author}
  {\bibfnamefont {K.}~\bibnamefont {Davidson}}, \bibinfo {author}
  {\bibfnamefont {X.}~\bibnamefont {Du}}, \bibinfo {author} {\bibfnamefont
  {K.}~\bibnamefont {Elliott}}, \bibinfo {author} {\bibfnamefont
  {B.}~\bibnamefont {Ewert}}, \bibinfo {author} {\bibfnamefont
  {A.}~\bibnamefont {Facco}}, \bibinfo {author} {\bibfnamefont
  {A.}~\bibnamefont {Fila}}, \bibinfo {author} {\bibfnamefont {K.}~\bibnamefont
  {Fukushima}}, \bibinfo {author} {\bibfnamefont {V.}~\bibnamefont {Ganni}},
  \bibinfo {author} {\bibfnamefont {A.}~\bibnamefont {Ganshyn}}, \bibinfo
  {author} {\bibfnamefont {J.}~\bibnamefont {Gao}}, \bibinfo {author}
  {\bibfnamefont {T.}~\bibnamefont {Glasmacher}}, \bibinfo {author}
  {\bibfnamefont {J.}~\bibnamefont {Guo}}, \bibinfo {author} {\bibfnamefont
  {Y.}~\bibnamefont {Hao}}, \bibinfo {author} {\bibfnamefont {W.}~\bibnamefont
  {Hartung}}, \bibinfo {author} {\bibfnamefont {N.}~\bibnamefont {Hasan}},
  \bibinfo {author} {\bibfnamefont {M.}~\bibnamefont {Hausmann}}, \bibinfo
  {author} {\bibfnamefont {K.}~\bibnamefont {Holland}}, \bibinfo {author}
  {\bibfnamefont {H.~C.}\ \bibnamefont {Hseuh}}, \bibinfo {author}
  {\bibfnamefont {M.}~\bibnamefont {Ikegami}}, \bibinfo {author} {\bibfnamefont
  {D.}~\bibnamefont {Jager}}, \bibinfo {author} {\bibfnamefont
  {S.}~\bibnamefont {Jones}}, \bibinfo {author} {\bibfnamefont
  {N.}~\bibnamefont {Joseph}}, \bibinfo {author} {\bibfnamefont
  {T.}~\bibnamefont {Kanemura}}, \bibinfo {author} {\bibfnamefont {S.-H.}\
  \bibnamefont {Kim}}, \bibinfo {author} {\bibfnamefont {P.}~\bibnamefont
  {Knudsen}}, \bibinfo {author} {\bibfnamefont {B.}~\bibnamefont {Kortum}},
  \bibinfo {author} {\bibfnamefont {E.}~\bibnamefont {Kwan}}, \bibinfo {author}
  {\bibfnamefont {T.}~\bibnamefont {Larter}}, \bibinfo {author} {\bibfnamefont
  {R.~E.}\ \bibnamefont {Laxdal}}, \bibinfo {author} {\bibfnamefont
  {M.}~\bibnamefont {Larmann}}, \bibinfo {author} {\bibfnamefont
  {K.}~\bibnamefont {Laturkar}}, \bibinfo {author} {\bibfnamefont
  {J.}~\bibnamefont {LeTourneau}}, \bibinfo {author} {\bibfnamefont {Z.-Y.}\
  \bibnamefont {Li}}, \bibinfo {author} {\bibfnamefont {S.}~\bibnamefont
  {Lidia}}, \bibinfo {author} {\bibfnamefont {G.}~\bibnamefont {Machicoane}},
  \bibinfo {author} {\bibfnamefont {C.}~\bibnamefont {Magsig}}, \bibinfo
  {author} {\bibfnamefont {P.}~\bibnamefont {Manwiller}}, \bibinfo {author}
  {\bibfnamefont {F.}~\bibnamefont {Marti}}, \bibinfo {author} {\bibfnamefont
  {T.}~\bibnamefont {Maruta}}, \bibinfo {author} {\bibfnamefont
  {A.}~\bibnamefont {McCartney}}, \bibinfo {author} {\bibfnamefont
  {E.}~\bibnamefont {Metzgar}}, \bibinfo {author} {\bibfnamefont
  {S.}~\bibnamefont {Miller}}, \bibinfo {author} {\bibfnamefont
  {Y.}~\bibnamefont {Momozaki}}, \bibinfo {author} {\bibfnamefont
  {D.}~\bibnamefont {Morris}}, \bibinfo {author} {\bibfnamefont
  {M.}~\bibnamefont {Mugerian}}, \bibinfo {author} {\bibfnamefont
  {I.}~\bibnamefont {Nesterenko}}, \bibinfo {author} {\bibfnamefont
  {C.}~\bibnamefont {Nguyen}}, \bibinfo {author} {\bibfnamefont
  {W.}~\bibnamefont {O’Brien}}, \bibinfo {author} {\bibfnamefont
  {K.}~\bibnamefont {Openlander}}, \bibinfo {author} {\bibfnamefont {P.~N.}\
  \bibnamefont {Ostroumov}}, \bibinfo {author} {\bibfnamefont {M.}~\bibnamefont
  {Patil}}, \bibinfo {author} {\bibfnamefont {A.~S.}\ \bibnamefont {Plastun}},
  \bibinfo {author} {\bibfnamefont {J.}~\bibnamefont {Popielarski}}, \bibinfo
  {author} {\bibfnamefont {L.}~\bibnamefont {Popielarski}}, \bibinfo {author}
  {\bibfnamefont {M.}~\bibnamefont {Portillo}}, \bibinfo {author}
  {\bibfnamefont {J.}~\bibnamefont {Priller}}, \bibinfo {author} {\bibfnamefont
  {X.}~\bibnamefont {Rao}}, \bibinfo {author} {\bibfnamefont {M.}~\bibnamefont
  {Reaume}}, \bibinfo {author} {\bibfnamefont {H.}~\bibnamefont {Ren}},
  \bibinfo {author} {\bibfnamefont {K.}~\bibnamefont {Saito}}, \bibinfo
  {author} {\bibfnamefont {M.}~\bibnamefont {Smith}}, \bibinfo {author}
  {\bibfnamefont {M.}~\bibnamefont {Steiner}}, \bibinfo {author} {\bibfnamefont
  {A.}~\bibnamefont {Stolz}}, \bibinfo {author} {\bibfnamefont {O.~B.}\
  \bibnamefont {Tarasov}}, \bibinfo {author} {\bibfnamefont {B.}~\bibnamefont
  {Tousignant}}, \bibinfo {author} {\bibfnamefont {R.}~\bibnamefont {Walker}},
  \bibinfo {author} {\bibfnamefont {X.}~\bibnamefont {Wang}}, \bibinfo {author}
  {\bibfnamefont {J.}~\bibnamefont {Wenstrom}}, \bibinfo {author}
  {\bibfnamefont {G.}~\bibnamefont {West}}, \bibinfo {author} {\bibfnamefont
  {K.}~\bibnamefont {Witgen}}, \bibinfo {author} {\bibfnamefont
  {M.}~\bibnamefont {Wright}}, \bibinfo {author} {\bibfnamefont
  {T.}~\bibnamefont {Xu}}, \bibinfo {author} {\bibfnamefont {Y.}~\bibnamefont
  {Xu}}, \bibinfo {author} {\bibfnamefont {Y.}~\bibnamefont {Yamazaki}},
  \bibinfo {author} {\bibfnamefont {T.}~\bibnamefont {Zhang}}, \bibinfo
  {author} {\bibfnamefont {Q.}~\bibnamefont {Zhao}}, \bibinfo {author}
  {\bibfnamefont {S.}~\bibnamefont {Zhao}}, \bibinfo {author} {\bibfnamefont
  {K.}~\bibnamefont {Dixon}}, \bibinfo {author} {\bibfnamefont
  {M.}~\bibnamefont {Wiseman}}, \bibinfo {author} {\bibfnamefont
  {M.}~\bibnamefont {Kelly}}, \bibinfo {author} {\bibfnamefont
  {K.}~\bibnamefont {Hosoyama}},\ and\ \bibinfo {author} {\bibfnamefont
  {S.}~\bibnamefont {Prestemon}},\ }\href
  {https://doi.org/10.1142/S0217732322300063} {\bibfield  {journal} {\bibinfo
  {journal} {Mod. Phys. Lett. A}\ }\textbf {\bibinfo {volume} {37}},\ \bibinfo
  {pages} {2230006} (\bibinfo {year} {2022})}\BibitemShut {NoStop}%
\bibitem [{\citenamefont {Hausmann}\ \emph {et~al.}(2013)\citenamefont
  {Hausmann}, \citenamefont {Aaron}, \citenamefont {Amthor}, \citenamefont
  {Avilov}, \citenamefont {Bandura}, \citenamefont {Bennett}, \citenamefont
  {Bollen}, \citenamefont {Borden}, \citenamefont {Burgess}, \citenamefont
  {Chouhan}, \citenamefont {Graves}, \citenamefont {Mittig}, \citenamefont
  {Morrissey}, \citenamefont {Pellemoine}, \citenamefont {Portillo},
  \citenamefont {Ronningen}, \citenamefont {Schein}, \citenamefont {Sherrill},\
  and\ \citenamefont {Zeller}}]{ARIS1}%
  \BibitemOpen
  \bibfield  {author} {\bibinfo {author} {\bibfnamefont {M.}~\bibnamefont
  {Hausmann}}, \bibinfo {author} {\bibfnamefont {A.~M.}\ \bibnamefont {Aaron}},
  \bibinfo {author} {\bibfnamefont {A.~M.}\ \bibnamefont {Amthor}}, \bibinfo
  {author} {\bibfnamefont {M.}~\bibnamefont {Avilov}}, \bibinfo {author}
  {\bibfnamefont {L.}~\bibnamefont {Bandura}}, \bibinfo {author} {\bibfnamefont
  {R.}~\bibnamefont {Bennett}}, \bibinfo {author} {\bibfnamefont
  {G.}~\bibnamefont {Bollen}}, \bibinfo {author} {\bibfnamefont
  {T.}~\bibnamefont {Borden}}, \bibinfo {author} {\bibfnamefont {T.~W.}\
  \bibnamefont {Burgess}}, \bibinfo {author} {\bibfnamefont {S.~S.}\
  \bibnamefont {Chouhan}}, \bibinfo {author} {\bibfnamefont {V.~B.}\
  \bibnamefont {Graves}}, \bibinfo {author} {\bibfnamefont {W.}~\bibnamefont
  {Mittig}}, \bibinfo {author} {\bibfnamefont {D.~J.}\ \bibnamefont
  {Morrissey}}, \bibinfo {author} {\bibfnamefont {F.}~\bibnamefont
  {Pellemoine}}, \bibinfo {author} {\bibfnamefont {M.}~\bibnamefont
  {Portillo}}, \bibinfo {author} {\bibfnamefont {R.~M.}\ \bibnamefont
  {Ronningen}}, \bibinfo {author} {\bibfnamefont {M.}~\bibnamefont {Schein}},
  \bibinfo {author} {\bibfnamefont {B.~M.}\ \bibnamefont {Sherrill}},\ and\
  \bibinfo {author} {\bibfnamefont {A.}~\bibnamefont {Zeller}},\ }\href
  {https://doi.org/10.1016/j.nimb.2013.06.042} {\bibfield  {journal} {\bibinfo
  {journal} {Nuclear Instruments and Methods in Physics Research Section B:
  Beam Interactions with Materials and Atoms}\ }\textbf {\bibinfo {volume}
  {317}},\ \bibinfo {pages} {349} (\bibinfo {year} {2013})}\BibitemShut
  {NoStop}%
\bibitem [{\citenamefont {Portillo}\ \emph {et~al.}(2023)\citenamefont
  {Portillo}, \citenamefont {Sherrill}, \citenamefont {Choi}, \citenamefont
  {Cortesi}, \citenamefont {Fukushima}, \citenamefont {Hausmann}, \citenamefont
  {Kwan}, \citenamefont {Lidia}, \citenamefont {Ostroumov}, \citenamefont
  {Ringle}, \citenamefont {Smith}, \citenamefont {Steiner}, \citenamefont
  {Tarasov}, \citenamefont {Villari},\ and\ \citenamefont {Zhang}}]{ARIS2}%
  \BibitemOpen
  \bibfield  {author} {\bibinfo {author} {\bibfnamefont {M.}~\bibnamefont
  {Portillo}}, \bibinfo {author} {\bibfnamefont {B.~M.}\ \bibnamefont
  {Sherrill}}, \bibinfo {author} {\bibfnamefont {Y.}~\bibnamefont {Choi}},
  \bibinfo {author} {\bibfnamefont {M.}~\bibnamefont {Cortesi}}, \bibinfo
  {author} {\bibfnamefont {K.}~\bibnamefont {Fukushima}}, \bibinfo {author}
  {\bibfnamefont {M.}~\bibnamefont {Hausmann}}, \bibinfo {author}
  {\bibfnamefont {E.}~\bibnamefont {Kwan}}, \bibinfo {author} {\bibfnamefont
  {S.}~\bibnamefont {Lidia}}, \bibinfo {author} {\bibfnamefont {P.~N.}\
  \bibnamefont {Ostroumov}}, \bibinfo {author} {\bibfnamefont {R.}~\bibnamefont
  {Ringle}}, \bibinfo {author} {\bibfnamefont {M.~K.}\ \bibnamefont {Smith}},
  \bibinfo {author} {\bibfnamefont {M.}~\bibnamefont {Steiner}}, \bibinfo
  {author} {\bibfnamefont {O.~B.}\ \bibnamefont {Tarasov}}, \bibinfo {author}
  {\bibfnamefont {A.~C.~C.}\ \bibnamefont {Villari}},\ and\ \bibinfo {author}
  {\bibfnamefont {T.}~\bibnamefont {Zhang}},\ }\href
  {https://doi.org/10.1016/j.nimb.2023.04.025} {\bibfield  {journal} {\bibinfo
  {journal} {Nuclear Instruments and Methods in Physics Research Section B:
  Beam Interactions with Materials and Atoms}\ }\textbf {\bibinfo {volume}
  {540}},\ \bibinfo {pages} {151} (\bibinfo {year} {2023})}\BibitemShut
  {NoStop}%
\bibitem [{\citenamefont {Bazin}\ \emph {et~al.}(2003)\citenamefont {Bazin},
  \citenamefont {Caggiano}, \citenamefont {Sherrill}, \citenamefont {Yurkon},\
  and\ \citenamefont {Zeller}}]{S800}%
  \BibitemOpen
  \bibfield  {author} {\bibinfo {author} {\bibfnamefont {D.}~\bibnamefont
  {Bazin}}, \bibinfo {author} {\bibfnamefont {J.~A.}\ \bibnamefont {Caggiano}},
  \bibinfo {author} {\bibfnamefont {B.~M.}\ \bibnamefont {Sherrill}}, \bibinfo
  {author} {\bibfnamefont {J.}~\bibnamefont {Yurkon}},\ and\ \bibinfo {author}
  {\bibfnamefont {A.}~\bibnamefont {Zeller}},\ }\href@noop {} {\bibfield
  {journal} {\bibinfo  {journal} {Nucl. Instrum. Methods Phys. Res. B}\
  }\textbf {\bibinfo {volume} {204}},\ \bibinfo {pages} {629} (\bibinfo {year}
  {2003})}\BibitemShut {NoStop}%
\bibitem [{\citenamefont {Force}\ \emph {et~al.}(2010)\citenamefont {Force},
  \citenamefont {Gr\'evy}, \citenamefont {Gaudefroy}, \citenamefont {Sorlin},
  \citenamefont {C\'aceres}, \citenamefont {Rotaru}, \citenamefont {Mrazek},
  \citenamefont {Achouri}, \citenamefont {Ang\'elique}, \citenamefont {Azaiez},
  \citenamefont {Bastin}, \citenamefont {Borcea}, \citenamefont {Buta},
  \citenamefont {Daugas}, \citenamefont {Dlouhy}, \citenamefont {Dombr\'adi},
  \citenamefont {De~Oliveira}, \citenamefont {Negoita}, \citenamefont
  {Penionzhkevich}, \citenamefont {Saint-Laurent}, \citenamefont {Sohler},
  \citenamefont {Stanoiu}, \citenamefont {Stefan}, \citenamefont {Stodel},\
  and\ \citenamefont {Nowacki}}]{Fo10}%
  \BibitemOpen
  \bibfield  {author} {\bibinfo {author} {\bibfnamefont {C.}~\bibnamefont
  {Force}}, \bibinfo {author} {\bibfnamefont {S.}~\bibnamefont {Gr\'evy}},
  \bibinfo {author} {\bibfnamefont {L.}~\bibnamefont {Gaudefroy}}, \bibinfo
  {author} {\bibfnamefont {O.}~\bibnamefont {Sorlin}}, \bibinfo {author}
  {\bibfnamefont {L.}~\bibnamefont {C\'aceres}}, \bibinfo {author}
  {\bibfnamefont {F.}~\bibnamefont {Rotaru}}, \bibinfo {author} {\bibfnamefont
  {J.}~\bibnamefont {Mrazek}}, \bibinfo {author} {\bibfnamefont {N.~L.}\
  \bibnamefont {Achouri}}, \bibinfo {author} {\bibfnamefont {J.~C.}\
  \bibnamefont {Ang\'elique}}, \bibinfo {author} {\bibfnamefont
  {F.}~\bibnamefont {Azaiez}}, \bibinfo {author} {\bibfnamefont
  {B.}~\bibnamefont {Bastin}}, \bibinfo {author} {\bibfnamefont
  {R.}~\bibnamefont {Borcea}}, \bibinfo {author} {\bibfnamefont
  {A.}~\bibnamefont {Buta}}, \bibinfo {author} {\bibfnamefont {J.~M.}\
  \bibnamefont {Daugas}}, \bibinfo {author} {\bibfnamefont {Z.}~\bibnamefont
  {Dlouhy}}, \bibinfo {author} {\bibfnamefont {Z.}~\bibnamefont {Dombr\'adi}},
  \bibinfo {author} {\bibfnamefont {F.}~\bibnamefont {De~Oliveira}}, \bibinfo
  {author} {\bibfnamefont {F.}~\bibnamefont {Negoita}}, \bibinfo {author}
  {\bibfnamefont {Y.}~\bibnamefont {Penionzhkevich}}, \bibinfo {author}
  {\bibfnamefont {M.~G.}\ \bibnamefont {Saint-Laurent}}, \bibinfo {author}
  {\bibfnamefont {D.}~\bibnamefont {Sohler}}, \bibinfo {author} {\bibfnamefont
  {M.}~\bibnamefont {Stanoiu}}, \bibinfo {author} {\bibfnamefont
  {I.}~\bibnamefont {Stefan}}, \bibinfo {author} {\bibfnamefont
  {C.}~\bibnamefont {Stodel}},\ and\ \bibinfo {author} {\bibfnamefont
  {F.}~\bibnamefont {Nowacki}},\ }\href
  {https://doi.org/10.1103/PhysRevLett.105.102501} {\bibfield  {journal}
  {\bibinfo  {journal} {Phys. Rev. Lett.}\ }\textbf {\bibinfo {volume} {105}},\
  \bibinfo {pages} {102501} (\bibinfo {year} {2010})}\BibitemShut {NoStop}%
\bibitem [{\citenamefont {Longfellow}()}]{LoPC}%
  \BibitemOpen
  \bibfield  {author} {\bibinfo {author} {\bibfnamefont {B.}~\bibnamefont
  {Longfellow}},\ }\href@noop {} {}\bibinfo {howpublished} {private
  communication}\BibitemShut {NoStop}%
\bibitem [{\citenamefont {Riley}\ \emph {et~al.}(2021)\citenamefont {Riley},
  \citenamefont {Weisshaar}, \citenamefont {Crawford}, \citenamefont
  {Agiorgousis}, \citenamefont {Campbell}, \citenamefont {Cromaz},
  \citenamefont {Fallon}, \citenamefont {Gade}, \citenamefont {Gregory},
  \citenamefont {Haldeman}, \citenamefont {Jarvis}, \citenamefont
  {Lawson-John}, \citenamefont {Roberts}, \citenamefont {Sadler},\ and\
  \citenamefont {Stine}}]{Ril21}%
  \BibitemOpen
  \bibfield  {author} {\bibinfo {author} {\bibfnamefont {L.~A.}\ \bibnamefont
  {Riley}}, \bibinfo {author} {\bibfnamefont {D.}~\bibnamefont {Weisshaar}},
  \bibinfo {author} {\bibfnamefont {H.~L.}\ \bibnamefont {Crawford}}, \bibinfo
  {author} {\bibfnamefont {M.~L.}\ \bibnamefont {Agiorgousis}}, \bibinfo
  {author} {\bibfnamefont {C.~M.}\ \bibnamefont {Campbell}}, \bibinfo {author}
  {\bibfnamefont {M.}~\bibnamefont {Cromaz}}, \bibinfo {author} {\bibfnamefont
  {P.}~\bibnamefont {Fallon}}, \bibinfo {author} {\bibfnamefont
  {A.}~\bibnamefont {Gade}}, \bibinfo {author} {\bibfnamefont {S.~D.}\
  \bibnamefont {Gregory}}, \bibinfo {author} {\bibfnamefont {E.~B.}\
  \bibnamefont {Haldeman}}, \bibinfo {author} {\bibfnamefont {L.~R.}\
  \bibnamefont {Jarvis}}, \bibinfo {author} {\bibfnamefont {E.~D.}\
  \bibnamefont {Lawson-John}}, \bibinfo {author} {\bibfnamefont
  {B.}~\bibnamefont {Roberts}}, \bibinfo {author} {\bibfnamefont {B.~V.}\
  \bibnamefont {Sadler}},\ and\ \bibinfo {author} {\bibfnamefont {C.~G.}\
  \bibnamefont {Stine}},\ }\href
  {https://doi.org/https://doi.org/10.1016/j.nima.2021.165305} {\bibfield
  {journal} {\bibinfo  {journal} {Nuclear Instruments and Methods in Physics
  Research Section A: Accelerators, Spectrometers, Detectors and Associated
  Equipment}\ }\textbf {\bibinfo {volume} {1003}},\ \bibinfo {pages} {165305}
  (\bibinfo {year} {2021})}\BibitemShut {NoStop}%
\bibitem [{\citenamefont {Paschalis}\ \emph {et~al.}(2013)\citenamefont
  {Paschalis}, \citenamefont {Lee}, \citenamefont {Macchiavelli}, \citenamefont
  {Campbell}, \citenamefont {Cromaz}, \citenamefont {Gros}, \citenamefont
  {Pavan}, \citenamefont {Qian}, \citenamefont {Clark}, \citenamefont
  {Crawford}, \citenamefont {Doering}, \citenamefont {Fallon}, \citenamefont
  {Lionberger}, \citenamefont {Loew}, \citenamefont {Petri}, \citenamefont
  {Stezelberger}, \citenamefont {Zimmermann}, \citenamefont {Radford},
  \citenamefont {Lagergren}, \citenamefont {Weisshaar}, \citenamefont
  {Winkler}, \citenamefont {Glasmacher}, \citenamefont {Anderson},\ and\
  \citenamefont {Beausang}}]{GRETINA}%
  \BibitemOpen
  \bibfield  {author} {\bibinfo {author} {\bibfnamefont {S.}~\bibnamefont
  {Paschalis}}, \bibinfo {author} {\bibfnamefont {I.~Y.}\ \bibnamefont {Lee}},
  \bibinfo {author} {\bibfnamefont {A.~O.}\ \bibnamefont {Macchiavelli}},
  \bibinfo {author} {\bibfnamefont {C.~M.}\ \bibnamefont {Campbell}}, \bibinfo
  {author} {\bibfnamefont {M.}~\bibnamefont {Cromaz}}, \bibinfo {author}
  {\bibfnamefont {S.}~\bibnamefont {Gros}}, \bibinfo {author} {\bibfnamefont
  {J.}~\bibnamefont {Pavan}}, \bibinfo {author} {\bibfnamefont
  {J.}~\bibnamefont {Qian}}, \bibinfo {author} {\bibfnamefont {R.~M.}\
  \bibnamefont {Clark}}, \bibinfo {author} {\bibfnamefont {H.~L.}\ \bibnamefont
  {Crawford}}, \bibinfo {author} {\bibfnamefont {D.}~\bibnamefont {Doering}},
  \bibinfo {author} {\bibfnamefont {P.}~\bibnamefont {Fallon}}, \bibinfo
  {author} {\bibfnamefont {C.}~\bibnamefont {Lionberger}}, \bibinfo {author}
  {\bibfnamefont {T.}~\bibnamefont {Loew}}, \bibinfo {author} {\bibfnamefont
  {M.}~\bibnamefont {Petri}}, \bibinfo {author} {\bibfnamefont
  {T.}~\bibnamefont {Stezelberger}}, \bibinfo {author} {\bibfnamefont
  {S.}~\bibnamefont {Zimmermann}}, \bibinfo {author} {\bibfnamefont {D.~C.}\
  \bibnamefont {Radford}}, \bibinfo {author} {\bibfnamefont {K.}~\bibnamefont
  {Lagergren}}, \bibinfo {author} {\bibfnamefont {D.}~\bibnamefont
  {Weisshaar}}, \bibinfo {author} {\bibfnamefont {R.}~\bibnamefont {Winkler}},
  \bibinfo {author} {\bibfnamefont {T.}~\bibnamefont {Glasmacher}}, \bibinfo
  {author} {\bibfnamefont {J.~T.}\ \bibnamefont {Anderson}},\ and\ \bibinfo
  {author} {\bibfnamefont {C.~W.}\ \bibnamefont {Beausang}},\ }\href@noop {}
  {\bibfield  {journal} {\bibinfo  {journal} {Nucl. Instrum. Methods Phys. Res.
  A}\ }\textbf {\bibinfo {volume} {709}},\ \bibinfo {pages} {44} (\bibinfo
  {year} {2013})}\BibitemShut {NoStop}%
\bibitem [{\citenamefont {Weisshaar}\ \emph {et~al.}(2017)\citenamefont
  {Weisshaar}, \citenamefont {Bazin}, \citenamefont {Bender}, \citenamefont
  {Campbell}, \citenamefont {Recchia}, \citenamefont {Bader}, \citenamefont
  {Baugher}, \citenamefont {Belarge}, \citenamefont {Carpenter}, \citenamefont
  {Crawford}, \citenamefont {Cromaz}, \citenamefont {Elman}, \citenamefont
  {Fallon}, \citenamefont {Forney}, \citenamefont {Gade}, \citenamefont
  {Harker}, \citenamefont {Kobayashi}, \citenamefont {Langer}, \citenamefont
  {Lauritsen}, \citenamefont {Lee}, \citenamefont {Lemasson}, \citenamefont
  {Longfellow}, \citenamefont {Lunderberg}, \citenamefont {Macchiavelli},
  \citenamefont {Miki}, \citenamefont {Momiyama}, \citenamefont {Noji},
  \citenamefont {Radford}, \citenamefont {Scott}, \citenamefont {Sethi},
  \citenamefont {Stroberg}, \citenamefont {Sullivan}, \citenamefont {Titus},
  \citenamefont {Wiens}, \citenamefont {Williams}, \citenamefont {Wimmer},\
  and\ \citenamefont {Zhu}}]{GRETINA2}%
  \BibitemOpen
  \bibfield  {author} {\bibinfo {author} {\bibfnamefont {D.}~\bibnamefont
  {Weisshaar}}, \bibinfo {author} {\bibfnamefont {D.}~\bibnamefont {Bazin}},
  \bibinfo {author} {\bibfnamefont {P.~C.}\ \bibnamefont {Bender}}, \bibinfo
  {author} {\bibfnamefont {C.~M.}\ \bibnamefont {Campbell}}, \bibinfo {author}
  {\bibfnamefont {F.}~\bibnamefont {Recchia}}, \bibinfo {author} {\bibfnamefont
  {V.}~\bibnamefont {Bader}}, \bibinfo {author} {\bibfnamefont
  {T.}~\bibnamefont {Baugher}}, \bibinfo {author} {\bibfnamefont
  {J.}~\bibnamefont {Belarge}}, \bibinfo {author} {\bibfnamefont {M.~P.}\
  \bibnamefont {Carpenter}}, \bibinfo {author} {\bibfnamefont {H.~L.}\
  \bibnamefont {Crawford}}, \bibinfo {author} {\bibfnamefont {M.}~\bibnamefont
  {Cromaz}}, \bibinfo {author} {\bibfnamefont {B.}~\bibnamefont {Elman}},
  \bibinfo {author} {\bibfnamefont {P.}~\bibnamefont {Fallon}}, \bibinfo
  {author} {\bibfnamefont {A.}~\bibnamefont {Forney}}, \bibinfo {author}
  {\bibfnamefont {A.}~\bibnamefont {Gade}}, \bibinfo {author} {\bibfnamefont
  {J.}~\bibnamefont {Harker}}, \bibinfo {author} {\bibfnamefont
  {N.}~\bibnamefont {Kobayashi}}, \bibinfo {author} {\bibfnamefont
  {C.}~\bibnamefont {Langer}}, \bibinfo {author} {\bibfnamefont
  {T.}~\bibnamefont {Lauritsen}}, \bibinfo {author} {\bibfnamefont {I.~Y.}\
  \bibnamefont {Lee}}, \bibinfo {author} {\bibfnamefont {A.}~\bibnamefont
  {Lemasson}}, \bibinfo {author} {\bibfnamefont {B.}~\bibnamefont
  {Longfellow}}, \bibinfo {author} {\bibfnamefont {E.}~\bibnamefont
  {Lunderberg}}, \bibinfo {author} {\bibfnamefont {A.~O.}\ \bibnamefont
  {Macchiavelli}}, \bibinfo {author} {\bibfnamefont {K.}~\bibnamefont {Miki}},
  \bibinfo {author} {\bibfnamefont {S.}~\bibnamefont {Momiyama}}, \bibinfo
  {author} {\bibfnamefont {S.}~\bibnamefont {Noji}}, \bibinfo {author}
  {\bibfnamefont {D.~C.}\ \bibnamefont {Radford}}, \bibinfo {author}
  {\bibfnamefont {M.}~\bibnamefont {Scott}}, \bibinfo {author} {\bibfnamefont
  {J.}~\bibnamefont {Sethi}}, \bibinfo {author} {\bibfnamefont {S.~R.}\
  \bibnamefont {Stroberg}}, \bibinfo {author} {\bibfnamefont {C.}~\bibnamefont
  {Sullivan}}, \bibinfo {author} {\bibfnamefont {R.}~\bibnamefont {Titus}},
  \bibinfo {author} {\bibfnamefont {A.}~\bibnamefont {Wiens}}, \bibinfo
  {author} {\bibfnamefont {S.}~\bibnamefont {Williams}}, \bibinfo {author}
  {\bibfnamefont {K.}~\bibnamefont {Wimmer}},\ and\ \bibinfo {author}
  {\bibfnamefont {S.}~\bibnamefont {Zhu}},\ }\href
  {https://doi.org/10.1016/j.nima.2016.12.001} {\bibfield  {journal} {\bibinfo
  {journal} {Nuclear Instruments and Methods in Physics Research Section A:
  Accelerators, Spectrometers, Detectors and Associated Equipment}\ }\textbf
  {\bibinfo {volume} {847}},\ \bibinfo {pages} {187} (\bibinfo {year}
  {2017})}\BibitemShut {NoStop}%
\bibitem [{\citenamefont {Winther}\ and\ \citenamefont {Alder}(1979)}]{Win79}%
  \BibitemOpen
  \bibfield  {author} {\bibinfo {author} {\bibfnamefont {A.}~\bibnamefont
  {Winther}}\ and\ \bibinfo {author} {\bibfnamefont {K.}~\bibnamefont
  {Alder}},\ }\href {https://doi.org/10.1016/0375-9474(79)90528-1} {\bibfield
  {journal} {\bibinfo  {journal} {Nuclear Physics A}\ }\textbf {\bibinfo
  {volume} {319}},\ \bibinfo {pages} {518} (\bibinfo {year}
  {1979})}\BibitemShut {NoStop}%
\bibitem [{\citenamefont {Takeuchi}\ \emph {et~al.}(2012)\citenamefont
  {Takeuchi}, \citenamefont {Matsushita}, \citenamefont {Aoi}, \citenamefont
  {Doornenbal}, \citenamefont {Li}, \citenamefont {Motobayashi}, \citenamefont
  {Scheit}, \citenamefont {Steppenbeck}, \citenamefont {Wang}, \citenamefont
  {Baba}, \citenamefont {Bazin}, \citenamefont {C\`aceres}, \citenamefont
  {Crawford}, \citenamefont {Fallon}, \citenamefont {Gernh\"auser},
  \citenamefont {Gibelin}, \citenamefont {Go}, \citenamefont {Gr\'evy},
  \citenamefont {Hinke}, \citenamefont {Hoffman}, \citenamefont {Hughes},
  \citenamefont {Ideguchi}, \citenamefont {Jenkins}, \citenamefont {Kobayashi},
  \citenamefont {Kondo}, \citenamefont {Kr\"ucken}, \citenamefont {Le~Bleis},
  \citenamefont {Lee}, \citenamefont {Lee}, \citenamefont {Matta},
  \citenamefont {Michimasa}, \citenamefont {Nakamura}, \citenamefont {Ota},
  \citenamefont {Petri}, \citenamefont {Sako}, \citenamefont {Sakurai},
  \citenamefont {Shimoura}, \citenamefont {Steiger}, \citenamefont {Takahashi},
  \citenamefont {Takechi}, \citenamefont {Togano}, \citenamefont {Winkler},\
  and\ \citenamefont {Yoneda}}]{Tak12}%
  \BibitemOpen
  \bibfield  {author} {\bibinfo {author} {\bibfnamefont {S.}~\bibnamefont
  {Takeuchi}}, \bibinfo {author} {\bibfnamefont {M.}~\bibnamefont
  {Matsushita}}, \bibinfo {author} {\bibfnamefont {N.}~\bibnamefont {Aoi}},
  \bibinfo {author} {\bibfnamefont {P.}~\bibnamefont {Doornenbal}}, \bibinfo
  {author} {\bibfnamefont {K.}~\bibnamefont {Li}}, \bibinfo {author}
  {\bibfnamefont {T.}~\bibnamefont {Motobayashi}}, \bibinfo {author}
  {\bibfnamefont {H.}~\bibnamefont {Scheit}}, \bibinfo {author} {\bibfnamefont
  {D.}~\bibnamefont {Steppenbeck}}, \bibinfo {author} {\bibfnamefont
  {H.}~\bibnamefont {Wang}}, \bibinfo {author} {\bibfnamefont {H.}~\bibnamefont
  {Baba}}, \bibinfo {author} {\bibfnamefont {D.}~\bibnamefont {Bazin}},
  \bibinfo {author} {\bibfnamefont {L.}~\bibnamefont {C\`aceres}}, \bibinfo
  {author} {\bibfnamefont {H.}~\bibnamefont {Crawford}}, \bibinfo {author}
  {\bibfnamefont {P.}~\bibnamefont {Fallon}}, \bibinfo {author} {\bibfnamefont
  {R.}~\bibnamefont {Gernh\"auser}}, \bibinfo {author} {\bibfnamefont
  {J.}~\bibnamefont {Gibelin}}, \bibinfo {author} {\bibfnamefont
  {S.}~\bibnamefont {Go}}, \bibinfo {author} {\bibfnamefont {S.}~\bibnamefont
  {Gr\'evy}}, \bibinfo {author} {\bibfnamefont {C.}~\bibnamefont {Hinke}},
  \bibinfo {author} {\bibfnamefont {C.~R.}\ \bibnamefont {Hoffman}}, \bibinfo
  {author} {\bibfnamefont {R.}~\bibnamefont {Hughes}}, \bibinfo {author}
  {\bibfnamefont {E.}~\bibnamefont {Ideguchi}}, \bibinfo {author}
  {\bibfnamefont {D.}~\bibnamefont {Jenkins}}, \bibinfo {author} {\bibfnamefont
  {N.}~\bibnamefont {Kobayashi}}, \bibinfo {author} {\bibfnamefont
  {Y.}~\bibnamefont {Kondo}}, \bibinfo {author} {\bibfnamefont
  {R.}~\bibnamefont {Kr\"ucken}}, \bibinfo {author} {\bibfnamefont
  {T.}~\bibnamefont {Le~Bleis}}, \bibinfo {author} {\bibfnamefont
  {J.}~\bibnamefont {Lee}}, \bibinfo {author} {\bibfnamefont {G.}~\bibnamefont
  {Lee}}, \bibinfo {author} {\bibfnamefont {A.}~\bibnamefont {Matta}}, \bibinfo
  {author} {\bibfnamefont {S.}~\bibnamefont {Michimasa}}, \bibinfo {author}
  {\bibfnamefont {T.}~\bibnamefont {Nakamura}}, \bibinfo {author}
  {\bibfnamefont {S.}~\bibnamefont {Ota}}, \bibinfo {author} {\bibfnamefont
  {M.}~\bibnamefont {Petri}}, \bibinfo {author} {\bibfnamefont
  {T.}~\bibnamefont {Sako}}, \bibinfo {author} {\bibfnamefont {H.}~\bibnamefont
  {Sakurai}}, \bibinfo {author} {\bibfnamefont {S.}~\bibnamefont {Shimoura}},
  \bibinfo {author} {\bibfnamefont {K.}~\bibnamefont {Steiger}}, \bibinfo
  {author} {\bibfnamefont {K.}~\bibnamefont {Takahashi}}, \bibinfo {author}
  {\bibfnamefont {M.}~\bibnamefont {Takechi}}, \bibinfo {author} {\bibfnamefont
  {Y.}~\bibnamefont {Togano}}, \bibinfo {author} {\bibfnamefont
  {R.}~\bibnamefont {Winkler}},\ and\ \bibinfo {author} {\bibfnamefont
  {K.}~\bibnamefont {Yoneda}},\ }\href
  {https://doi.org/10.1103/PhysRevLett.109.182501} {\bibfield  {journal}
  {\bibinfo  {journal} {Phys. Rev. Lett.}\ }\textbf {\bibinfo {volume} {109}},\
  \bibinfo {pages} {182501} (\bibinfo {year} {2012})}\BibitemShut {NoStop}%
\bibitem [{\citenamefont {Agostinelli}\ \emph {et~al.}(2003)\citenamefont
  {Agostinelli}, \citenamefont {Allison}, \citenamefont {Amako}, \citenamefont
  {Apostolakis}, \citenamefont {Araujo}, \citenamefont {Arce}, \citenamefont
  {Asai}, \citenamefont {Axen}, \citenamefont {Banerjee}, \citenamefont
  {Barrand}, \citenamefont {Behner}, \citenamefont {Bellagamba}, \citenamefont
  {Boudreau}, \citenamefont {Broglia}, \citenamefont {Brunengo}, \citenamefont
  {Burkhardt}, \citenamefont {Chauvie}, \citenamefont {Chuma}, \citenamefont
  {Chytracek}, \citenamefont {Cooperman}, \citenamefont {Cosmo}, \citenamefont
  {Degtyarenko}, \citenamefont {Dell'Acqua}, \citenamefont {Depaola},
  \citenamefont {Dietrich}, \citenamefont {Enami}, \citenamefont {Feliciello},
  \citenamefont {Ferguson}, \citenamefont {Fesefeldt}, \citenamefont {Folger},
  \citenamefont {Foppiano}, \citenamefont {Forti}, \citenamefont {Garelli},
  \citenamefont {Giani}, \citenamefont {Giannitrapani}, \citenamefont {Gibin},
  \citenamefont {{Gómez Cadenas}}, \citenamefont {González}, \citenamefont
  {{Gracia Abril}}, \citenamefont {Greeniaus}, \citenamefont {Greiner},
  \citenamefont {Grichine}, \citenamefont {Grossheim}, \citenamefont
  {Guatelli}, \citenamefont {Gumplinger}, \citenamefont {Hamatsu},
  \citenamefont {Hashimoto}, \citenamefont {Hasui}, \citenamefont {Heikkinen},
  \citenamefont {Howard}, \citenamefont {Ivanchenko}, \citenamefont {Johnson},
  \citenamefont {Jones}, \citenamefont {Kallenbach}, \citenamefont {Kanaya},
  \citenamefont {Kawabata}, \citenamefont {Kawabata}, \citenamefont {Kawaguti},
  \citenamefont {Kelner}, \citenamefont {Kent}, \citenamefont {Kimura},
  \citenamefont {Kodama}, \citenamefont {Kokoulin}, \citenamefont {Kossov},
  \citenamefont {Kurashige}, \citenamefont {Lamanna}, \citenamefont {Lampén},
  \citenamefont {Lara}, \citenamefont {Lefebure}, \citenamefont {Lei},
  \citenamefont {Liendl}, \citenamefont {Lockman}, \citenamefont {Longo},
  \citenamefont {Magni}, \citenamefont {Maire}, \citenamefont {Medernach},
  \citenamefont {Minamimoto}, \citenamefont {{Mora de Freitas}}, \citenamefont
  {Morita}, \citenamefont {Murakami}, \citenamefont {Nagamatu}, \citenamefont
  {Nartallo}, \citenamefont {Nieminen}, \citenamefont {Nishimura},
  \citenamefont {Ohtsubo}, \citenamefont {Okamura}, \citenamefont {O'Neale},
  \citenamefont {Oohata}, \citenamefont {Paech}, \citenamefont {Perl},
  \citenamefont {Pfeiffer}, \citenamefont {Pia}, \citenamefont {Ranjard},
  \citenamefont {Rybin}, \citenamefont {Sadilov}, \citenamefont {{Di Salvo}},
  \citenamefont {Santin}, \citenamefont {Sasaki}, \citenamefont {Savvas},
  \citenamefont {Sawada}, \citenamefont {Scherer}, \citenamefont {Sei},
  \citenamefont {Sirotenko}, \citenamefont {Smith}, \citenamefont {Starkov},
  \citenamefont {Stoecker}, \citenamefont {Sulkimo}, \citenamefont {Takahata},
  \citenamefont {Tanaka}, \citenamefont {Tcherniaev}, \citenamefont {{Safai
  Tehrani}}, \citenamefont {Tropeano}, \citenamefont {Truscott}, \citenamefont
  {Uno}, \citenamefont {Urban}, \citenamefont {Urban}, \citenamefont {Verderi},
  \citenamefont {Walkden}, \citenamefont {Wander}, \citenamefont {Weber},
  \citenamefont {Wellisch}, \citenamefont {Wenaus}, \citenamefont {Williams},
  \citenamefont {Wright}, \citenamefont {Yamada}, \citenamefont {Yoshida},\
  and\ \citenamefont {Zschiesche}}]{Geant4}%
  \BibitemOpen
  \bibfield  {author} {\bibinfo {author} {\bibfnamefont {S.}~\bibnamefont
  {Agostinelli}}, \bibinfo {author} {\bibfnamefont {J.}~\bibnamefont
  {Allison}}, \bibinfo {author} {\bibfnamefont {K.}~\bibnamefont {Amako}},
  \bibinfo {author} {\bibfnamefont {J.}~\bibnamefont {Apostolakis}}, \bibinfo
  {author} {\bibfnamefont {H.}~\bibnamefont {Araujo}}, \bibinfo {author}
  {\bibfnamefont {P.}~\bibnamefont {Arce}}, \bibinfo {author} {\bibfnamefont
  {M.}~\bibnamefont {Asai}}, \bibinfo {author} {\bibfnamefont {D.}~\bibnamefont
  {Axen}}, \bibinfo {author} {\bibfnamefont {S.}~\bibnamefont {Banerjee}},
  \bibinfo {author} {\bibfnamefont {G.}~\bibnamefont {Barrand}}, \bibinfo
  {author} {\bibfnamefont {F.}~\bibnamefont {Behner}}, \bibinfo {author}
  {\bibfnamefont {L.}~\bibnamefont {Bellagamba}}, \bibinfo {author}
  {\bibfnamefont {J.}~\bibnamefont {Boudreau}}, \bibinfo {author}
  {\bibfnamefont {L.}~\bibnamefont {Broglia}}, \bibinfo {author} {\bibfnamefont
  {A.}~\bibnamefont {Brunengo}}, \bibinfo {author} {\bibfnamefont
  {H.}~\bibnamefont {Burkhardt}}, \bibinfo {author} {\bibfnamefont
  {S.}~\bibnamefont {Chauvie}}, \bibinfo {author} {\bibfnamefont
  {J.}~\bibnamefont {Chuma}}, \bibinfo {author} {\bibfnamefont
  {R.}~\bibnamefont {Chytracek}}, \bibinfo {author} {\bibfnamefont
  {G.}~\bibnamefont {Cooperman}}, \bibinfo {author} {\bibfnamefont
  {G.}~\bibnamefont {Cosmo}}, \bibinfo {author} {\bibfnamefont
  {P.}~\bibnamefont {Degtyarenko}}, \bibinfo {author} {\bibfnamefont
  {A.}~\bibnamefont {Dell'Acqua}}, \bibinfo {author} {\bibfnamefont
  {G.}~\bibnamefont {Depaola}}, \bibinfo {author} {\bibfnamefont
  {D.}~\bibnamefont {Dietrich}}, \bibinfo {author} {\bibfnamefont
  {R.}~\bibnamefont {Enami}}, \bibinfo {author} {\bibfnamefont
  {A.}~\bibnamefont {Feliciello}}, \bibinfo {author} {\bibfnamefont
  {C.}~\bibnamefont {Ferguson}}, \bibinfo {author} {\bibfnamefont
  {H.}~\bibnamefont {Fesefeldt}}, \bibinfo {author} {\bibfnamefont
  {G.}~\bibnamefont {Folger}}, \bibinfo {author} {\bibfnamefont
  {F.}~\bibnamefont {Foppiano}}, \bibinfo {author} {\bibfnamefont
  {A.}~\bibnamefont {Forti}}, \bibinfo {author} {\bibfnamefont
  {S.}~\bibnamefont {Garelli}}, \bibinfo {author} {\bibfnamefont
  {S.}~\bibnamefont {Giani}}, \bibinfo {author} {\bibfnamefont
  {R.}~\bibnamefont {Giannitrapani}}, \bibinfo {author} {\bibfnamefont
  {D.}~\bibnamefont {Gibin}}, \bibinfo {author} {\bibfnamefont
  {J.}~\bibnamefont {{Gómez Cadenas}}}, \bibinfo {author} {\bibfnamefont
  {I.}~\bibnamefont {González}}, \bibinfo {author} {\bibfnamefont
  {G.}~\bibnamefont {{Gracia Abril}}}, \bibinfo {author} {\bibfnamefont
  {G.}~\bibnamefont {Greeniaus}}, \bibinfo {author} {\bibfnamefont
  {W.}~\bibnamefont {Greiner}}, \bibinfo {author} {\bibfnamefont
  {V.}~\bibnamefont {Grichine}}, \bibinfo {author} {\bibfnamefont
  {A.}~\bibnamefont {Grossheim}}, \bibinfo {author} {\bibfnamefont
  {S.}~\bibnamefont {Guatelli}}, \bibinfo {author} {\bibfnamefont
  {P.}~\bibnamefont {Gumplinger}}, \bibinfo {author} {\bibfnamefont
  {R.}~\bibnamefont {Hamatsu}}, \bibinfo {author} {\bibfnamefont
  {K.}~\bibnamefont {Hashimoto}}, \bibinfo {author} {\bibfnamefont
  {H.}~\bibnamefont {Hasui}}, \bibinfo {author} {\bibfnamefont
  {A.}~\bibnamefont {Heikkinen}}, \bibinfo {author} {\bibfnamefont
  {A.}~\bibnamefont {Howard}}, \bibinfo {author} {\bibfnamefont
  {V.}~\bibnamefont {Ivanchenko}}, \bibinfo {author} {\bibfnamefont
  {A.}~\bibnamefont {Johnson}}, \bibinfo {author} {\bibfnamefont
  {F.}~\bibnamefont {Jones}}, \bibinfo {author} {\bibfnamefont
  {J.}~\bibnamefont {Kallenbach}}, \bibinfo {author} {\bibfnamefont
  {N.}~\bibnamefont {Kanaya}}, \bibinfo {author} {\bibfnamefont
  {M.}~\bibnamefont {Kawabata}}, \bibinfo {author} {\bibfnamefont
  {Y.}~\bibnamefont {Kawabata}}, \bibinfo {author} {\bibfnamefont
  {M.}~\bibnamefont {Kawaguti}}, \bibinfo {author} {\bibfnamefont
  {S.}~\bibnamefont {Kelner}}, \bibinfo {author} {\bibfnamefont
  {P.}~\bibnamefont {Kent}}, \bibinfo {author} {\bibfnamefont {A.}~\bibnamefont
  {Kimura}}, \bibinfo {author} {\bibfnamefont {T.}~\bibnamefont {Kodama}},
  \bibinfo {author} {\bibfnamefont {R.}~\bibnamefont {Kokoulin}}, \bibinfo
  {author} {\bibfnamefont {M.}~\bibnamefont {Kossov}}, \bibinfo {author}
  {\bibfnamefont {H.}~\bibnamefont {Kurashige}}, \bibinfo {author}
  {\bibfnamefont {E.}~\bibnamefont {Lamanna}}, \bibinfo {author} {\bibfnamefont
  {T.}~\bibnamefont {Lampén}}, \bibinfo {author} {\bibfnamefont
  {V.}~\bibnamefont {Lara}}, \bibinfo {author} {\bibfnamefont {V.}~\bibnamefont
  {Lefebure}}, \bibinfo {author} {\bibfnamefont {F.}~\bibnamefont {Lei}},
  \bibinfo {author} {\bibfnamefont {M.}~\bibnamefont {Liendl}}, \bibinfo
  {author} {\bibfnamefont {W.}~\bibnamefont {Lockman}}, \bibinfo {author}
  {\bibfnamefont {F.}~\bibnamefont {Longo}}, \bibinfo {author} {\bibfnamefont
  {S.}~\bibnamefont {Magni}}, \bibinfo {author} {\bibfnamefont
  {M.}~\bibnamefont {Maire}}, \bibinfo {author} {\bibfnamefont
  {E.}~\bibnamefont {Medernach}}, \bibinfo {author} {\bibfnamefont
  {K.}~\bibnamefont {Minamimoto}}, \bibinfo {author} {\bibfnamefont
  {P.}~\bibnamefont {{Mora de Freitas}}}, \bibinfo {author} {\bibfnamefont
  {Y.}~\bibnamefont {Morita}}, \bibinfo {author} {\bibfnamefont
  {K.}~\bibnamefont {Murakami}}, \bibinfo {author} {\bibfnamefont
  {M.}~\bibnamefont {Nagamatu}}, \bibinfo {author} {\bibfnamefont
  {R.}~\bibnamefont {Nartallo}}, \bibinfo {author} {\bibfnamefont
  {P.}~\bibnamefont {Nieminen}}, \bibinfo {author} {\bibfnamefont
  {T.}~\bibnamefont {Nishimura}}, \bibinfo {author} {\bibfnamefont
  {K.}~\bibnamefont {Ohtsubo}}, \bibinfo {author} {\bibfnamefont
  {M.}~\bibnamefont {Okamura}}, \bibinfo {author} {\bibfnamefont
  {S.}~\bibnamefont {O'Neale}}, \bibinfo {author} {\bibfnamefont
  {Y.}~\bibnamefont {Oohata}}, \bibinfo {author} {\bibfnamefont
  {K.}~\bibnamefont {Paech}}, \bibinfo {author} {\bibfnamefont
  {J.}~\bibnamefont {Perl}}, \bibinfo {author} {\bibfnamefont {A.}~\bibnamefont
  {Pfeiffer}}, \bibinfo {author} {\bibfnamefont {M.}~\bibnamefont {Pia}},
  \bibinfo {author} {\bibfnamefont {F.}~\bibnamefont {Ranjard}}, \bibinfo
  {author} {\bibfnamefont {A.}~\bibnamefont {Rybin}}, \bibinfo {author}
  {\bibfnamefont {S.}~\bibnamefont {Sadilov}}, \bibinfo {author} {\bibfnamefont
  {E.}~\bibnamefont {{Di Salvo}}}, \bibinfo {author} {\bibfnamefont
  {G.}~\bibnamefont {Santin}}, \bibinfo {author} {\bibfnamefont
  {T.}~\bibnamefont {Sasaki}}, \bibinfo {author} {\bibfnamefont
  {N.}~\bibnamefont {Savvas}}, \bibinfo {author} {\bibfnamefont
  {Y.}~\bibnamefont {Sawada}}, \bibinfo {author} {\bibfnamefont
  {S.}~\bibnamefont {Scherer}}, \bibinfo {author} {\bibfnamefont
  {S.}~\bibnamefont {Sei}}, \bibinfo {author} {\bibfnamefont {V.}~\bibnamefont
  {Sirotenko}}, \bibinfo {author} {\bibfnamefont {D.}~\bibnamefont {Smith}},
  \bibinfo {author} {\bibfnamefont {N.}~\bibnamefont {Starkov}}, \bibinfo
  {author} {\bibfnamefont {H.}~\bibnamefont {Stoecker}}, \bibinfo {author}
  {\bibfnamefont {J.}~\bibnamefont {Sulkimo}}, \bibinfo {author} {\bibfnamefont
  {M.}~\bibnamefont {Takahata}}, \bibinfo {author} {\bibfnamefont
  {S.}~\bibnamefont {Tanaka}}, \bibinfo {author} {\bibfnamefont
  {E.}~\bibnamefont {Tcherniaev}}, \bibinfo {author} {\bibfnamefont
  {E.}~\bibnamefont {{Safai Tehrani}}}, \bibinfo {author} {\bibfnamefont
  {M.}~\bibnamefont {Tropeano}}, \bibinfo {author} {\bibfnamefont
  {P.}~\bibnamefont {Truscott}}, \bibinfo {author} {\bibfnamefont
  {H.}~\bibnamefont {Uno}}, \bibinfo {author} {\bibfnamefont {L.}~\bibnamefont
  {Urban}}, \bibinfo {author} {\bibfnamefont {P.}~\bibnamefont {Urban}},
  \bibinfo {author} {\bibfnamefont {M.}~\bibnamefont {Verderi}}, \bibinfo
  {author} {\bibfnamefont {A.}~\bibnamefont {Walkden}}, \bibinfo {author}
  {\bibfnamefont {W.}~\bibnamefont {Wander}}, \bibinfo {author} {\bibfnamefont
  {H.}~\bibnamefont {Weber}}, \bibinfo {author} {\bibfnamefont
  {J.}~\bibnamefont {Wellisch}}, \bibinfo {author} {\bibfnamefont
  {T.}~\bibnamefont {Wenaus}}, \bibinfo {author} {\bibfnamefont
  {D.}~\bibnamefont {Williams}}, \bibinfo {author} {\bibfnamefont
  {D.}~\bibnamefont {Wright}}, \bibinfo {author} {\bibfnamefont
  {T.}~\bibnamefont {Yamada}}, \bibinfo {author} {\bibfnamefont
  {H.}~\bibnamefont {Yoshida}},\ and\ \bibinfo {author} {\bibfnamefont
  {D.}~\bibnamefont {Zschiesche}},\ }\href
  {https://doi.org/https://doi.org/10.1016/S0168-9002(03)01368-8} {\bibfield
  {journal} {\bibinfo  {journal} {Nuclear Instruments and Methods in Physics
  Research Section A: Accelerators, Spectrometers, Detectors and Associated
  Equipment}\ }\textbf {\bibinfo {volume} {506}},\ \bibinfo {pages} {250}
  (\bibinfo {year} {2003})}\BibitemShut {NoStop}%
\bibitem [{\citenamefont {Pritychenko}\ \emph {et~al.}(2016)\citenamefont
  {Pritychenko}, \citenamefont {Birch}, \citenamefont {Singh},\ and\
  \citenamefont {Horoi}}]{Pri16}%
  \BibitemOpen
  \bibfield  {author} {\bibinfo {author} {\bibfnamefont {B.}~\bibnamefont
  {Pritychenko}}, \bibinfo {author} {\bibfnamefont {M.}~\bibnamefont {Birch}},
  \bibinfo {author} {\bibfnamefont {B.}~\bibnamefont {Singh}},\ and\ \bibinfo
  {author} {\bibfnamefont {M.}~\bibnamefont {Horoi}},\ }\href
  {https://doi.org/10.1016/j.adt.2015.10.001} {\bibfield  {journal} {\bibinfo
  {journal} {Atomic Data and Nuclear Data Tables}\ }\textbf {\bibinfo {volume}
  {107}},\ \bibinfo {pages} {1} (\bibinfo {year} {2016})}\BibitemShut {NoStop}%
\bibitem [{\citenamefont {Chen}\ and\ \citenamefont {Singh}(2023)}]{Che23}%
  \BibitemOpen
  \bibfield  {author} {\bibinfo {author} {\bibfnamefont {J.}~\bibnamefont
  {Chen}}\ and\ \bibinfo {author} {\bibfnamefont {B.}~\bibnamefont {Singh}},\
  }\href@noop {} {\bibfield  {journal} {\bibinfo  {journal} {Nuclear Data
  Sheets}\ }\textbf {\bibinfo {volume} {190}},\ \bibinfo {pages} {1} (\bibinfo
  {year} {2023})}\BibitemShut {NoStop}%
\bibitem [{\citenamefont {Lemmon}\ \emph {et~al.}(2024)\citenamefont {Lemmon},
  \citenamefont {Bell}, \citenamefont {Huber},\ and\ \citenamefont
  {McLinden}}]{Lem24}%
  \BibitemOpen
  \bibfield  {author} {\bibinfo {author} {\bibfnamefont {E.~W.}\ \bibnamefont
  {Lemmon}}, \bibinfo {author} {\bibfnamefont {I.~H.}\ \bibnamefont {Bell}},
  \bibinfo {author} {\bibfnamefont {M.~L.}\ \bibnamefont {Huber}},\ and\
  \bibinfo {author} {\bibfnamefont {M.~O.}\ \bibnamefont {McLinden}},\
  }\bibinfo {title} {Nist chemistry webbook, nist standard reference database
  number 69}\ (\bibinfo  {publisher} {National Institute of Standards and
  Technology},\ \bibinfo {year} {2024})\ Chap.\ \bibinfo {chapter}
  {Thermophysical Properties of Fluid System},\ \bibinfo {note}
  {\url{https://doi.org/10.18434/T4D303}(retrieved 2024-03-26)}\BibitemShut
  {NoStop}%
\bibitem [{\citenamefont {James}(1981)}]{Jam81}%
  \BibitemOpen
  \bibfield  {author} {\bibinfo {author} {\bibfnamefont {F.}~\bibnamefont
  {James}},\ }\href@noop {} {\emph {\bibinfo {title} {Determining the
  statistical significance of experimental results}}},\ \bibinfo {type} {Tech.
  Rep.}\ \bibinfo {number} {Technical Report DD/81/02}\ (\bibinfo
  {institution} {CERN},\ \bibinfo {year} {1981})\BibitemShut {NoStop}%
\bibitem [{\citenamefont {Thompson}(1988)}]{FRESCO}%
  \BibitemOpen
  \bibfield  {author} {\bibinfo {author} {\bibfnamefont {I.~J.}\ \bibnamefont
  {Thompson}},\ }\href@noop {} {\bibfield  {journal} {\bibinfo  {journal}
  {Computer Physics Reports}\ }\textbf {\bibinfo {volume} {7}},\ \bibinfo
  {pages} {167} (\bibinfo {year} {1988})}\BibitemShut {NoStop}%
\bibitem [{\citenamefont {Furumoto}\ \emph {et~al.}(2012)\citenamefont
  {Furumoto}, \citenamefont {Horiuchi}, \citenamefont {Takashina},
  \citenamefont {Yamamoto},\ and\ \citenamefont {Sakuragi}}]{Fur12}%
  \BibitemOpen
  \bibfield  {author} {\bibinfo {author} {\bibfnamefont {T.}~\bibnamefont
  {Furumoto}}, \bibinfo {author} {\bibfnamefont {W.}~\bibnamefont {Horiuchi}},
  \bibinfo {author} {\bibfnamefont {M.}~\bibnamefont {Takashina}}, \bibinfo
  {author} {\bibfnamefont {Y.}~\bibnamefont {Yamamoto}},\ and\ \bibinfo
  {author} {\bibfnamefont {Y.}~\bibnamefont {Sakuragi}},\ }\href
  {https://doi.org/10.1103/PhysRevC.85.044607} {\bibfield  {journal} {\bibinfo
  {journal} {Phys. Rev. C}\ }\textbf {\bibinfo {volume} {85}},\ \bibinfo
  {pages} {044607} (\bibinfo {year} {2012})}\BibitemShut {NoStop}%
\bibitem [{\citenamefont {Koning}\ and\ \citenamefont
  {Delaroche}(2003)}]{Kon03}%
  \BibitemOpen
  \bibfield  {author} {\bibinfo {author} {\bibfnamefont {A.~J.}\ \bibnamefont
  {Koning}}\ and\ \bibinfo {author} {\bibfnamefont {J.-P.}\ \bibnamefont
  {Delaroche}},\ }\href@noop {} {\bibfield  {journal} {\bibinfo  {journal}
  {Nucl. Phys. A}\ }\textbf {\bibinfo {volume} {713}},\ \bibinfo {pages} {231}
  (\bibinfo {year} {2003})}\BibitemShut {NoStop}%
\bibitem [{\citenamefont {Bernstein}\ \emph {et~al.}(1983)\citenamefont
  {Bernstein}, \citenamefont {Brown},\ and\ \citenamefont {Madsen}}]{Be83}%
  \BibitemOpen
  \bibfield  {author} {\bibinfo {author} {\bibfnamefont {A.~M.}\ \bibnamefont
  {Bernstein}}, \bibinfo {author} {\bibfnamefont {V.~R.}\ \bibnamefont
  {Brown}},\ and\ \bibinfo {author} {\bibfnamefont {V.~A.}\ \bibnamefont
  {Madsen}},\ }\href@noop {} {\bibfield  {journal} {\bibinfo  {journal}
  {Comments Nucl. Part. Phys.}\ }\textbf {\bibinfo {volume} {11}},\ \bibinfo
  {pages} {203} (\bibinfo {year} {1983})}\BibitemShut {NoStop}%
\bibitem [{\citenamefont {Raynal}()}]{ECIS97}%
  \BibitemOpen
  \bibfield  {author} {\bibinfo {author} {\bibfnamefont {J.}~\bibnamefont
  {Raynal}},\ }\href@noop {} {\bibinfo {title} {\textsc{ecis97}}},\ \bibinfo
  {howpublished} {(unpublished)}\BibitemShut {NoStop}%
\bibitem [{\citenamefont {Barrette}\ \emph {et~al.}(1988)\citenamefont
  {Barrette}, \citenamefont {Alamanos}, \citenamefont {Auger}, \citenamefont
  {Fernandez}, \citenamefont {Gillibert}, \citenamefont {Horen}, \citenamefont
  {Beene}, \citenamefont {Bertrand}, \citenamefont {Auble}, \citenamefont
  {Burks}, \citenamefont {Gomez Del~Campo}, \citenamefont {Halbert},
  \citenamefont {Sayer}, \citenamefont {Mittig}, \citenamefont {Schutz},
  \citenamefont {Haas},\ and\ \citenamefont {Vivien}}]{Bar88}%
  \BibitemOpen
  \bibfield  {author} {\bibinfo {author} {\bibfnamefont {J.}~\bibnamefont
  {Barrette}}, \bibinfo {author} {\bibfnamefont {N.}~\bibnamefont {Alamanos}},
  \bibinfo {author} {\bibfnamefont {F.}~\bibnamefont {Auger}}, \bibinfo
  {author} {\bibfnamefont {B.}~\bibnamefont {Fernandez}}, \bibinfo {author}
  {\bibfnamefont {A.}~\bibnamefont {Gillibert}}, \bibinfo {author}
  {\bibfnamefont {D.~J.}\ \bibnamefont {Horen}}, \bibinfo {author}
  {\bibfnamefont {J.~R.}\ \bibnamefont {Beene}}, \bibinfo {author}
  {\bibfnamefont {F.~E.}\ \bibnamefont {Bertrand}}, \bibinfo {author}
  {\bibfnamefont {R.~L.}\ \bibnamefont {Auble}}, \bibinfo {author}
  {\bibfnamefont {B.~L.}\ \bibnamefont {Burks}}, \bibinfo {author}
  {\bibfnamefont {J.}~\bibnamefont {Gomez Del~Campo}}, \bibinfo {author}
  {\bibfnamefont {M.~L.}\ \bibnamefont {Halbert}}, \bibinfo {author}
  {\bibfnamefont {R.~O.}\ \bibnamefont {Sayer}}, \bibinfo {author}
  {\bibfnamefont {W.}~\bibnamefont {Mittig}}, \bibinfo {author} {\bibfnamefont
  {Y.}~\bibnamefont {Schutz}}, \bibinfo {author} {\bibfnamefont
  {B.}~\bibnamefont {Haas}},\ and\ \bibinfo {author} {\bibfnamefont {J.~P.}\
  \bibnamefont {Vivien}},\ }\href
  {https://doi.org/10.1016/0370-2693(88)90929-X} {\bibfield  {journal}
  {\bibinfo  {journal} {Physics Letters B}\ }\textbf {\bibinfo {volume}
  {209}},\ \bibinfo {pages} {182} (\bibinfo {year} {1988})}\BibitemShut
  {NoStop}%
\bibitem [{\citenamefont {Mermaz}\ \emph {et~al.}(1987)\citenamefont {Mermaz},
  \citenamefont {Berthier}, \citenamefont {Barrette}, \citenamefont
  {Gastebois}, \citenamefont {Gillibert}, \citenamefont {Lucas}, \citenamefont
  {Matuszek}, \citenamefont {Miczaika}, \citenamefont {Van~Renterghem},
  \citenamefont {Suomij\"{a}rvi}, \citenamefont {Boucenna}, \citenamefont
  {Disdier}, \citenamefont {Gorodetzky}, \citenamefont {Kraus}, \citenamefont
  {Linck}, \citenamefont {Lott}, \citenamefont {Rauch}, \citenamefont
  {Rebmeister}, \citenamefont {Scheibling}, \citenamefont {Schulz},
  \citenamefont {Sens}, \citenamefont {Grunberg},\ and\ \citenamefont
  {Mittig}}]{Mer87}%
  \BibitemOpen
  \bibfield  {author} {\bibinfo {author} {\bibfnamefont {M.~C.}\ \bibnamefont
  {Mermaz}}, \bibinfo {author} {\bibfnamefont {B.}~\bibnamefont {Berthier}},
  \bibinfo {author} {\bibfnamefont {J.}~\bibnamefont {Barrette}}, \bibinfo
  {author} {\bibfnamefont {J.}~\bibnamefont {Gastebois}}, \bibinfo {author}
  {\bibfnamefont {A.}~\bibnamefont {Gillibert}}, \bibinfo {author}
  {\bibfnamefont {R.}~\bibnamefont {Lucas}}, \bibinfo {author} {\bibfnamefont
  {J.}~\bibnamefont {Matuszek}}, \bibinfo {author} {\bibfnamefont
  {A.}~\bibnamefont {Miczaika}}, \bibinfo {author} {\bibfnamefont
  {E.}~\bibnamefont {Van~Renterghem}}, \bibinfo {author} {\bibfnamefont
  {T.}~\bibnamefont {Suomij\"{a}rvi}}, \bibinfo {author} {\bibfnamefont
  {A.}~\bibnamefont {Boucenna}}, \bibinfo {author} {\bibfnamefont
  {D.}~\bibnamefont {Disdier}}, \bibinfo {author} {\bibfnamefont
  {P.}~\bibnamefont {Gorodetzky}}, \bibinfo {author} {\bibfnamefont
  {L.}~\bibnamefont {Kraus}}, \bibinfo {author} {\bibfnamefont
  {I.}~\bibnamefont {Linck}}, \bibinfo {author} {\bibfnamefont
  {B.}~\bibnamefont {Lott}}, \bibinfo {author} {\bibfnamefont {V.}~\bibnamefont
  {Rauch}}, \bibinfo {author} {\bibfnamefont {R.}~\bibnamefont {Rebmeister}},
  \bibinfo {author} {\bibfnamefont {F.}~\bibnamefont {Scheibling}}, \bibinfo
  {author} {\bibfnamefont {N.}~\bibnamefont {Schulz}}, \bibinfo {author}
  {\bibfnamefont {J.~C.}\ \bibnamefont {Sens}}, \bibinfo {author}
  {\bibfnamefont {C.}~\bibnamefont {Grunberg}},\ and\ \bibinfo {author}
  {\bibfnamefont {W.}~\bibnamefont {Mittig}},\ }\href
  {https://doi.org/10.1007/BF01289537} {\bibfield  {journal} {\bibinfo
  {journal} {Z. Physik A - Atomic Nuclei}\ }\textbf {\bibinfo {volume} {326}},\
  \bibinfo {pages} {353} (\bibinfo {year} {1987})}\BibitemShut {NoStop}%
\bibitem [{\citenamefont {Suomij\"{a}rvi}\ \emph {et~al.}(1990)\citenamefont
  {Suomij\"{a}rvi}, \citenamefont {Beaumel}, \citenamefont {Blumenfeld},
  \citenamefont {Chomaz}, \citenamefont {Frascaria}, \citenamefont {Garron},
  \citenamefont {Roynette}, \citenamefont {Scarpaci}, \citenamefont {Barrette},
  \citenamefont {Fernandez}, \citenamefont {Gastebois},\ and\ \citenamefont
  {Mittig}}]{Suo90}%
  \BibitemOpen
  \bibfield  {author} {\bibinfo {author} {\bibfnamefont {T.}~\bibnamefont
  {Suomij\"{a}rvi}}, \bibinfo {author} {\bibfnamefont {D.}~\bibnamefont
  {Beaumel}}, \bibinfo {author} {\bibfnamefont {Y.}~\bibnamefont {Blumenfeld}},
  \bibinfo {author} {\bibfnamefont {P.}~\bibnamefont {Chomaz}}, \bibinfo
  {author} {\bibfnamefont {N.}~\bibnamefont {Frascaria}}, \bibinfo {author}
  {\bibfnamefont {J.~P.}\ \bibnamefont {Garron}}, \bibinfo {author}
  {\bibfnamefont {J.~C.}\ \bibnamefont {Roynette}}, \bibinfo {author}
  {\bibfnamefont {J.~A.}\ \bibnamefont {Scarpaci}}, \bibinfo {author}
  {\bibfnamefont {J.}~\bibnamefont {Barrette}}, \bibinfo {author}
  {\bibfnamefont {B.}~\bibnamefont {Fernandez}}, \bibinfo {author}
  {\bibfnamefont {J.}~\bibnamefont {Gastebois}},\ and\ \bibinfo {author}
  {\bibfnamefont {W.}~\bibnamefont {Mittig}},\ }\href
  {https://doi.org/10.1016/0375-9474(90)90428-O} {\bibfield  {journal}
  {\bibinfo  {journal} {Nuclear Physics A}\ }\textbf {\bibinfo {volume}
  {509}},\ \bibinfo {pages} {369} (\bibinfo {year} {1990})}\BibitemShut
  {NoStop}%
\bibitem [{\citenamefont {Khan}(2022)}]{Kh22}%
  \BibitemOpen
  \bibfield  {author} {\bibinfo {author} {\bibfnamefont {E.}~\bibnamefont
  {Khan}},\ }\href {https://doi.org/10.1103/PhysRevC.105.014306} {\bibfield
  {journal} {\bibinfo  {journal} {Phys. Rev. C}\ }\textbf {\bibinfo {volume}
  {105}},\ \bibinfo {pages} {014306} (\bibinfo {year} {2022})}\BibitemShut
  {NoStop}%
\bibitem [{\citenamefont {Vaquero~Soto}(2018)}]{Vaq18}%
  \BibitemOpen
  \bibfield  {author} {\bibinfo {author} {\bibfnamefont {V.}~\bibnamefont
  {Vaquero~Soto}},\ }\href@noop {} {\bibfield  {journal} {\bibinfo  {journal}
  {Ph.D. {T}hesis, Universidad Aut\'{o}noma de Madrid}\ } (\bibinfo {year}
  {2018})}\BibitemShut {NoStop}%
\bibitem [{\citenamefont {Takeuchi}\ \emph {et~al.}(2014)\citenamefont
  {Takeuchi}, \citenamefont {Motobayashi}, \citenamefont {Togano},
  \citenamefont {Matsushita}, \citenamefont {Aoi}, \citenamefont {Demichi},
  \citenamefont {Hasegawa},\ and\ \citenamefont {Murakami}}]{Tak14}%
  \BibitemOpen
  \bibfield  {author} {\bibinfo {author} {\bibfnamefont {S.}~\bibnamefont
  {Takeuchi}}, \bibinfo {author} {\bibfnamefont {T.}~\bibnamefont
  {Motobayashi}}, \bibinfo {author} {\bibfnamefont {Y.}~\bibnamefont {Togano}},
  \bibinfo {author} {\bibfnamefont {M.}~\bibnamefont {Matsushita}}, \bibinfo
  {author} {\bibfnamefont {N.}~\bibnamefont {Aoi}}, \bibinfo {author}
  {\bibfnamefont {K.}~\bibnamefont {Demichi}}, \bibinfo {author} {\bibfnamefont
  {H.}~\bibnamefont {Hasegawa}},\ and\ \bibinfo {author} {\bibfnamefont
  {H.}~\bibnamefont {Murakami}},\ }\href
  {https://doi.org/10.1016/j.nima.2014.06.087} {\bibfield  {journal} {\bibinfo
  {journal} {Nuclear Instruments and Methods in Physics Research Section A:
  Accelerators, Spectrometers, Detectors and Associated Equipment}\ }\textbf
  {\bibinfo {volume} {763}},\ \bibinfo {pages} {596} (\bibinfo {year}
  {2014})}\BibitemShut {NoStop}%
\bibitem [{\citenamefont {Riley}\ \emph {et~al.}(2019)\citenamefont {Riley},
  \citenamefont {Bazin}, \citenamefont {Belarge}, \citenamefont {Bender},
  \citenamefont {Brown}, \citenamefont {Cottle}, \citenamefont {Elman},
  \citenamefont {Gade}, \citenamefont {Gregory}, \citenamefont {Haldeman},
  \citenamefont {Kemper}, \citenamefont {Klybor}, \citenamefont {Liggett},
  \citenamefont {Lipschutz}, \citenamefont {Longfellow}, \citenamefont
  {Lunderberg}, \citenamefont {Mijatovic}, \citenamefont {Pereira},
  \citenamefont {Skiles}, \citenamefont {Titus}, \citenamefont {Volya},
  \citenamefont {Weisshaar}, \citenamefont {Zamora},\ and\ \citenamefont
  {Zegers}}]{Ril19}%
  \BibitemOpen
  \bibfield  {author} {\bibinfo {author} {\bibfnamefont {L.~A.}\ \bibnamefont
  {Riley}}, \bibinfo {author} {\bibfnamefont {D.}~\bibnamefont {Bazin}},
  \bibinfo {author} {\bibfnamefont {J.}~\bibnamefont {Belarge}}, \bibinfo
  {author} {\bibfnamefont {P.~C.}\ \bibnamefont {Bender}}, \bibinfo {author}
  {\bibfnamefont {B.~A.}\ \bibnamefont {Brown}}, \bibinfo {author}
  {\bibfnamefont {P.~D.}\ \bibnamefont {Cottle}}, \bibinfo {author}
  {\bibfnamefont {B.}~\bibnamefont {Elman}}, \bibinfo {author} {\bibfnamefont
  {A.}~\bibnamefont {Gade}}, \bibinfo {author} {\bibfnamefont {S.~D.}\
  \bibnamefont {Gregory}}, \bibinfo {author} {\bibfnamefont {E.~B.}\
  \bibnamefont {Haldeman}}, \bibinfo {author} {\bibfnamefont {K.~W.}\
  \bibnamefont {Kemper}}, \bibinfo {author} {\bibfnamefont {B.~R.}\
  \bibnamefont {Klybor}}, \bibinfo {author} {\bibfnamefont {M.~A.}\
  \bibnamefont {Liggett}}, \bibinfo {author} {\bibfnamefont {S.}~\bibnamefont
  {Lipschutz}}, \bibinfo {author} {\bibfnamefont {B.}~\bibnamefont
  {Longfellow}}, \bibinfo {author} {\bibfnamefont {E.}~\bibnamefont
  {Lunderberg}}, \bibinfo {author} {\bibfnamefont {T.}~\bibnamefont
  {Mijatovic}}, \bibinfo {author} {\bibfnamefont {J.}~\bibnamefont {Pereira}},
  \bibinfo {author} {\bibfnamefont {L.~M.}\ \bibnamefont {Skiles}}, \bibinfo
  {author} {\bibfnamefont {R.}~\bibnamefont {Titus}}, \bibinfo {author}
  {\bibfnamefont {A.}~\bibnamefont {Volya}}, \bibinfo {author} {\bibfnamefont
  {D.}~\bibnamefont {Weisshaar}}, \bibinfo {author} {\bibfnamefont {J.~C.}\
  \bibnamefont {Zamora}},\ and\ \bibinfo {author} {\bibfnamefont {R.~G.~T.}\
  \bibnamefont {Zegers}},\ }\href {https://doi.org/10.1103/PhysRevC.100.044312}
  {\bibfield  {journal} {\bibinfo  {journal} {Phys. Rev. C}\ }\textbf {\bibinfo
  {volume} {100}},\ \bibinfo {pages} {044312} (\bibinfo {year}
  {2019})}\BibitemShut {NoStop}%
\bibitem [{\citenamefont {Grévy}\ \emph {et~al.}(2004)\citenamefont {Grévy},
  \citenamefont {Angélique}, \citenamefont {Baumann}, \citenamefont {Borcea},
  \citenamefont {Buta}, \citenamefont {Canchel}, \citenamefont {Catford},
  \citenamefont {Courtin}, \citenamefont {Daugas}, \citenamefont {{de
  Oliveira}}, \citenamefont {Dessagne}, \citenamefont {Dlouhy}, \citenamefont
  {Knipper}, \citenamefont {Kratz}, \citenamefont {Lecolley}, \citenamefont
  {Lecouey}, \citenamefont {Lehrsenneau}, \citenamefont {Lewitowicz},
  \citenamefont {Liénard}, \citenamefont {Lukyanov}, \citenamefont
  {Maréchal}, \citenamefont {Miehé}, \citenamefont {Mrazek}, \citenamefont
  {Negoita}, \citenamefont {Orr}, \citenamefont {Pantelica}, \citenamefont
  {Penionzhkevich}, \citenamefont {Péter}, \citenamefont {Pfeiffer},
  \citenamefont {Pietri}, \citenamefont {Poirier}, \citenamefont {Sorlin},
  \citenamefont {Stanoiu}, \citenamefont {Stefan}, \citenamefont {Stodel},\
  and\ \citenamefont {Timis}}]{Gr04}%
  \BibitemOpen
  \bibfield  {author} {\bibinfo {author} {\bibfnamefont {S.}~\bibnamefont
  {Grévy}}, \bibinfo {author} {\bibfnamefont {J.}~\bibnamefont {Angélique}},
  \bibinfo {author} {\bibfnamefont {P.}~\bibnamefont {Baumann}}, \bibinfo
  {author} {\bibfnamefont {C.}~\bibnamefont {Borcea}}, \bibinfo {author}
  {\bibfnamefont {A.}~\bibnamefont {Buta}}, \bibinfo {author} {\bibfnamefont
  {G.}~\bibnamefont {Canchel}}, \bibinfo {author} {\bibfnamefont
  {W.}~\bibnamefont {Catford}}, \bibinfo {author} {\bibfnamefont
  {S.}~\bibnamefont {Courtin}}, \bibinfo {author} {\bibfnamefont
  {J.}~\bibnamefont {Daugas}}, \bibinfo {author} {\bibfnamefont
  {F.}~\bibnamefont {{de Oliveira}}}, \bibinfo {author} {\bibfnamefont
  {P.}~\bibnamefont {Dessagne}}, \bibinfo {author} {\bibfnamefont
  {Z.}~\bibnamefont {Dlouhy}}, \bibinfo {author} {\bibfnamefont
  {A.}~\bibnamefont {Knipper}}, \bibinfo {author} {\bibfnamefont
  {K.}~\bibnamefont {Kratz}}, \bibinfo {author} {\bibfnamefont
  {F.}~\bibnamefont {Lecolley}}, \bibinfo {author} {\bibfnamefont
  {J.}~\bibnamefont {Lecouey}}, \bibinfo {author} {\bibfnamefont
  {G.}~\bibnamefont {Lehrsenneau}}, \bibinfo {author} {\bibfnamefont
  {M.}~\bibnamefont {Lewitowicz}}, \bibinfo {author} {\bibfnamefont
  {E.}~\bibnamefont {Liénard}}, \bibinfo {author} {\bibfnamefont
  {S.}~\bibnamefont {Lukyanov}}, \bibinfo {author} {\bibfnamefont
  {F.}~\bibnamefont {Maréchal}}, \bibinfo {author} {\bibfnamefont
  {C.}~\bibnamefont {Miehé}}, \bibinfo {author} {\bibfnamefont
  {J.}~\bibnamefont {Mrazek}}, \bibinfo {author} {\bibfnamefont
  {F.}~\bibnamefont {Negoita}}, \bibinfo {author} {\bibfnamefont
  {N.}~\bibnamefont {Orr}}, \bibinfo {author} {\bibfnamefont {D.}~\bibnamefont
  {Pantelica}}, \bibinfo {author} {\bibfnamefont {Y.}~\bibnamefont
  {Penionzhkevich}}, \bibinfo {author} {\bibfnamefont {J.}~\bibnamefont
  {Péter}}, \bibinfo {author} {\bibfnamefont {B.}~\bibnamefont {Pfeiffer}},
  \bibinfo {author} {\bibfnamefont {S.}~\bibnamefont {Pietri}}, \bibinfo
  {author} {\bibfnamefont {E.}~\bibnamefont {Poirier}}, \bibinfo {author}
  {\bibfnamefont {O.}~\bibnamefont {Sorlin}}, \bibinfo {author} {\bibfnamefont
  {M.}~\bibnamefont {Stanoiu}}, \bibinfo {author} {\bibfnamefont
  {I.}~\bibnamefont {Stefan}}, \bibinfo {author} {\bibfnamefont
  {C.}~\bibnamefont {Stodel}},\ and\ \bibinfo {author} {\bibfnamefont
  {C.}~\bibnamefont {Timis}},\ }\href
  {https://doi.org/https://doi.org/10.1016/j.physletb.2004.06.005} {\bibfield
  {journal} {\bibinfo  {journal} {Physics Letters B}\ }\textbf {\bibinfo
  {volume} {594}},\ \bibinfo {pages} {252} (\bibinfo {year}
  {2004})}\BibitemShut {NoStop}%
\bibitem [{\citenamefont {Fridmann}\ \emph {et~al.}(2005)\citenamefont
  {Fridmann}, \citenamefont {Wiedenh{\"o}ver}, \citenamefont {Gade},
  \citenamefont {Baby}, \citenamefont {Bazin}, \citenamefont {Brown},
  \citenamefont {Campbell}, \citenamefont {Cook}, \citenamefont {Cottle},
  \citenamefont {Diffenderfer}, \citenamefont {Dinca}, \citenamefont
  {Glasmacher}, \citenamefont {Hansen}, \citenamefont {Kemper}, \citenamefont
  {Lecouey}, \citenamefont {Mueller}, \citenamefont {Olliver}, \citenamefont
  {Rodriguez-Vieitez}, \citenamefont {Terry}, \citenamefont {Tostevin},\ and\
  \citenamefont {Yoneda}}]{Fr05}%
  \BibitemOpen
  \bibfield  {author} {\bibinfo {author} {\bibfnamefont {J.}~\bibnamefont
  {Fridmann}}, \bibinfo {author} {\bibfnamefont {I.}~\bibnamefont
  {Wiedenh{\"o}ver}}, \bibinfo {author} {\bibfnamefont {A.}~\bibnamefont
  {Gade}}, \bibinfo {author} {\bibfnamefont {L.~T.}\ \bibnamefont {Baby}},
  \bibinfo {author} {\bibfnamefont {D.}~\bibnamefont {Bazin}}, \bibinfo
  {author} {\bibfnamefont {B.~A.}\ \bibnamefont {Brown}}, \bibinfo {author}
  {\bibfnamefont {C.~M.}\ \bibnamefont {Campbell}}, \bibinfo {author}
  {\bibfnamefont {J.~M.}\ \bibnamefont {Cook}}, \bibinfo {author}
  {\bibfnamefont {P.~D.}\ \bibnamefont {Cottle}}, \bibinfo {author}
  {\bibfnamefont {E.}~\bibnamefont {Diffenderfer}}, \bibinfo {author}
  {\bibfnamefont {D.-C.}\ \bibnamefont {Dinca}}, \bibinfo {author}
  {\bibfnamefont {T.}~\bibnamefont {Glasmacher}}, \bibinfo {author}
  {\bibfnamefont {P.~G.}\ \bibnamefont {Hansen}}, \bibinfo {author}
  {\bibfnamefont {K.~W.}\ \bibnamefont {Kemper}}, \bibinfo {author}
  {\bibfnamefont {J.~L.}\ \bibnamefont {Lecouey}}, \bibinfo {author}
  {\bibfnamefont {W.~F.}\ \bibnamefont {Mueller}}, \bibinfo {author}
  {\bibfnamefont {H.}~\bibnamefont {Olliver}}, \bibinfo {author} {\bibfnamefont
  {E.}~\bibnamefont {Rodriguez-Vieitez}}, \bibinfo {author} {\bibfnamefont
  {J.~R.}\ \bibnamefont {Terry}}, \bibinfo {author} {\bibfnamefont {J.~A.}\
  \bibnamefont {Tostevin}},\ and\ \bibinfo {author} {\bibfnamefont
  {K.}~\bibnamefont {Yoneda}},\ }\href@noop {} {\bibfield  {journal} {\bibinfo
  {journal} {Nature}\ }\textbf {\bibinfo {volume} {435}},\ \bibinfo {pages}
  {922} (\bibinfo {year} {2005})}\BibitemShut {NoStop}%
\bibitem [{\citenamefont {Fridmann}\ \emph {et~al.}(2006)\citenamefont
  {Fridmann}, \citenamefont {Wiedenh\"over}, \citenamefont {Gade},
  \citenamefont {Baby}, \citenamefont {Bazin}, \citenamefont {Brown},
  \citenamefont {Campbell}, \citenamefont {Cook}, \citenamefont {Cottle},
  \citenamefont {Diffenderfer}, \citenamefont {Dinca}, \citenamefont
  {Glasmacher}, \citenamefont {Hansen}, \citenamefont {Kemper}, \citenamefont
  {Lecouey}, \citenamefont {Mueller}, \citenamefont {Rodriguez-Vieitez},
  \citenamefont {Terry}, \citenamefont {Tostevin}, \citenamefont {Yoneda},\
  and\ \citenamefont {Zwahlen}}]{Fr06}%
  \BibitemOpen
  \bibfield  {author} {\bibinfo {author} {\bibfnamefont {J.}~\bibnamefont
  {Fridmann}}, \bibinfo {author} {\bibfnamefont {I.}~\bibnamefont
  {Wiedenh\"over}}, \bibinfo {author} {\bibfnamefont {A.}~\bibnamefont {Gade}},
  \bibinfo {author} {\bibfnamefont {L.~T.}\ \bibnamefont {Baby}}, \bibinfo
  {author} {\bibfnamefont {D.}~\bibnamefont {Bazin}}, \bibinfo {author}
  {\bibfnamefont {B.~A.}\ \bibnamefont {Brown}}, \bibinfo {author}
  {\bibfnamefont {C.~M.}\ \bibnamefont {Campbell}}, \bibinfo {author}
  {\bibfnamefont {J.~M.}\ \bibnamefont {Cook}}, \bibinfo {author}
  {\bibfnamefont {P.~D.}\ \bibnamefont {Cottle}}, \bibinfo {author}
  {\bibfnamefont {E.}~\bibnamefont {Diffenderfer}}, \bibinfo {author}
  {\bibfnamefont {D.-C.}\ \bibnamefont {Dinca}}, \bibinfo {author}
  {\bibfnamefont {T.}~\bibnamefont {Glasmacher}}, \bibinfo {author}
  {\bibfnamefont {P.~G.}\ \bibnamefont {Hansen}}, \bibinfo {author}
  {\bibfnamefont {K.~W.}\ \bibnamefont {Kemper}}, \bibinfo {author}
  {\bibfnamefont {J.~L.}\ \bibnamefont {Lecouey}}, \bibinfo {author}
  {\bibfnamefont {W.~F.}\ \bibnamefont {Mueller}}, \bibinfo {author}
  {\bibfnamefont {E.}~\bibnamefont {Rodriguez-Vieitez}}, \bibinfo {author}
  {\bibfnamefont {J.~R.}\ \bibnamefont {Terry}}, \bibinfo {author}
  {\bibfnamefont {J.~A.}\ \bibnamefont {Tostevin}}, \bibinfo {author}
  {\bibfnamefont {K.}~\bibnamefont {Yoneda}},\ and\ \bibinfo {author}
  {\bibfnamefont {H.}~\bibnamefont {Zwahlen}},\ }\href
  {https://doi.org/10.1103/PhysRevC.74.034313} {\bibfield  {journal} {\bibinfo
  {journal} {Phys. Rev. C}\ }\textbf {\bibinfo {volume} {74}},\ \bibinfo
  {pages} {034313} (\bibinfo {year} {2006})}\BibitemShut {NoStop}%
\bibitem [{\citenamefont {Tostevin}\ \emph {et~al.}(2013)\citenamefont
  {Tostevin}, \citenamefont {Brown},\ and\ \citenamefont {Simpson}}]{To13}%
  \BibitemOpen
  \bibfield  {author} {\bibinfo {author} {\bibfnamefont {J.~A.}\ \bibnamefont
  {Tostevin}}, \bibinfo {author} {\bibfnamefont {B.~A.}\ \bibnamefont
  {Brown}},\ and\ \bibinfo {author} {\bibfnamefont {E.~C.}\ \bibnamefont
  {Simpson}},\ }\href {https://doi.org/10.1103/PhysRevC.87.027601} {\bibfield
  {journal} {\bibinfo  {journal} {Phys. Rev. C}\ }\textbf {\bibinfo {volume}
  {87}},\ \bibinfo {pages} {027601} (\bibinfo {year} {2013})}\BibitemShut
  {NoStop}%
\bibitem [{\citenamefont {Sohler}\ \emph {et~al.}(2002)\citenamefont {Sohler},
  \citenamefont {Dombr\'adi}, \citenamefont {Tim\'ar}, \citenamefont {Sorlin},
  \citenamefont {Azaiez}, \citenamefont {Amorini}, \citenamefont {Belleguic},
  \citenamefont {Bourgeois}, \citenamefont {Donzaud}, \citenamefont {Duprat},
  \citenamefont {Guillemaud-Mueller}, \citenamefont {Ibrahim}, \citenamefont
  {Scarpaci}, \citenamefont {Stanoiu}, \citenamefont {Lopez}, \citenamefont
  {Saint-Laurent}, \citenamefont {Becker}, \citenamefont {Sarazin},
  \citenamefont {Stodel}, \citenamefont {Voltolini}, \citenamefont {Lukyanov},
  \citenamefont {Maslov}, \citenamefont {Penionzhkevich}, \citenamefont
  {Girod}, \citenamefont {P\'eru}, \citenamefont {Nowacki}, \citenamefont
  {Sletten}, \citenamefont {Lucas}, \citenamefont {Theisen}, \citenamefont
  {Baiborodin}, \citenamefont {Dlouhy}, \citenamefont {Mrazek}, \citenamefont
  {Borcea}, \citenamefont {Bauchet}, \citenamefont {Moore},\ and\ \citenamefont
  {Taylor}}]{So02}%
  \BibitemOpen
  \bibfield  {author} {\bibinfo {author} {\bibfnamefont {D.}~\bibnamefont
  {Sohler}}, \bibinfo {author} {\bibfnamefont {Z.}~\bibnamefont {Dombr\'adi}},
  \bibinfo {author} {\bibfnamefont {J.}~\bibnamefont {Tim\'ar}}, \bibinfo
  {author} {\bibfnamefont {O.}~\bibnamefont {Sorlin}}, \bibinfo {author}
  {\bibfnamefont {F.}~\bibnamefont {Azaiez}}, \bibinfo {author} {\bibfnamefont
  {F.}~\bibnamefont {Amorini}}, \bibinfo {author} {\bibfnamefont
  {M.}~\bibnamefont {Belleguic}}, \bibinfo {author} {\bibfnamefont
  {C.}~\bibnamefont {Bourgeois}}, \bibinfo {author} {\bibfnamefont
  {C.}~\bibnamefont {Donzaud}}, \bibinfo {author} {\bibfnamefont
  {J.}~\bibnamefont {Duprat}}, \bibinfo {author} {\bibfnamefont
  {D.}~\bibnamefont {Guillemaud-Mueller}}, \bibinfo {author} {\bibfnamefont
  {F.}~\bibnamefont {Ibrahim}}, \bibinfo {author} {\bibfnamefont {J.~A.}\
  \bibnamefont {Scarpaci}}, \bibinfo {author} {\bibfnamefont {M.}~\bibnamefont
  {Stanoiu}}, \bibinfo {author} {\bibfnamefont {M.~J.}\ \bibnamefont {Lopez}},
  \bibinfo {author} {\bibfnamefont {M.~G.}\ \bibnamefont {Saint-Laurent}},
  \bibinfo {author} {\bibfnamefont {F.}~\bibnamefont {Becker}}, \bibinfo
  {author} {\bibfnamefont {F.}~\bibnamefont {Sarazin}}, \bibinfo {author}
  {\bibfnamefont {C.}~\bibnamefont {Stodel}}, \bibinfo {author} {\bibfnamefont
  {G.}~\bibnamefont {Voltolini}}, \bibinfo {author} {\bibfnamefont {S.~M.}\
  \bibnamefont {Lukyanov}}, \bibinfo {author} {\bibfnamefont {V.}~\bibnamefont
  {Maslov}}, \bibinfo {author} {\bibfnamefont {Y.-E.}\ \bibnamefont
  {Penionzhkevich}}, \bibinfo {author} {\bibfnamefont {M.}~\bibnamefont
  {Girod}}, \bibinfo {author} {\bibfnamefont {S.}~\bibnamefont {P\'eru}},
  \bibinfo {author} {\bibfnamefont {F.}~\bibnamefont {Nowacki}}, \bibinfo
  {author} {\bibfnamefont {G.}~\bibnamefont {Sletten}}, \bibinfo {author}
  {\bibfnamefont {R.}~\bibnamefont {Lucas}}, \bibinfo {author} {\bibfnamefont
  {C.}~\bibnamefont {Theisen}}, \bibinfo {author} {\bibfnamefont
  {D.}~\bibnamefont {Baiborodin}}, \bibinfo {author} {\bibfnamefont
  {Z.}~\bibnamefont {Dlouhy}}, \bibinfo {author} {\bibfnamefont
  {J.}~\bibnamefont {Mrazek}}, \bibinfo {author} {\bibfnamefont
  {C.}~\bibnamefont {Borcea}}, \bibinfo {author} {\bibfnamefont
  {A.}~\bibnamefont {Bauchet}}, \bibinfo {author} {\bibfnamefont {C.~J.}\
  \bibnamefont {Moore}},\ and\ \bibinfo {author} {\bibfnamefont {M.~J.}\
  \bibnamefont {Taylor}},\ }\href {https://doi.org/10.1103/PhysRevC.66.054302}
  {\bibfield  {journal} {\bibinfo  {journal} {Phys. Rev. C}\ }\textbf {\bibinfo
  {volume} {66}},\ \bibinfo {pages} {054302} (\bibinfo {year}
  {2002})}\BibitemShut {NoStop}%
\bibitem [{\citenamefont {Santiago-Gonzalez}\ \emph {et~al.}(2011)\citenamefont
  {Santiago-Gonzalez}, \citenamefont {Wiedenh\"over}, \citenamefont
  {Abramkina}, \citenamefont {Avila}, \citenamefont {Baugher}, \citenamefont
  {Bazin}, \citenamefont {Brown}, \citenamefont {Cottle}, \citenamefont {Gade},
  \citenamefont {Glasmacher}, \citenamefont {Kemper}, \citenamefont {McDaniel},
  \citenamefont {Rojas}, \citenamefont {Ratkiewicz}, \citenamefont
  {Meharchand}, \citenamefont {Simpson}, \citenamefont {Tostevin},
  \citenamefont {Volya},\ and\ \citenamefont {Weisshaar}}]{Sa11}%
  \BibitemOpen
  \bibfield  {author} {\bibinfo {author} {\bibfnamefont {D.}~\bibnamefont
  {Santiago-Gonzalez}}, \bibinfo {author} {\bibfnamefont {I.}~\bibnamefont
  {Wiedenh\"over}}, \bibinfo {author} {\bibfnamefont {V.}~\bibnamefont
  {Abramkina}}, \bibinfo {author} {\bibfnamefont {M.~L.}\ \bibnamefont
  {Avila}}, \bibinfo {author} {\bibfnamefont {T.}~\bibnamefont {Baugher}},
  \bibinfo {author} {\bibfnamefont {D.}~\bibnamefont {Bazin}}, \bibinfo
  {author} {\bibfnamefont {B.~A.}\ \bibnamefont {Brown}}, \bibinfo {author}
  {\bibfnamefont {P.~D.}\ \bibnamefont {Cottle}}, \bibinfo {author}
  {\bibfnamefont {A.}~\bibnamefont {Gade}}, \bibinfo {author} {\bibfnamefont
  {T.}~\bibnamefont {Glasmacher}}, \bibinfo {author} {\bibfnamefont {K.~W.}\
  \bibnamefont {Kemper}}, \bibinfo {author} {\bibfnamefont {S.}~\bibnamefont
  {McDaniel}}, \bibinfo {author} {\bibfnamefont {A.}~\bibnamefont {Rojas}},
  \bibinfo {author} {\bibfnamefont {A.}~\bibnamefont {Ratkiewicz}}, \bibinfo
  {author} {\bibfnamefont {R.}~\bibnamefont {Meharchand}}, \bibinfo {author}
  {\bibfnamefont {E.~C.}\ \bibnamefont {Simpson}}, \bibinfo {author}
  {\bibfnamefont {J.~A.}\ \bibnamefont {Tostevin}}, \bibinfo {author}
  {\bibfnamefont {A.}~\bibnamefont {Volya}},\ and\ \bibinfo {author}
  {\bibfnamefont {D.}~\bibnamefont {Weisshaar}},\ }\href
  {https://doi.org/10.1103/PhysRevC.83.061305} {\bibfield  {journal} {\bibinfo
  {journal} {Phys. Rev. C}\ }\textbf {\bibinfo {volume} {83}},\ \bibinfo
  {pages} {061305} (\bibinfo {year} {2011})}\BibitemShut {NoStop}%
\bibitem [{\citenamefont {Parker}\ \emph {et~al.}(2017)\citenamefont {Parker},
  \citenamefont {Wiedenh\"over}, \citenamefont {Cottle}, \citenamefont {Baker},
  \citenamefont {McPherson}, \citenamefont {Riley}, \citenamefont
  {Santiago-Gonzalez}, \citenamefont {Volya}, \citenamefont {Bader},
  \citenamefont {Baugher}, \citenamefont {Bazin}, \citenamefont {Gade},
  \citenamefont {Ginter}, \citenamefont {Iwasaki}, \citenamefont {Loelius},
  \citenamefont {Morse}, \citenamefont {Recchia}, \citenamefont {Smalley},
  \citenamefont {Stroberg}, \citenamefont {Whitmore}, \citenamefont
  {Weisshaar}, \citenamefont {Lemasson}, \citenamefont {Crawford},
  \citenamefont {Macchiavelli},\ and\ \citenamefont {Wimmer}}]{Pa17}%
  \BibitemOpen
  \bibfield  {author} {\bibinfo {author} {\bibfnamefont {J.~J.}\ \bibnamefont
  {Parker}}, \bibinfo {author} {\bibfnamefont {I.}~\bibnamefont
  {Wiedenh\"over}}, \bibinfo {author} {\bibfnamefont {P.~D.}\ \bibnamefont
  {Cottle}}, \bibinfo {author} {\bibfnamefont {J.}~\bibnamefont {Baker}},
  \bibinfo {author} {\bibfnamefont {D.}~\bibnamefont {McPherson}}, \bibinfo
  {author} {\bibfnamefont {M.~A.}\ \bibnamefont {Riley}}, \bibinfo {author}
  {\bibfnamefont {D.}~\bibnamefont {Santiago-Gonzalez}}, \bibinfo {author}
  {\bibfnamefont {A.}~\bibnamefont {Volya}}, \bibinfo {author} {\bibfnamefont
  {V.~M.}\ \bibnamefont {Bader}}, \bibinfo {author} {\bibfnamefont
  {T.}~\bibnamefont {Baugher}}, \bibinfo {author} {\bibfnamefont
  {D.}~\bibnamefont {Bazin}}, \bibinfo {author} {\bibfnamefont
  {A.}~\bibnamefont {Gade}}, \bibinfo {author} {\bibfnamefont {T.}~\bibnamefont
  {Ginter}}, \bibinfo {author} {\bibfnamefont {H.}~\bibnamefont {Iwasaki}},
  \bibinfo {author} {\bibfnamefont {C.}~\bibnamefont {Loelius}}, \bibinfo
  {author} {\bibfnamefont {C.}~\bibnamefont {Morse}}, \bibinfo {author}
  {\bibfnamefont {F.}~\bibnamefont {Recchia}}, \bibinfo {author} {\bibfnamefont
  {D.}~\bibnamefont {Smalley}}, \bibinfo {author} {\bibfnamefont {S.~R.}\
  \bibnamefont {Stroberg}}, \bibinfo {author} {\bibfnamefont {K.}~\bibnamefont
  {Whitmore}}, \bibinfo {author} {\bibfnamefont {D.}~\bibnamefont {Weisshaar}},
  \bibinfo {author} {\bibfnamefont {A.}~\bibnamefont {Lemasson}}, \bibinfo
  {author} {\bibfnamefont {H.~L.}\ \bibnamefont {Crawford}}, \bibinfo {author}
  {\bibfnamefont {A.~O.}\ \bibnamefont {Macchiavelli}},\ and\ \bibinfo {author}
  {\bibfnamefont {K.}~\bibnamefont {Wimmer}},\ }\href
  {https://doi.org/10.1103/PhysRevLett.118.052501} {\bibfield  {journal}
  {\bibinfo  {journal} {Phys. Rev. Lett.}\ }\textbf {\bibinfo {volume} {118}},\
  \bibinfo {pages} {052501} (\bibinfo {year} {2017})}\BibitemShut {NoStop}%
\bibitem [{\citenamefont {Suzuki}\ and\ \citenamefont {Kimura}(2021)}]{Su21}%
  \BibitemOpen
  \bibfield  {author} {\bibinfo {author} {\bibfnamefont {Y.}~\bibnamefont
  {Suzuki}}\ and\ \bibinfo {author} {\bibfnamefont {M.}~\bibnamefont
  {Kimura}},\ }\href {https://doi.org/10.1103/PhysRevC.104.024327} {\bibfield
  {journal} {\bibinfo  {journal} {Phys. Rev. C}\ }\textbf {\bibinfo {volume}
  {104}},\ \bibinfo {pages} {024327} (\bibinfo {year} {2021})}\BibitemShut
  {NoStop}%
\bibitem [{\citenamefont {Macchiavelli}\ \emph {et~al.}(2022)\citenamefont
  {Macchiavelli}, \citenamefont {Crawford}, \citenamefont {Campbell},
  \citenamefont {Clark}, \citenamefont {Cromaz}, \citenamefont {Fallon},
  \citenamefont {Lee}, \citenamefont {Gade}, \citenamefont {Poves},\ and\
  \citenamefont {Rice}}]{Ma22}%
  \BibitemOpen
  \bibfield  {author} {\bibinfo {author} {\bibfnamefont {A.~O.}\ \bibnamefont
  {Macchiavelli}}, \bibinfo {author} {\bibfnamefont {H.~L.}\ \bibnamefont
  {Crawford}}, \bibinfo {author} {\bibfnamefont {C.~M.}\ \bibnamefont
  {Campbell}}, \bibinfo {author} {\bibfnamefont {R.~M.}\ \bibnamefont {Clark}},
  \bibinfo {author} {\bibfnamefont {M.}~\bibnamefont {Cromaz}}, \bibinfo
  {author} {\bibfnamefont {P.}~\bibnamefont {Fallon}}, \bibinfo {author}
  {\bibfnamefont {I.~Y.}\ \bibnamefont {Lee}}, \bibinfo {author} {\bibfnamefont
  {A.}~\bibnamefont {Gade}}, \bibinfo {author} {\bibfnamefont {A.}~\bibnamefont
  {Poves}},\ and\ \bibinfo {author} {\bibfnamefont {E.}~\bibnamefont {Rice}},\
  }\href {https://doi.org/10.1103/PhysRevC.105.014309} {\bibfield  {journal}
  {\bibinfo  {journal} {Phys. Rev. C}\ }\textbf {\bibinfo {volume} {105}},\
  \bibinfo {pages} {014309} (\bibinfo {year} {2022})}\BibitemShut {NoStop}%
\bibitem [{\citenamefont {Suzuki}\ \emph {et~al.}(2022)\citenamefont {Suzuki},
  \citenamefont {Horiuchi},\ and\ \citenamefont {Kimura}}]{Su22}%
  \BibitemOpen
  \bibfield  {author} {\bibinfo {author} {\bibfnamefont {Y.}~\bibnamefont
  {Suzuki}}, \bibinfo {author} {\bibfnamefont {W.}~\bibnamefont {Horiuchi}},\
  and\ \bibinfo {author} {\bibfnamefont {M.}~\bibnamefont {Kimura}},\ }\href
  {https://doi.org/10.1093/ptep/ptac071} {\bibfield  {journal} {\bibinfo
  {journal} {Progress of Theoretical and Experimental Physics}\ }\textbf
  {\bibinfo {volume} {2022}},\ \bibinfo {pages} {063D02} (\bibinfo {year}
  {2022})},\ \Eprint
  {https://arxiv.org/abs/https://academic.oup.com/ptep/article-pdf/2022/6/063D02/44062687/ptac071.pdf}
  {https://academic.oup.com/ptep/article-pdf/2022/6/063D02/44062687/ptac071.pdf}
  \BibitemShut {NoStop}%
\bibitem [{\citenamefont {Washiyama}\ and\ \citenamefont
  {Yoshida}(2023)}]{Wa23}%
  \BibitemOpen
  \bibfield  {author} {\bibinfo {author} {\bibfnamefont {K.}~\bibnamefont
  {Washiyama}}\ and\ \bibinfo {author} {\bibfnamefont {K.}~\bibnamefont
  {Yoshida}},\ }\href {https://doi.org/10.1103/PhysRevC.108.014323} {\bibfield
  {journal} {\bibinfo  {journal} {Phys. Rev. C}\ }\textbf {\bibinfo {volume}
  {108}},\ \bibinfo {pages} {014323} (\bibinfo {year} {2023})}\BibitemShut
  {NoStop}%
\bibitem [{\citenamefont {Yuan}\ \emph {et~al.}(2024)\citenamefont {Yuan},
  \citenamefont {Li},\ and\ \citenamefont {Li}}]{Yu24}%
  \BibitemOpen
  \bibfield  {author} {\bibinfo {author} {\bibfnamefont {Q.}~\bibnamefont
  {Yuan}}, \bibinfo {author} {\bibfnamefont {J.}~\bibnamefont {Li}},\ and\
  \bibinfo {author} {\bibfnamefont {H.}~\bibnamefont {Li}},\ }\href
  {https://doi.org/https://doi.org/10.1016/j.physletb.2023.138331} {\bibfield
  {journal} {\bibinfo  {journal} {Physics Letters B}\ }\textbf {\bibinfo
  {volume} {848}},\ \bibinfo {pages} {138331} (\bibinfo {year}
  {2024})}\BibitemShut {NoStop}%
\bibitem [{\citenamefont {Maheshwari}\ and\ \citenamefont
  {Nomura}(2024)}]{Ma24}%
  \BibitemOpen
  \bibfield  {author} {\bibinfo {author} {\bibfnamefont {B.}~\bibnamefont
  {Maheshwari}}\ and\ \bibinfo {author} {\bibfnamefont {K.}~\bibnamefont
  {Nomura}},\ }\href {https://doi.org/10.1088/1361-6471/ad6170} {\bibfield
  {journal} {\bibinfo  {journal} {Journal of Physics G: Nuclear and Particle
  Physics}\ }\textbf {\bibinfo {volume} {51}},\ \bibinfo {pages} {095101}
  (\bibinfo {year} {2024})}\BibitemShut {NoStop}%
\bibitem [{\citenamefont {Utsuno}\ \emph {et~al.}(2015)\citenamefont {Utsuno},
  \citenamefont {Shimizu}, \citenamefont {Otsuka}, \citenamefont {Yoshida},\
  and\ \citenamefont {Tsunoda}}]{Ut15}%
  \BibitemOpen
  \bibfield  {author} {\bibinfo {author} {\bibfnamefont {Y.}~\bibnamefont
  {Utsuno}}, \bibinfo {author} {\bibfnamefont {N.}~\bibnamefont {Shimizu}},
  \bibinfo {author} {\bibfnamefont {T.}~\bibnamefont {Otsuka}}, \bibinfo
  {author} {\bibfnamefont {T.}~\bibnamefont {Yoshida}},\ and\ \bibinfo {author}
  {\bibfnamefont {Y.}~\bibnamefont {Tsunoda}},\ }\href
  {https://doi.org/10.1103/PhysRevLett.114.032501} {\bibfield  {journal}
  {\bibinfo  {journal} {Phys. Rev. Lett.}\ }\textbf {\bibinfo {volume} {114}},\
  \bibinfo {pages} {032501} (\bibinfo {year} {2015})}\BibitemShut {NoStop}%
\bibitem [{\citenamefont {Mar\'echal}\ \emph {et~al.}(1999)\citenamefont
  {Mar\'echal}, \citenamefont {Suomij\"arvi}, \citenamefont {Blumenfeld},
  \citenamefont {Azhari}, \citenamefont {Bauge}, \citenamefont {Bazin},
  \citenamefont {Brown}, \citenamefont {Cottle}, \citenamefont {Delaroche},
  \citenamefont {Fauerbach}, \citenamefont {Girod}, \citenamefont {Glasmacher},
  \citenamefont {Hirzebruch}, \citenamefont {Jewell}, \citenamefont {Kelley},
  \citenamefont {Kemper}, \citenamefont {Mantica}, \citenamefont {Morrissey},
  \citenamefont {Riley}, \citenamefont {Scarpaci}, \citenamefont {Scheit},\
  and\ \citenamefont {Steiner}}]{Ma99}%
  \BibitemOpen
  \bibfield  {author} {\bibinfo {author} {\bibfnamefont {F.}~\bibnamefont
  {Mar\'echal}}, \bibinfo {author} {\bibfnamefont {T.}~\bibnamefont
  {Suomij\"arvi}}, \bibinfo {author} {\bibfnamefont {Y.}~\bibnamefont
  {Blumenfeld}}, \bibinfo {author} {\bibfnamefont {A.}~\bibnamefont {Azhari}},
  \bibinfo {author} {\bibfnamefont {E.}~\bibnamefont {Bauge}}, \bibinfo
  {author} {\bibfnamefont {D.}~\bibnamefont {Bazin}}, \bibinfo {author}
  {\bibfnamefont {J.~A.}\ \bibnamefont {Brown}}, \bibinfo {author}
  {\bibfnamefont {P.~D.}\ \bibnamefont {Cottle}}, \bibinfo {author}
  {\bibfnamefont {J.~P.}\ \bibnamefont {Delaroche}}, \bibinfo {author}
  {\bibfnamefont {M.}~\bibnamefont {Fauerbach}}, \bibinfo {author}
  {\bibfnamefont {M.}~\bibnamefont {Girod}}, \bibinfo {author} {\bibfnamefont
  {T.}~\bibnamefont {Glasmacher}}, \bibinfo {author} {\bibfnamefont {S.~E.}\
  \bibnamefont {Hirzebruch}}, \bibinfo {author} {\bibfnamefont {J.~K.}\
  \bibnamefont {Jewell}}, \bibinfo {author} {\bibfnamefont {J.~H.}\
  \bibnamefont {Kelley}}, \bibinfo {author} {\bibfnamefont {K.~W.}\
  \bibnamefont {Kemper}}, \bibinfo {author} {\bibfnamefont {P.~F.}\
  \bibnamefont {Mantica}}, \bibinfo {author} {\bibfnamefont {D.~J.}\
  \bibnamefont {Morrissey}}, \bibinfo {author} {\bibfnamefont {L.~A.}\
  \bibnamefont {Riley}}, \bibinfo {author} {\bibfnamefont {J.~A.}\ \bibnamefont
  {Scarpaci}}, \bibinfo {author} {\bibfnamefont {H.}~\bibnamefont {Scheit}},\
  and\ \bibinfo {author} {\bibfnamefont {M.}~\bibnamefont {Steiner}},\ }\href
  {https://doi.org/10.1103/PhysRevC.60.034615} {\bibfield  {journal} {\bibinfo
  {journal} {Phys. Rev. C}\ }\textbf {\bibinfo {volume} {60}},\ \bibinfo
  {pages} {034615} (\bibinfo {year} {1999})}\BibitemShut {NoStop}%
\bibitem [{\citenamefont {Campbell}\ \emph {et~al.}(2007)\citenamefont
  {Campbell}, \citenamefont {Aoi}, \citenamefont {Bazin}, \citenamefont
  {Bowen}, \citenamefont {Brown}, \citenamefont {Cook}, \citenamefont {Dinca},
  \citenamefont {Gade}, \citenamefont {Glasmacher}, \citenamefont {Horoi},
  \citenamefont {Kanno}, \citenamefont {Motobayashi}, \citenamefont {Riley},
  \citenamefont {Sagawa}, \citenamefont {Sakurai}, \citenamefont {Starosta},
  \citenamefont {Suzuki}, \citenamefont {Takeuchi}, \citenamefont {Terry},
  \citenamefont {Yoneda},\ and\ \citenamefont {Zwahlen}}]{Ca07}%
  \BibitemOpen
  \bibfield  {author} {\bibinfo {author} {\bibfnamefont {C.}~\bibnamefont
  {Campbell}}, \bibinfo {author} {\bibfnamefont {N.}~\bibnamefont {Aoi}},
  \bibinfo {author} {\bibfnamefont {D.}~\bibnamefont {Bazin}}, \bibinfo
  {author} {\bibfnamefont {M.}~\bibnamefont {Bowen}}, \bibinfo {author}
  {\bibfnamefont {B.}~\bibnamefont {Brown}}, \bibinfo {author} {\bibfnamefont
  {J.}~\bibnamefont {Cook}}, \bibinfo {author} {\bibfnamefont {D.-C.}\
  \bibnamefont {Dinca}}, \bibinfo {author} {\bibfnamefont {A.}~\bibnamefont
  {Gade}}, \bibinfo {author} {\bibfnamefont {T.}~\bibnamefont {Glasmacher}},
  \bibinfo {author} {\bibfnamefont {M.}~\bibnamefont {Horoi}}, \bibinfo
  {author} {\bibfnamefont {S.}~\bibnamefont {Kanno}}, \bibinfo {author}
  {\bibfnamefont {T.}~\bibnamefont {Motobayashi}}, \bibinfo {author}
  {\bibfnamefont {L.}~\bibnamefont {Riley}}, \bibinfo {author} {\bibfnamefont
  {H.}~\bibnamefont {Sagawa}}, \bibinfo {author} {\bibfnamefont
  {H.}~\bibnamefont {Sakurai}}, \bibinfo {author} {\bibfnamefont
  {K.}~\bibnamefont {Starosta}}, \bibinfo {author} {\bibfnamefont
  {H.}~\bibnamefont {Suzuki}}, \bibinfo {author} {\bibfnamefont
  {S.}~\bibnamefont {Takeuchi}}, \bibinfo {author} {\bibfnamefont
  {J.}~\bibnamefont {Terry}}, \bibinfo {author} {\bibfnamefont
  {K.}~\bibnamefont {Yoneda}},\ and\ \bibinfo {author} {\bibfnamefont
  {H.}~\bibnamefont {Zwahlen}},\ }\href
  {https://doi.org/https://doi.org/10.1016/j.physletb.2007.07.005} {\bibfield
  {journal} {\bibinfo  {journal} {Physics Letters B}\ }\textbf {\bibinfo
  {volume} {652}},\ \bibinfo {pages} {169} (\bibinfo {year}
  {2007})}\BibitemShut {NoStop}%
\bibitem [{\citenamefont {Lubna}\ \emph {et~al.}(2020)\citenamefont {Lubna},
  \citenamefont {Kravvaris}, \citenamefont {Tabor}, \citenamefont {Tripathi},
  \citenamefont {Rubino},\ and\ \citenamefont {Volya}}]{Lu20}%
  \BibitemOpen
  \bibfield  {author} {\bibinfo {author} {\bibfnamefont {R.~S.}\ \bibnamefont
  {Lubna}}, \bibinfo {author} {\bibfnamefont {K.}~\bibnamefont {Kravvaris}},
  \bibinfo {author} {\bibfnamefont {S.~L.}\ \bibnamefont {Tabor}}, \bibinfo
  {author} {\bibfnamefont {V.}~\bibnamefont {Tripathi}}, \bibinfo {author}
  {\bibfnamefont {E.}~\bibnamefont {Rubino}},\ and\ \bibinfo {author}
  {\bibfnamefont {A.}~\bibnamefont {Volya}},\ }\href
  {https://doi.org/10.1103/PhysRevResearch.2.043342} {\bibfield  {journal}
  {\bibinfo  {journal} {Phys. Rev. Res.}\ }\textbf {\bibinfo {volume} {2}},\
  \bibinfo {pages} {043342} (\bibinfo {year} {2020})}\BibitemShut {NoStop}%
\bibitem [{\citenamefont {Lubna}\ \emph {et~al.}(2019)\citenamefont {Lubna},
  \citenamefont {Kravvaris}, \citenamefont {Tabor}, \citenamefont {Tripathi},
  \citenamefont {Volya}, \citenamefont {Rubino}, \citenamefont {Allmond},
  \citenamefont {Abromeit}, \citenamefont {Baby},\ and\ \citenamefont
  {Hensley}}]{Lu19}%
  \BibitemOpen
  \bibfield  {author} {\bibinfo {author} {\bibfnamefont {R.~S.}\ \bibnamefont
  {Lubna}}, \bibinfo {author} {\bibfnamefont {K.}~\bibnamefont {Kravvaris}},
  \bibinfo {author} {\bibfnamefont {S.~L.}\ \bibnamefont {Tabor}}, \bibinfo
  {author} {\bibfnamefont {V.}~\bibnamefont {Tripathi}}, \bibinfo {author}
  {\bibfnamefont {A.}~\bibnamefont {Volya}}, \bibinfo {author} {\bibfnamefont
  {E.}~\bibnamefont {Rubino}}, \bibinfo {author} {\bibfnamefont {J.~M.}\
  \bibnamefont {Allmond}}, \bibinfo {author} {\bibfnamefont {B.}~\bibnamefont
  {Abromeit}}, \bibinfo {author} {\bibfnamefont {L.~T.}\ \bibnamefont {Baby}},\
  and\ \bibinfo {author} {\bibfnamefont {T.~C.}\ \bibnamefont {Hensley}},\
  }\href {https://doi.org/10.1103/PhysRevC.100.034308} {\bibfield  {journal}
  {\bibinfo  {journal} {Phys. Rev. C}\ }\textbf {\bibinfo {volume} {100}},\
  \bibinfo {pages} {034308} (\bibinfo {year} {2019})}\BibitemShut {NoStop}%
\bibitem [{\citenamefont {Volya}(2023)}]{cosmo}%
  \BibitemOpen
  \bibfield  {author} {\bibinfo {author} {\bibfnamefont {A.}~\bibnamefont
  {Volya}},\ }\href {https://doi.org/10.5281/zenodo.14198442} {\bibinfo {title}
  {{{CoSMo}}: Shell model code}} (\bibinfo {year} {2023})\BibitemShut {NoStop}%
\bibitem [{\citenamefont {Brown}(2022)}]{Brown2022}%
  \BibitemOpen
  \bibfield  {author} {\bibinfo {author} {\bibfnamefont {B.~A.}\ \bibnamefont
  {Brown}},\ }\href {https://doi.org/10.3390/physics4020035} {\bibfield
  {journal} {\bibinfo  {journal} {Physics}\ }\textbf {\bibinfo {volume} {4}},\
  \bibinfo {pages} {525} (\bibinfo {year} {2022})}\BibitemShut {NoStop}%
\bibitem [{\citenamefont {Brown}\ and\ \citenamefont
  {Richter}(2006)}]{Brown2006}%
  \BibitemOpen
  \bibfield  {author} {\bibinfo {author} {\bibfnamefont {B.~A.}\ \bibnamefont
  {Brown}}\ and\ \bibinfo {author} {\bibfnamefont {W.~A.}\ \bibnamefont
  {Richter}},\ }\href {https://doi.org/10.1103/PhysRevC.74.034315} {\bibfield
  {journal} {\bibinfo  {journal} {Phys. Rev. C}\ }\textbf {\bibinfo {volume}
  {74}},\ \bibinfo {pages} {034315} (\bibinfo {year} {2006})}\BibitemShut
  {NoStop}%
\bibitem [{\citenamefont {Zelevinsky}\ and\ \citenamefont
  {Volya}(2017)}]{Zelevinsky2017}%
  \BibitemOpen
  \bibfield  {author} {\bibinfo {author} {\bibfnamefont {V.}~\bibnamefont
  {Zelevinsky}}\ and\ \bibinfo {author} {\bibfnamefont {A.}~\bibnamefont
  {Volya}},\ }\href@noop {} {\emph {\bibinfo {title} {Physics of Atomic
  Nuclei}}}\ (\bibinfo  {publisher} {Wiley-VCH},\ \bibinfo {year}
  {2017})\BibitemShut {NoStop}%
\bibitem [{\citenamefont {Utsuno}\ \emph {et~al.}(2012)\citenamefont {Utsuno},
  \citenamefont {Otsuka}, \citenamefont {Brown}, \citenamefont {Honma},
  \citenamefont {Mizusaki},\ and\ \citenamefont {Shimizu}}]{Ut12}%
  \BibitemOpen
  \bibfield  {author} {\bibinfo {author} {\bibfnamefont {Y.}~\bibnamefont
  {Utsuno}}, \bibinfo {author} {\bibfnamefont {T.}~\bibnamefont {Otsuka}},
  \bibinfo {author} {\bibfnamefont {B.~A.}\ \bibnamefont {Brown}}, \bibinfo
  {author} {\bibfnamefont {M.}~\bibnamefont {Honma}}, \bibinfo {author}
  {\bibfnamefont {T.}~\bibnamefont {Mizusaki}},\ and\ \bibinfo {author}
  {\bibfnamefont {N.}~\bibnamefont {Shimizu}},\ }\href
  {https://doi.org/10.1103/PhysRevC.86.051301} {\bibfield  {journal} {\bibinfo
  {journal} {Phys. Rev. C}\ }\textbf {\bibinfo {volume} {86}},\ \bibinfo
  {pages} {051301} (\bibinfo {year} {2012})}\BibitemShut {NoStop}%
\end{thebibliography}

\providecommand{\noopsort}[1]{}\providecommand{\singleletter}[1]{#1}%

\end{document}